\documentclass[fleqn]{iopart}

\usepackage{iopams}  
\expandafter\let\csname equation*\endcsname\relax

\expandafter\let\csname endequation*\endcsname\relax

\usepackage{vmargin}
\usepackage{natbib}  
\usepackage{algorithmicx}
\usepackage{algorithm}
\usepackage{graphicx}

\usepackage{amsfonts}
\usepackage{stackrel}
\usepackage{amssymb}
\usepackage{mathrsfs}
\usepackage{algpascal}
\usepackage{color}
\usepackage{epsfig}

\makeatletter
\def\footnoterule{\kern-3\p@
  \hrule \@width 2in \kern 2.6\p@} % the \hrule is .4pt high
\makeatother

\begin{document}

\title[Optimal transport FWI with the multimensional Kantorovich-Rubinstein norm]{Optimal transport in full-waveform inversion: Analysis and practice of the multidimensional Kantorovich-Rubinstein norm}

\author{
J{\'e}r{\'e}mie Messud$^1$, Rapha\"el Poncet$^2$, Gilles Lambar\'e$^1$
}

\address{
$^{1}$ CGG, 27 avenue Carnot, 91341 Massy (France)\\
$^{2}$ Formerly CGG
}
%\ead{submissions@iop.org}

\vspace{10pt}
\begin{indented}
\item[]January 8, 2021
\end{indented}

\begin{abstract}
In the last ten years, full-waveform inversion (FWI) has emerged as a robust and efficient
high-resolution subsurface model-building tool for seismic imaging, with the unique
ability to invert for complex models.
FWI is based on the minimization of a cost function between observed and
modelled data, the data space consisting in collections of time-series.
Originally considering a least-squares cost function, the method suffered from high sensitivity to local minimums and was therefore of poor efficiency in handling large time shifts between
observed and modelled
data events. 
To tackle this problem, a common practice is to
start the inversion using the low temporal frequencies of the data and selecting
specific data events called diving waves. Complementary to this, the use of other cost functions has been investigated.
Among these, cost functions based on optimal transport appeared
appealing to possibly handle large time shifts between observed and modelled
data events. Several strategies inspired by optimal transport have been proposed,
taking into account the specificities of seismic data.
Among them,
the approach based on the Kantorovich-Rubinstein norm offers the possibility
of the direct use of seismic data and an efficient numerical implementation allowing
for a multidimensional (data coordinate space) application.

We present here an analysis of the Kantorovich-Rubinstein norm, discussing its theoretical and
practical aspects. A key component of our analysis is the adjoint-source
or data-space gradient of the cost function
(converted into the model-space gradient within FWI). We highlight its piecewise linearity, analyze its frequency
content and amplitude, and emphasize the benefit of having a multidimensional
implementation. We give practical rules for setting the tuning
parameters. Our set of synthetic and
field data examples demonstrate the improvements brought by the use of the Kantorovich-Rubinstein norm
over least-squares FWI, and highlight the improvements brought
by the multidimensional approach over the one-dimensional one.
\end{abstract}


\submitto{\IP}
%

%\maketitle

%\small
\section{Introduction}

Full-waveform inversion (FWI) was proposed in the early 80's
as a data fitting process aimed at inverting for a subsurface model \cite[]{Lailly1983,Tarantola1984}.
It is based on a cost function between observed and
modelled data, initially the least-squares (LSQ) one, minimized through a non-linear iterative local optimization scheme.
The observed data consists in collections of time-series or ``traces''
(generated by a source and measured at the Earth's surface after propagation in the subsurface), called common shot gathers. The modelled data is computed by solving the wave-equation in a given subsurface model \cite[]{Virieux2009}.

After initial successes, FWI encountered difficulties related to its computational cost and the building of the long spatial wavelength components of the subsurface model from seismic data events called reflected waves
\cite[]{Tarantola2005}.
The investigations continued, moving from the inversion of reflected waves to the inversion of other seismic data events called transmitted and diving waves, and 
starting the inversion using the low temporal frequencies of the data
and frequently an offset (or traces) selection \cite[]{Pratt1990,Pratt1999,Bunks1995,Sirgue2004}. 
From the mid-2000's the successes and potential of these approaches
gave industrial perspectives for complex subsurface imaging
\cite[]{Operto2004,Brenders2007,Virieux2009}.
The computational cost remained a blocking factor in the case of 3D subsurface models, but this was rapidly overcome thanks to specific implementations and the increase in computational capabilities
\cite[]{Vigh2007,Sirgue2009,Plessix2009}.
Since the late-2000's, FWI has been established as an essential component of the subsurface (or ``velocity'') model building toolbox used in the seismic imaging industry. Its capability to recover complex and high-resolution subsurface models has made it a favored tool. 

However, issues still remained. 
Firstly, the LSQ cost function is very sensitive to local minimums because of the oscillatory nature of seismic data, so that the local optimization can easily get trapped into a local minimum
\cite[]{Virieux2009}.
This is called the cycle-skipping problem.
Secondly, the LSQ cost function is very sensitive to amplitudes so that spurious inversion results can be  obtained when the assumptions used in the modelling do not make it possible to ``quantitatively reproduce'' the amplitudes in the data (most implementations of FWI are currently based for example on an acoustic modelling while the subsurface is elastic). 
As mentioned above, starting the optimization from a good initial model, inverting first the low temporal frequency components of the data \cite[]{Bunks1995} and using mainly diving waves reduces the sensitivity to cycle-skipping \cite[]{Tarantola2005},
while data preprocessing (in particular trace normalization)
reduces the sensitivity to amplitude.
However, these ad-hoc strategies 
do not completely solve the problem and require a careful implementation.
As a consequence, a search for complementary solutions, in particular in terms of more suitable cost functions, has emerged as a very active research field.
The expected features of alternative cost functions are:
an increased sensitivity to the kinematic information contained in the data (or time shifts between observed and modelled data events) and a relaxed sensitivity to the amplitude information. 
Many proposals were made around the early-2010's
\cite[]{Shin2008,VanLeeuwen2010,Luo2011,Warner2016},
following some pioneering work \cite[]{Luo1991}.
The first proposal of a cost function inspired from optimal transport (OT) was made by \cite{Engquist2014}.

\cite{Metivier2016} provide the state-of-the-art of OT-based cost functions and their first usage in FWI.
They mention the original work of Gaspard Monge (18th century) aiming at finding the optimal way to transport piles of sand by minimizing the expended energy or cost \cite[]{Monge1781}.
Mathematically, the distribution of masses of initial and final piles of sand are described by probability density functions (PDFs) (positive, with unit ``mass''). The transportation is described as a mapping between the initial and final PDFs. 
Minimizing the transportation cost involves the Monge-Amp\`ere equation in the general case~\cite[]{Villani2008}.
Various reformulations of the original problem exist, that can usually be related to Wasserstein distances.
Regarding FWI, the approach proposed by \cite{Engquist2014,Engquist2016,Yang2018b,Yang2018} and recently applied to field data by \cite{Wang2019} is based on a 2-Wasserstein distance.
Because of the computational cost of the resolution of the Monge-Amp\`ere problem,
the approach currently only affords for a trace-by-trace comparison in the 
2-Wasserstein distance  for large scale applications, not a comparison of a set of traces as in a common shot gather.
%This would not make a difference for the LSQ cost function but makes a difference for OT-based cost functions, as they can
So, the formulation does not use the property of OT-based cost functions to eventually be multidimensional in the data coordinate space
(not to be confused with the dimensionality of the subsurface model),
i.e. to account for correlations between traces (multidimensional) and not only within one trace (one-dimensional).
Also, within this formulation, the requisite for positive values and mass conservation imposes ad-hoc transformations of seismic traces.

In this context, \cite{Metivier2016TLE,Metivier2016,Metivier2016a} proposed an alternative to this first family of OT-based cost functions, allowing the ad-hoc transformations of seismic traces to be bypassed.
They started from the dual formulation of the 1-Wasserstein distance, called the Kantorovich-Rubinstein (KR) formulation, that leads to a maximization problem over 1-Lipschitz functions.
More specifically, they proposed to compute a distance between seismic data 
using the so-called KR norm \cite[]{Hanin1992,Villani2003,Lellmann2014}.
Adding a bounding constraint allows the direct use of the seismic data
without any transformation.
%(getting rid of the need for mass conservation).
An important feature is that a version of the KR norm exists that can be very efficiently computed using the simultaneous descent method of multipliers (SDMM) iterative algorithm \cite[]{Combettes2011,Metivier2016}.
As a consequence, the use of the KR norm is in practice not limited to a trace-per-trace comparison.
A multidimensional formulation in the data coordinate space is feasible for large scale applications,
which can account for correlations in the offset direction between traces in a common shot gather,
which is not the case with an approach based on a 2-Wasserstein distance as \cite{Engquist2014}.
As shown in \cite{Metivier2016TLE} the KR norm and its numerical approximation does not lead to a full convexity when comparing two shifted Ricker wavelets.
It however becomes more convex when the number of iterations of the SDMM algorithm increases (the secondary minima tend to be pushed out) and outperforms the convexity of LSQ while keeping a sharp valley of attraction
(it can tend at convergence to a cost function valley almost twice wider than the LSQ one).
Recently, this approach led to numerous successful industrial FWI applications, see e.g.
\cite{Poncet2018, Messud2019, Sedova2019, Hermant2019, Carotti2020, Hermant2020}.
Compared to LSQ FWI, an interesting reduction in sensitivity to cycle-skipping has been observed, together with an improved structural consistency in inverted subsurface models.
These behaviors can be related to the specific nature of the corresponding ``adjoint-source'' or data-space gradient of the cost function
(converted within FWI into the subsurface model-space gradient through back-propagation \cite[]{Virieux2009}).
Compared to the LSQ FWI adjoint-source, the KR FWI adjoint-source
exhibits an enhancement in low frequencies, balancing of amplitudes and an increase in coherency along the events (i.e. along a direction called ``moveout'') when using the multidimensional formulation \cite[]{Messud2019}.

This paper deals with the use of the KR norm in FWI, in the continuation of the work of \cite{Metivier2016TLE,Metivier2016, Metivier2016a}.
We firstly focus the analysis on the KR adjoint-source.
We clarify theoretical aspects related to the establishment of the expression of the adjoint-source.
We rigorously define the ``texture'' of the KR adjoint-source (piecewise linearity, lower frequency content, reduced amplitude dynamics) and emphasize the benefit of having a multidimensional implementation. 
We demonstrate the interest of the KR adjoint-source for FWI in terms of ``physical'' characteristics,
%frequency content and amplitude balancing, and emphasize the benefit of a multidimensional implementation to improve the coherency of the events.
explain the meaning of the tuning parameters (critical for a successful implementation)
and give a set of practical rules for setting these parameters.
Then, we present a set of synthetic and field data FWI examples demonstrating the effectiveness of the approach over LSQ FWI,
as well as the benefit of the multidimensional KR FWI approach over the one-dimensional one.
Finally, to overcome some limitations, we discuss the possibility to combine the approach with a kinematic transformation applied to data, like the ``graph-space" OT one proposed by~\cite{Metivier2019}.

Before presenting our contributions, we start with reminders 
on FWI and OT formalisms (\S \ref{sec:Theory}).

\section{Theory}
\label{sec:Theory}

\subsection{FWI and the challenge of cycle-skipping}
\label{sec:FWI-prob}

We denote by $\mathbb{R}^{(X)}$ the family of real functions $f:X\rightarrow \mathbb{R}$ indexed on a space $X$ (finite or infinite dimensional) that is measurable for a measure $\mu$.
A seismic survey corresponds to a series of recorded shots,
each shot producing a collection of traces called common shot gather.
The data space related 
to one common shot gather is
%(shot indexes are implicit in the following)
%
\begin{eqnarray}
\label{eq:XY}
&&
D(X)\subset
\mathbb{R}^{(X)}
%\mathbb{R}^{H_{xl}}\times\mathbb{R}^{H_{inl}}\times\mathbb{R}^{T}
.
\end{eqnarray}
It is indexed on $X=[H_{xl}^{min},H_{xl}^{max}]\times[H_{inl}^{min},H_{inl}^{max}]\times[0,T]
\subset \mathbb{R}^3$
that represents the set related to the positions in a common shot data.
$T$ denotes the maximum recording time of each trace.
$H_{xl}^{min},H_{xl}^{max}$ and $H_{inl}^{min},H_{inl}^{max}$ denote the minimum and maximum receiver positions, respectively in the crossline and inline directions, that index the position of the traces within the common shot data.
$x=[x_{xl},x_{inl},x_t]^t\in X$ represents a 3 dimensional (3D) vector, where $^t$ denotes the transpose and $x_{xl}, x_{inl}, x_t$ are scalars.
In the following, we call $X$ the data ``coordinate space'',
considered as continuous, with the associated Lebesgue measure denoted by $\mu(x)$.
However, the considerations in this article generalize to discrete sets $X$ simply considering $\mu(x)$ to be the counting measure
\footnote[1]{
\samepage
For instance taking
%$[1,N_{xl}]\times[1,N_{inl}]\times[1,N_t]\subset \mathbb{N}^3$ and
$X=\{H_{xl}(i_{xl}),i_{xl}=1..N_{xl}\}\times\{H_{inl}(i_{inl}),i_{inl}=1..N_{inl}\}\times\{T(i_{t}),i_{t}=1..N_{t}\}$, with 
$H_{xl}(1)=H_{xl}^{min}$, $H_{xl}(N_{xl})=H_{xl}^{max}$, 
$H_{inl}(1)=H_{inl}^{min}$, $H_{inl}(N_{inl})=H_{inl}^{max}$,
$T(0)=0$ and $T(N_t)=T$.
$N_{xl}\times N_{inl}$ then represents the total number of traces in the considered common shot data.
$N_{xl}\times N_{inl}\times N_t$ represents the total number of samples in the common shot data, i.e. the dimensionality of the data space $D(X)$, typically equal to $10^6-10^9$.
The associated counting measure is defined by
$
\mu(x)
=
\sum_{i_{xl}=1}^{N_{xl}}
\sum_{i_{inl}=1}^{N_{inl}}
\sum_{x_t=1}^{N_t}
\delta(x_{xl}-H_{xl}(i_{xl}))
\delta(x_{inl}-H_{inl}(i_{inl}))
\delta(x_t-T(i_t))
$.
}.

A seismic observed shot data, here considered as a scalar field measured at the Earth's surface during a seismic experiment, is described by $f_{obs}:X\rightarrow \mathbb{R}$,
$f_{obs}\in D(X)$.
In FWI, modelled shot data is produced by solving the wave-equation in a given subsurface model $m\in M$,
where $M$ denotes the model space
\footnote{
Typically a grid with $10^6-10^8$ samples.
},
and extracting the result at the Earth's surface.
Such a modelled shot data is described by $f[m]:X\rightarrow \mathbb{R}$, $f[m]\in D(X)$.
The FWI problem consists in finding $m^*\in M$
which best explains all the observed common shot data,
i.e. makes  $f[m^*]$ as close as possible to $f_{obs}$ for all shots from the point of view of a given similarity measure in the data space \cite[]{Tarantola2005}.
We denote such a similarity measure, also called cost function, by a functional
$J:[{D}(X)\times {D}(X)]^{N_{shots}}\rightarrow\mathbb{R}^+$, where $N_{shot}$ represents the number of shots; the latter dependency is implicit in the following to lighten the notations.
We resolve
\begin{equation}
m^\ast = \stackrel[m \in {M}]{}{\mathrm{argmin}} \quad J(f[m], f_{obs})
.
\end{equation}
The model space dimensionality requires
to use an iterative local optimization scheme, based on the 
model-space gradient, i.e.
$
\partial J(f[m], f_{obs})/\partial m
=
\int_X
\partial f[m](x)/\partial m
\times
\partial J(f, f_{obs})/\partial f(x)\big|_{f=f[m]}
d\mu(x)
$.
Firstly, a forward propagation consists of resolving the wave-equation to compute
the forward-propagated wavefield, extracted at the Earth's surface to give $f[m]$.
Secondly, the gradient in the data-space or ``adjoint-source'' is computed, i.e.
\begin{equation}
\frac{\partial J(f, f_{obs})}{\partial f(x)}\Big|_{f=f[m]}
,
\label{eq:adj-s}
\end{equation}
that represents the sensitivity of the cost function to the data.
Thirdly, the adjoint-source is ``converted'' into the model-space gradient $\partial J(f[m], f_{obs})/\partial m$ .
To that aim, as the dimensionality of the problem (in data and model spaces) 
makes a $\partial f[m](x)/\partial m$-based conversion unfeasible,
FWI uses the adjoint-state method \cite[]{plessix-2006-adjoint,Virieux2009}.
The adjoint-source is back-propagated using the time-reversed wave-equation
to obtain the back-propagated wavefield.
Next, the forward-propagated wavefield and the second temporal derivative of the 
back-propagated wavefield are zero-lag cross-correlated in the time direction
(when the wave-equation is considered in the time domain).
This gives as a result the 
model-space gradient \cite[]{plessix-2006-adjoint}.
The latter may then be preconditioned and finally a line-search is applied to give the correct magnitude to the resulting descent direction.
These are the main features of the FWI descent direction computation \cite[]{plessix-2006-adjoint,Virieux2009}.
The important point for this article is that
the choice of the cost function $J$ affects the method directly only through the adjoint-source computation,
eq. (\ref{eq:adj-s}).
So, to understand the interest of a given cost function,
interpreting how the adjoint-source is impacted is a key.

Common FWI cost functions (and thus adjoint-source forms) are based on $L_p$ norms ($p\ge 1$) through
\begin{eqnarray}
%&&
%\forall f\in :
%\nonumber\\
&&
J(f[m], f_{obs})
=
||f[m] - f_{obs}||_{p}^p
\quad\mathrm{with}\quad
||f||_{p}
=
\Big(
\int_{X}
|f(x)|^p d\mu(x)
\Big)^{1/p}
\label{eq:norm_pixels_2}\\
&&
\Rightarrow
\frac{\partial J(f, f_{obs})}{\partial f(x)}
=
\mathrm{sign}(f(x)-f_{obs}(x))\times |f(x)- f_{obs}(x)|^{p-1}
,
\nonumber
\end{eqnarray}
for $D(X)\subseteq L^p(X)$ where
$L^p(X)$ denotes the space of functions whose moments of order $p$ are integrable.
At the limit $p\rightarrow\infty$, the ``uniform norm'' is defined by $||f||_{\infty}=\max_{x\in X} |f(x)|$.
The $p=2$ case leads to the most common FWI cost function, called least-squares (LSQ).
The LSQ adjoint-source is equal to $f[m]-f_{obs}$ and called the ``residual''.

Let us come back to the (data) coordinate space $X$.
As already mentioned, the dimensionality of $X$ is $3$ for common shot data (time direction, and receiver crossline and inline directions).
However, if an application utilizes the data trace-per-trace independently,
the {effective} dimensionality of $X$ becomes $1$ (time direction only enters into the computation).
If an application utilizes the data inline-per-inline independently,
the {effective} dimensionality of $X$ becomes $2$ (inline and time directions only enter into the computation).
Hereafter, when we mention multidimensionality (multiD), we refer to the effective dimensionality of the (data) coordinate space $X$
(not to be confused with the dimensionality of the data space $D(X)$ or of the model space).

It is obvious from from eq. (\ref{eq:norm_pixels_2}) that
$L_p$-based cost functions measure a point-wise similarity
so that their effective dimensionality in the coordinate space is $0$;
in other words, no correlation between different positions in the time-direction are explicitly accounted for in the cost function
(of course, some correlations are implicitly accounted for by inverting first the low temporal frequency components of the data and progressively adding higher frequencies).
This is a fundamental reason for the high sensitivity to cycle-skipping of the $L_p$-based cost functions.
%% i.e. there are many local minima  where the local optimization can get trapped in the $L_p$-based cost function valleys \cite[]{Virieux2009}.
%Reducing this issue requires to start the optimization from a good initial model and often careful data selection and processing
%(for instance starting from the low frequencies~\cite[]{bunks-1995}, using diving waves only \cite[]{Tarantola2005}...).
%
Another potential issue is that these cost functions tend to be very sensitive to amplitudes;
%This may lead to
%spurious inversion results when the amplitudes are not fundamentally describable by the modelling or outlayers are present in the data.
this sensitivity can somewhat be reduced by making $p$ very close to $1$.
Indeed, all amplitudes are equalized in the $L_1$ adjoint-source, $\mathrm{sign}(f[m]-f_{obs})$, eq.~(\ref{eq:norm_pixels_2}).
But this can lead to convergence difficulties and is still plagued by cycle-skipping.

%A major challenge consists in finding cost functions:
%\begin{itemize}
%\item
%That would enlarge the valley of the global minimum,
%i.e. make the problem more linear,
%to allows to start iterating from models that are more ``distant" from $m^*$ with less data selection and processing.
%\item
%That can alleviate the amplitude dependency in a controllable way.
%\item
%Whose related free (or user-defined) parameters are few and physically interpretable.
%\end{itemize}

In contrast, OT-based cost functions are explicitly at least 1D in the coordinate space, 
allowing them to explicitly account for correlation at least in the time direction,
i.e. between an observed full trace and the corresponding modelled full trace.
This provides more sensitivity to time shifts between the data events, thus more robustness to cycle-skipping
\cite[]{Engquist2014, Yang2018, Metivier2016a, Poncet2018, Messud2019}.
Furthermore, these cost functions can even be multiD (2D or 3D) in the coordinate space,
allowing the exploitation of the correlations between many observed traces 
and the corresponding modelled traces,
i.e. the coherency in the moveout direction;
%\cite[]{Metivier2016a, Poncet2018, Messud2019};
also, OT-based cost functions are less sensitive to amplitudes
\cite[]{Metivier2016a, Poncet2018, Messud2019}.
They thus seem a very good candidate to overcome the limitations of $L_p$-based cost functions.

However, there are two possible obstacles to the application of OT
to large scale FWI problems.
OT is originally designed to compare probability distributions,
not signed and oscillatory functions like the seismic data.
Also, the OT algorithm originally proposed for FWI \cite[]{Engquist2014} is 
for now viable in an industrial context only for a 1D coordinate space,
i.e. trace-per-trace comparison.
In the next section, we review 
the general OT formalism and a formulation recently proposed by \cite{Metivier2016, Metivier2016a} that
can overcome these obstacles. 

\subsection{From optimal transport to the Kantorovich-Rubinstein norm}
\label{sec:ot-rem}

We plan to use OT within the FWI framework. The latter is deterministic
whereas OT is originally a statistical formalism that measures
a similarity between probability distributions~\cite[]{Ambrosio2005}.
So, let us start with some probability-related considerations.
We consider probability distributions that are absolutely continuous with respect to the measure $\mu(x)$.
%that represents a slight restriction to the most general OT formalism.
These probability distributions can be represented by PDFs, denoted by $\rho:X\rightarrow \mathbb{R}^+$.
They have a unit ``mass'': $\int_X \rho(x)d\mu(x)=1$.

We provide the coordinate space $X$ with a metric $d: X\times X\rightarrow \mathbb{R}^+$.
We consider $d^p(x,y)$ to represent the cost of displacing information from $x$ to $y$ in the coordinate space.
%coordinate space ``costs'' given by a power of the distance $d(x,y)$,
$\mathcal{P}^p(X)$
denotes the space of PDFs with finite $d$-moment of order $p$,
i.e. $\int_X d^p(x',x) \rho(x)d\mu(x) < +\infty$
for some (thus any) $x'\in X$ \cite[]{Villani2008}.
OT allows the computation of 
a cost of ``transporting'' the PDF $\rho_1\in\mathcal{P}^p(X)$ onto the PDF $\rho_2\in\mathcal{P}^p(X)$
based on the coordinate space cost $d^p$.
Kantorovich's formulation represents the most general OT form \cite[]{Kantorovich1942,Villani2008,Ambrosio2009}:
\begin{eqnarray}
W_{d^p}^p(\rho_1,\rho_2)
=
\min_{\pi\in\Pi(\rho_1,\rho_2)}
\int_{X\times X}
d^p(x,y)\pi(x,y)d\mu(x) d\mu(y)
,
\label{eq:OT_2}
\end{eqnarray}
where $\pi$ represent ``transference plans''
belonging to the set of PDFs on $X\times X$ that have marginals $\rho_1$ and $\rho_2$:
\begin{eqnarray}
\Pi(\rho_1,\rho_2)
=
\Big\{
&&
\pi\in \mathcal{P}^p(X\times X);
\label{eq:OT_1}\\
&&
\forall x\in X:
\int_X \pi(x,y) d\mu(y) = \rho_1(x)
,
\forall y\in {Y}:
\int_X \pi(x,y) d\mu(x) = \rho_2(y)
\Big\}
.
\nonumber
\end{eqnarray}
$\pi$ contains information on how to transport $\rho_1$ onto $\rho_2$.
%$d^p(x,y)$ represents the cost of displacing information from $x$ to $y$ in the $X$ (pixel) space. 
$W_{d^p}(\rho_1,\rho_2)$ defines for $p\ge 1$ a distance between the PDFs $\rho_1$ and $\rho_2$,
% in the $\mathcal{P}^p(X)$ space,
called ``$p$-Wasserstein" distance for the metric $d$.
Eq.~(\ref{eq:OT_2}) represents
a linear programming problem (minimization of a linear functional with 
linear equality constraints).
Despite the linearity, this problem can easilly become intractable because it implies a search in the space $\mathcal{P}^p(X\times X)$ that has large dimensionality in the seismic case.
% (remind section \ref{sec:FWI-prob}).
So, let us discuss more tractable reformulations.

Monge's problem (the original OT formulation) is a restriction of Kantorovich's problem to transference plans of the form
$\pi(x,y)=\rho_1(x)\delta(y-T(x))$ where $T:X\rightarrow X$ is an invertible function, giving \cite[]{Villani2008,Ambrosio2009}:
%(keeping the $W_{d^p}^p$ notation by slight abuse of notation) \cite[]{Monge1781,Villani2008}
%
\begin{eqnarray}
\widehat{W}_{d^p}^p(\rho_1,\rho_2)
=
\min_{T\in\Sigma(\rho_1,\rho_2)}
\int_{X}
d^p(x,T(x))\rho_1(x)d\mu(x)
,
\label{eq:OT_3}
\end{eqnarray}
where the map $T$ is constrained to transport $\rho_1$ onto $\rho_2$:
\begin{eqnarray}
\Sigma(\rho_1,\rho_2)
=
\Big\{
T:X\rightarrow X;
%T\#\rho_1=\rho_2
\forall x\in X: \rho_1[T^{-1}(x)]=\rho_2(x)
\Big\}
.
\label{eq:OT_4}
\end{eqnarray}
Monge's problem represents a restriction of Kantorovich's problem
and cannot always be solved.
%
%Any solution of the Monge problem
%is a solution of the Kantorovich problem but, contrarily to the Kantorovich problem, the Monge problem
%The Kantorovich formulation thus generalizes the Monge one.
%
However, it implies a search in the space of bijective functions defined on $X$,
that is smaller than the Kantorovich search space.
Despite this advantage, the constraint in eq. (\ref{eq:OT_4}) is non-linear and difficult to enforce as soon as the coordinate space is multiD,
because it then involves resolving the Monge-Amp\`ere equation~\cite[]{Villani2008}.
In practice, only the 1D coordinate space version
has been used for industrial FWI applications, considering a trace-per-trace comparison,
i.e. explicit correlations only in the time-direction.
Used together with a transformation of each seismic trace into a PDF, here called a ``PDF-transformation'',
this formulation proved to reduce the cycle-skipping issue in FWI
\cite[]{Engquist2016, Yang2018b, Yang2018,Wang2019}.
The limitations of this scheme are that in practice it does not allow the exploitation of multidimensionality in the coordinate space and that using ``PDF-transformations'' may lead to difficulties with field data \cite[]{Metivier2018}.
%a loss of information present in seismic data and to give same mass to data with different number of events \JMcomm{(need Refs.)}.

A simplification of the Kantorovich problem occurs
when the metric $d$ is lower semi-continuous and $p=1$.
This allows to switch to the Kantorovich-Rubinstein (KR) dual formulation, defined
for $\rho_1$ and $\rho_2$ in the $\mathcal{P}^1(X)$ space by
\cite[]{Villani2008}:
\begin{eqnarray}
W_d(\rho_1,\rho_2)
=
\max_{\varphi\in \mathrm{Lip}(d,1)}
\int_{X}
\varphi(x)(\rho_1(x)-\rho_2(x))d\mu(x)
,
\label{eq:OT_5}
\end{eqnarray}
where we denote by $\mathrm{Lip}(d,\alpha)$ the set of
$\alpha$-Lipschitz functions on $X$
with respect to the distance $d$:
\begin{eqnarray}
\mathrm{Lip}(d,\alpha)
=
\Big\{
\varphi:X\rightarrow \mathbb{R};
\max_{x\ne y}\frac{|\varphi(x)-\varphi(y)|}{d(x,y)}\le \alpha,
\varphi\in L^1\big(|\rho_1-\rho_2|d\mu\big)
\Big\}
.
\label{eq:OT_6}
\end{eqnarray}
$\varphi$ is constrained to be ``slowly'' varying,
the absolute value of its ``local slopes'' being bounded by $\alpha=1$.
The KR problem implies a search in the space of 1-Lipschitz functions on $X$
(that must be integrable for the measure $|\rho_1-\rho_2|d\mu$, hence belong to $L^1\big(|\rho_1-\rho_2|d\mu\big)$).
The Lipschitz constraint can be recast into linear constraints and leads to a linear programming problem that is numerically manageable even in the multiD case.
Another nice feature of this formulation is that it is well defined not only for $(\rho_1,\rho_2)\in \mathcal{P}^1(X)\times\mathcal{P}^1(X)$, but also for any function $(f_1,f_2)\in L^1(X)\times L^1(X)$ provided
they have the same ``mass", i.e. $\int_{X}f_1(x)d\mu(x)=\int_{X}f_2(x)d\mu(x)$.
This is still too constraining for seismic data 
(even if oscillating, seismic data cannot be considered to have approximately null mass within FWI,
especially as many applications use the low frequencies and mutes).
When $f_1$ and $f_2$ do not have same mass, 
the $\varphi$ inverted in eq. (\ref{eq:OT_6}) becomes singular;
%A simple solution to remed\mu(y) this problem
%%
%%Apply a transformation to rescale the observed and modelled data masses (not necessarilly making them positive)
%%
%is to add a supplementary constraint to avoid the singularity,
%i.e. a $||\varphi||_\infty<\lambda$ term to prevent large $\varphi$ values.
%
%to remove the mass restriction and avoid a singularity,
%(singularities in $\varphi$),
a solution is to add a bounding constraint to $\varphi$
\cite[]{Hanin1992,Lellmann2014}.
Introducing the ``$\lambda$-bounded $\alpha$-Lipschitz'' set
\begin{eqnarray}
\mathrm{BLip}(d,\alpha,\lambda)
=
\Big\{
&&
\varphi:X\rightarrow \mathbb{R};
\label{eq:OT_7}\\
&&
\max_{x\ne y}\frac{|\varphi(x)-\varphi(y)|}{d(x,y)}\le \alpha,
||\varphi||_{\infty}\le\lambda,
\varphi\in L^1\big(|f_1-f_2|d\mu\big)
\Big\}
,
\nonumber
\end{eqnarray}
we now resolve for any function $(f_1,f_2)\in L^1(X)\times L^1(X)$
\begin{eqnarray}
\tilde{W}_d(f_1,f_2)
=
\max_{\varphi\in \mathrm{BLip}(d,1,\lambda)}
\int_{X}
\varphi(x)(f_1(x)-f_2(x))d\mu(x)
.
\label{eq:OT_8}
\end{eqnarray}
The absolute value of the ``local slopes'' of $\varphi$ is as above constrained to be bounded by $\alpha$,
and in addition the absolute value of $\varphi$ is constrained to be thresholded by $\lambda$.
$\tilde{W}_d$ still defines a distance between $f_1$ and $f_2$,
induced by a norm called the ``KR norm'';
$\tilde{W}_d$ represents a generalization of ${W}_d$ and
some relationship between both can be found in
\cite[]{Hanin1992,Villani2003,Lellmann2014}.
Solving for eq. (\ref{eq:OT_8}) is still a linear programming problem.

Let us now consider coordinate space distances $d$ induced by a norm,
i.e. $d(x,y)=||x-y||$;
$(X,d)$ then becomes a Banach space \cite[]{Rudin1991,Brezis1983}.
$d$ is typically induced by a Mahalanobis-like $L_p^{(X)}$ norm ($p\ge 1$)
\footnote[1]{
The coordinate space $L_p^{(X)}$-norm $|| x||_p^{(X)}$, eq.~(\ref{eq:lq_omega}), is not to be confused with the data space $L_p$-norm $||f||^{ }_{p}$, eq.~(\ref{eq:norm_pixels_2}), hence the superscript $(X)$ to make this explicit.
}:
\begin{eqnarray}
\label{eq:lq_omega}
||x||\rightarrow|| x||_p^{(X)}
=
%\Big(
%\sum_{i=1}^{N}
%\frac{1}{\sigma_i^p}|x_i|^p
%\Big)^{1/p}
\Big(
\frac{|x_{xl}|^p}{\sigma_{xl}^p}
+
\frac{|x_{inl}|^p}{\sigma_{inl}^p}
+
\frac{|x_t|^p}{\sigma_t^p}
\Big)^{1/p}
.
\end{eqnarray}
%
%where the 3D coordinate space case has been considered.
The $\sigma$, with $\infty>1/\sigma>0$, denote standard-deviation-like weights
that can account for uncertainties and rescale the different physical dimensions between crossline and inline positions,
and time.
At the limit $p\rightarrow\infty$, the coordinate space ``uniform norm'' is defined by $||x||_{\infty}^{(X)}=\max(\frac{|x_{xl}|}{\sigma_{xl}},\frac{|x_{inl}|}{\sigma_{inl}},\frac{|x_t|}{\sigma_t})$.

\begin{figure}[H]
\centering
\includegraphics[width=0.5\linewidth]{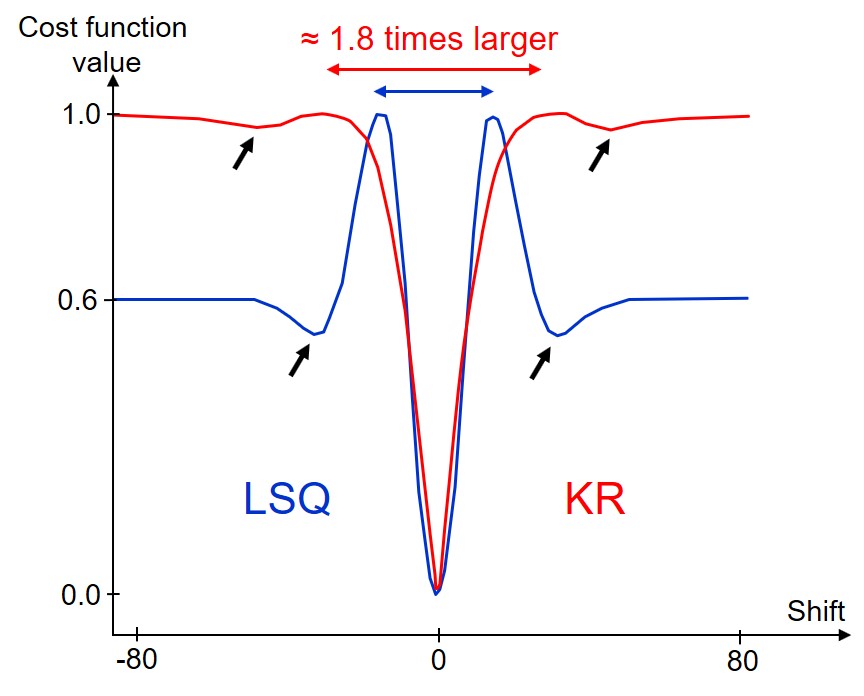}
\caption{
Comparison of the KR norm and LSQ cost function values for 2 shifted Ricker wavelets, the horizontal axis being the shift value. The global minimum valley is almost twice the width
(even if the number of local minima is the same, two, highlighted by the black arrows), which is very interesting for FWI especially when dealing with the low frequencies.
}
\label{fig:fig-kr-loss_valley}
\end{figure}

The use of the KR norm within FWI has been proposed by \cite{Metivier2016TLE,Metivier2016, Metivier2016a},
with the specific choice $d\rightarrow ||.||_1^{(X)}$ that they argued to be the most interesting computationally speaking; this choice will be implicit in the following.
Theses authors 
proposed to use the SDMM algorithm to efficiently resolve the problem and underlined the many interesting properties of the KR norm.
It allows the direct use of seismic data
without application of a PDF-transformation
as required by the approaches 
derived from \cite{Engquist2014}.
Reduced sensitivity to cycle-skipping 
and amplitudes are satisfyingly achieved,
mostly thanks to the 1-Lipschitz constraint that brings low frequencies and tends to equalize the amplitudes in the adjoint-source
\cite[]{Metivier2016,Metivier2016a,Poncet2018,Messud2019}.
Fig. \ref{fig:fig-kr-loss_valley} illustrates that the KR norm global minimum valley
outperforms the convexity of LSQ, giving a global minimum valley that is almost twice the width (even if the number of local minima, i.e. two, is the same).
So far, it has been the only OT-based scheme implemented numerically in multiD in an industrial context,
allowing a study of the advantage of taking into account the correlations of the events within common shot data;
%\cite[]{Poncet2018, Messud2019, Sedova2019};
also, the corresponding tuning parameters are few and have a clear physical interpretation \cite[]{Poncet2018, Messud2019}.
Before detailing these points, we firstly clarify formal aspects related to the 
KR norm adjoint-source computation and precisely define the KR norm adjoint-source ``texture'',
that represent the keys to better understanding the interest regarding the cycle-skipping problem.

\section{Analysis of the exact KR FWI adjoint-source}

\subsection{Form of the adjoint-source}
\label{sec:ot-as}

We consider a given shot data $f_{obs}\in L^1(X)$.
Eq. (\ref{eq:OT_8}) with $d\rightarrow ||.||_1^{(X)}$ can be rewritten for all $f\in L^1(X)$
\begin{eqnarray}
&&
\tilde{W}_d(f,f_{obs})
=
\int_{X}
\varphi_{max}[f](x)(f(x)-f_{obs}(x))d\mu(x)
\label{eq:OT_10}\\
&&
\varphi_{max}[f]=
\stackrel[\varphi\in \mathrm{BLip}(||.||_1^{(X)},1,\lambda)]{}{\mathrm{argmax}}
\int_{X}
\varphi(x)(f(x)-f_{obs}(x))d\mu(x)
.
\nonumber
\end{eqnarray}
The KR adjoint-source, eq. (\ref{eq:adj-s}), is thus defined by
\begin{eqnarray}
\frac{\partial \tilde{W}_d(f,f_{obs})}{\partial f(x)}
%&=&
%\frac{\partial }{\partial f(x)}
%\int_{X}
%\varphi_{max}[f](x')(f(x')-f_{obs}(x'))d\mu(x)'
%\nonumber\\
&=&
\varphi_{max}[f](x)
+
\int_{X}
\frac{\partial \varphi_{max}[f](x')}{\partial f(x)}
(f(x')-f_{obs}(x'))d\mu(x')
.
\label{eq:OT_11}
\end{eqnarray}
The second term, $\int_{X}\frac{\partial \varphi_{max}[f](x')}{\partial f(x)}
(f(x')-f_{obs}(x'))d\mu(x')$, 
is defined provided $\varphi_{max}[f]$ is differentiable
and
has been neglected in \cite{Metivier2016TLE,Metivier2016,Metivier2016a}.
In this section, we demonstrate that
the differential $\partial \varphi_{max}[f](x')/\partial f(x)$ exists and can be neglected, 
fundamentally justifying the following form for the KR adjoint-source:
\begin{eqnarray}
\label{eq:OT_12bis}
\frac{\partial \tilde{W}_d(f,f_{obs})}{\partial f(x)}
=
\varphi_{max}[f](x)
.
\end{eqnarray}
To achieve this goal, it is sufficient to prove
\begin{eqnarray}
&&
\exists \gamma >0,
\forall h\in L^1(X) \mathrm{\hspace{1.5mm}such\hspace{1.5mm}as\hspace{1mm}} ||h||_1 < \gamma:
%, \forall x\in X_{sub}:
\quad
\varphi_{max}[f+h](x)
=
\varphi_{max}[f](x)
,
\label{eq:OT_12_0}
\end{eqnarray}
where the equality needs to hold almost everywhere ``only'' and
\begin{eqnarray}
\label{eq:OT_12}
&&
\varphi_{max}[f+h]
=
\stackrel[
\varphi\in \mathrm{BLip}(||.||_1^{(X)},1,\lambda)
]{}{\mathrm{argmax}}
\int_{X}
\varphi(x)(f(x)-f_{obs}(x)+h(x))d\mu(x)
.
\end{eqnarray}
In other terms, if $\varphi_{max}$ is invariant under any infinitesimal
perturbations of the residual $f-f_{obs}$,
the differential $\partial \varphi_{max}[f](x')/\partial f(x)$ exists and is null.
Intuitively, as we consider a maximization problem under constraints,
eq. (\ref{eq:OT_12_0}) would not be satisfied where the problem ``hesitates'' between different constraints saturations, so that infinitesimal perturbations of the residual may
lead to ``jumps'' in the inverted $\varphi_{max}$.
Our following goal is to qualify subsets where this behavior can occur, i.e. where eq. (\ref{eq:OT_12_0})
cannot be satisfied nor the differential defined.

To simplify, we firstly define a subset $X_{sub}\subseteq X$ where
it is the most pertinent to demonstrate eq.~(\ref{eq:OT_12_0}), called the subset that ``matters''.
%must be the case
%as the spatial variations of $\varphi_{max}$ should be quite driven by the sign of the residual (due to the maximisation principle in eq. (\ref{eq:OT_10})) and the $\mathrm{BLip}(d,1,\lambda)$ constraint should be able to saturate at most positions.
%Difficulties can occur only where infinitesimal perturbations of the residual
%
%Introducing 
%$\forall f\in L^1(X): X_{(f>0)}=\{ x\in X; f(x)>0 \}$, $X_{(f<0)}=\{ x\in X; f(x)<0 \}$
%and $X_{(f=0)}=\{ x\in X; f(x)=0 \}$,
%we have
%%
%\begin{eqnarray}
%X_{(f-f_{obs}+h>0)}
%\stackrel{||h||_1^{(X)}\rightarrow 0}{\longrightarrow}
%X_{(f-f_{obs}>0)}
%,\quad
%X_{(f-f_{obs}+h<0)}
%\stackrel{||h||_1^{(X)}\rightarrow 0}{\longrightarrow}
%X_{(f-f_{obs}<0)}
%,
%\end{eqnarray}

\subsection{The subset that ``matters''}
\label{sec:ot-as-2}

To simplify the following notations, we denote the residual by
\begin{eqnarray}
\Delta f = f-f_{obs}
.
\end{eqnarray}
We denote by $X_{null}^{(\Delta f)}$ the subset of $X$ where the residual is infinitesimal:
\begin{eqnarray}
X_{null}^{(\Delta f)}=\{ x\in X;
\forall \varepsilon >0: |\Delta f(x)|\le \varepsilon \}
,
\label{eq:OT_15}
\end{eqnarray}
and by $X_{null'}^{(\Delta f)}\subseteq X_{null}^{(\Delta f)}$ where the spatial gradient of the residual is also infinitesimal:
\begin{eqnarray}
X_{null'}^{(\Delta f)}=\{ x\in X_{null}^{(\Delta f)};
\forall \varepsilon >0: \Big|\Big|\frac{\partial}{\partial x}\Delta f(x)\Big|\Big|_\infty^{(X)}\le \varepsilon \}
%\subset X_{null}
.
\label{eq:OT_17}
\end{eqnarray}

\begin{figure}[H]
\centering
\includegraphics[width=1.1\linewidth]{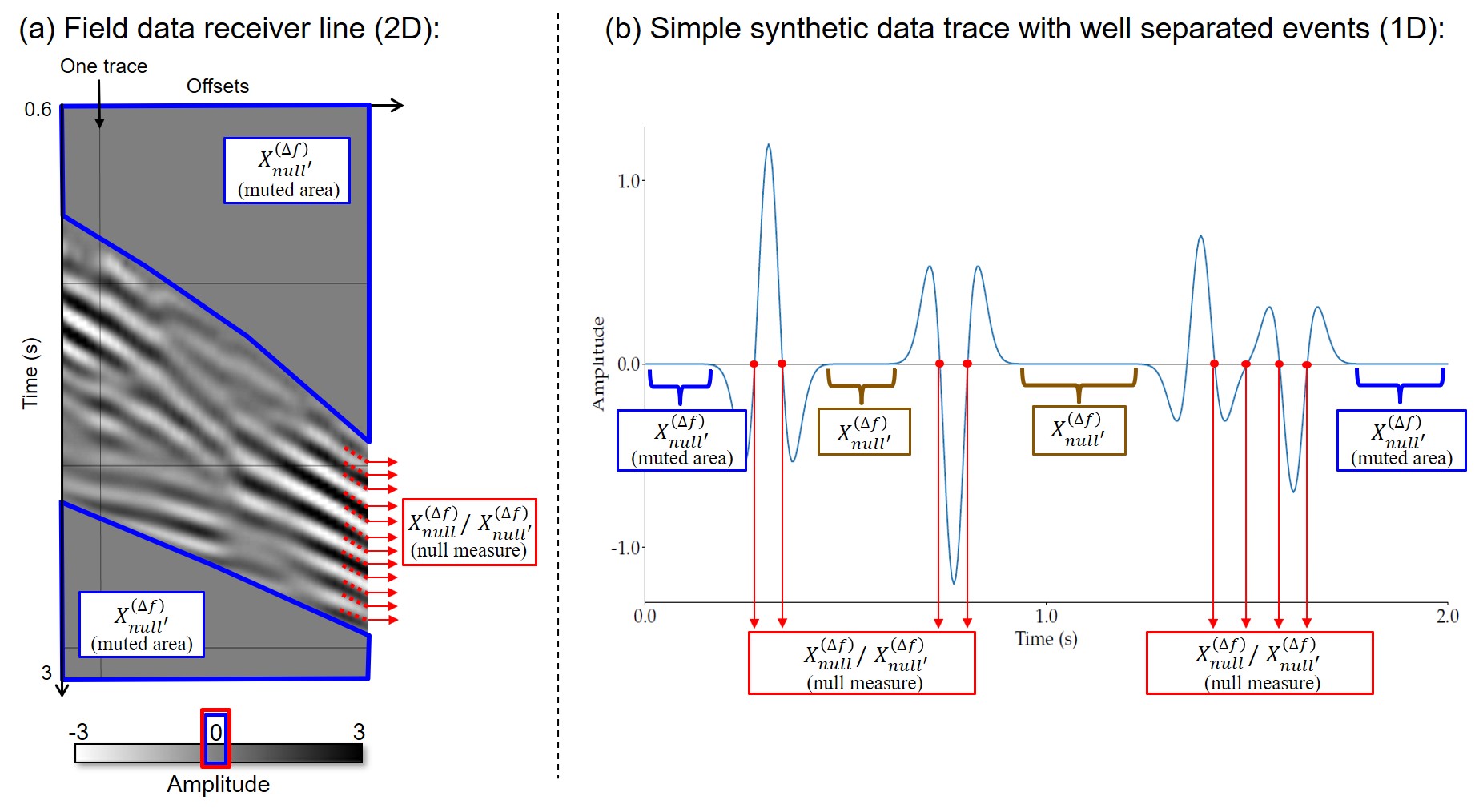}
\caption{
LSQ residual $\Delta f$ for
(a) a field data receiver line (2D) and (b) a simple synthetic data trace (1D)
(where $f_{obs}$ contains two Ricker wavelets and $f$ contains the same Ricker wavelets but shifted).
The various corresponding subsets of $X$ are explicited in each case.
The various contributions to $X_{null'}^{(\Delta f)}$ are represented by the blue and brown areas;
the blue areas contribution corresponds to the muted (or edge) areas;
the brown areas contribution corresponds to ``deaf zones'' which very rarely occur with realistic seismic data.
The various contributions to $X_{null}^{(\Delta f)}\backslash X_{null'}^{(\Delta f)}$ have a null Lebesgue measure in $X$; they tend to be lines in the 2D case (partly represented in (a) by the dashed red curves) and points in the 1D case (indidated in (b) by the red points).
}
\label{fig:fig-subest}
\end{figure}

For realistic seismic data, $X_{null'}^{(\Delta f)}$ mostly corresponds to the area before the first arrival and the muted areas (that help the inversion to concentrate on the most relevant data like the diving waves).
%i.e. ``edge areas'' where $\varphi_{max}$ is usually tapered to zero.
The blue areas in Fig.~\ref{fig:fig-subest}a give an illustration for a field data.

Fig. \ref{fig:fig-subest}b gives an illustration on a simple synthetic trace.
We observe that an additional contribution to the muted areas occurs in $X_{null'}^{(\Delta f)}$, represented by the brown areas, i.e. ``deaf zones'' where the residual stays very small while not on the edges.
These contributions very rarely occur with realistic seismic data (unless specific subsurface structure or acquisition problems),
as signal is usually permanently recorded between minimum and maximum experiment time.

Also, for any seismic data, $X_{null}^{(\Delta f)}\backslash X_{null'}^{(\Delta f)}$ 
(i.e. the complement of $X_{null'}^{(\Delta f)}$ in $X_{null}^{(\Delta f)}$) has a null Lebesgue measure in $X_{null}^{(\Delta f)}$ (thus in $X$) as it ``just'' represents where the oscillatory seismic signal passes through $0$.
Corresponding contributions are partly represented by the dashed red curves in Fig.~\ref{fig:fig-subest}a, for a field 2D data, and by red points in Fig.~\ref{fig:fig-subest}b, for a simple synthetic 1D trace.

An important aspect with realistic FWI data and applications is that what happens 
in the $X_{null}^{(\Delta f)}$ subset has no real importance.
%, whatever the FWI adjoint-source (LSQ residual $\Delta f$ or beyond).
Indeed,  before being back-propagated, any FWI adjoint-source is in practice muted (i.e. equalled to zero with possible rapid tapering) in the muted areas (or ``blue areas''),
and the rest of the $X_{null}^{(\Delta f)}$ contributions (``dashed red curves'') has null measure.
%
%Indeed, the FWI residual $\Delta f$ contains information on the events modelled from the forward-propagated wavefield and extracted at the Earth surface, $f$, as well as on the observed seismic data events, $f_{obs}$.
%The adjoint-source in the general case if computed from $f$ and $f_{obs}$ (and from $\Delta f=f-f_{obs}$ in the KR case using eq.~(\ref{eq:OT_10})) and back-propagated.
%The second temporal derivative of the obtained wavefield is zero-lag cross-correlated in time with the forward-propagated wavefield to obtain the model space gradient as described in \S\ref{sec:FWI-prob}.
%These steps being linear with respect to the adjoint-source, only only its non-null components can contribute,
%which are by construction related to the support of the synthetic events in $\Delta f$.
%This de facto excludes $X_{null'}^{(\Delta f)}$ to contribute to the model space gradient.
%
A consequence for KR adjoint-source considerations it that is sufficient to ensure that eq. (\ref{eq:OT_12_0}) is satisfied $\forall x\in X\backslash X_{null}^{(\Delta f)}$,
the latter subset being called the subset that ``matters'' in the following~
\footnote[1]{
In the following, we will demonstrate formal results on the KR adjoint-source by reasoning in $X\backslash X_{null}^{(\Delta f)}$.
However, from tests on simple synthetic data containing ``deaf zones'', we will observe that the demonstrated results tend to remain true also in the ``brown areas'' contribution to $X_{null}^{(\Delta f)}$.
This is due to the strong continuity constraint in the KR maximization problem.
}.
Note that this subset remains stable 
under infinitesimal perturbations of $\Delta f$.

\subsection{Rewriting the 1-Lipschitz contraint}
\label{sec:ot-as-2}

To ease the next formal considerations, we rewrite the 1-Lipschitz constraint of eq. (\ref{eq:OT_7})
in a form that is valid almost everywhere.
We consider $d$ induced by the $L_1^{(X)}$ norm,
i.e $p=1$ in eq. (\ref{eq:lq_omega}).
\ref{app:proofs_0}
demonstrates that, for a differentiable $\varphi(x)$,
the 1-Lipschitz constraint can be rewritten
\begin{eqnarray}
\label{eq:lip_1D_text0}
&&
\max_{x\ne y}\frac{|\varphi(x)-\varphi(y)|}{||x-y||_1^{(X)}}
=
\max_{x}\Big|\Big|{\frac{\partial\varphi(x)}{\partial x}}\Big|\Big|_{\infty}^{(X)}
\le 1,
\\
&&
\quad\quad\mathrm{with}\quad\quad
\Big|\Big|{\frac{\partial\varphi(x)}{\partial x}}\Big|\Big|_{\infty}^{(X)}
=
\max\Big(
\sigma_{xl}\Big|\frac{\partial\varphi(x)}{\partial x_{xl}}\Big|
,
\sigma_{inl}\Big|\frac{\partial\varphi(x)}{\partial x_{inl}}\Big|
,
\sigma_{t}\Big|\frac{\partial\varphi(x)}{\partial x_{t}}\Big|
\Big)
.
\nonumber
\end{eqnarray}
Thus, when the $L_1^{(X)}$ norm is considered in the 1-Lipschitz constraint,
the corresponding dual norm representation must be considered for the derivative $\partial\varphi(x)/\partial x$,
i.e. the $L_\infty^{(X)}$ norm
with non-inverted ``$\sigma$''.

As the 1-Lipschitz constraint implies strong continuity, thus differentiability of $\varphi(x)$ almost everywhere, and as our next considerations ``just'' need to be valid almost everywhere, we propose to consider $\partial\varphi(x)/\partial x$ in the following.
Then, the 1-Lipschitz constraint implies almost everywhere in $X$
(which is implicit from now)
\begin{eqnarray}
\label{eq:lip_1D_text}
%x\in X:
%\quad
&&
\max\Big(
\sigma_{xl}\Big|\frac{\partial\varphi(x)}{\partial x_{xl}}\Big|
,
\sigma_{inl}\Big|\frac{\partial\varphi(x)}{\partial x_{inl}}\Big|
,
\sigma_{t}\Big|\frac{\partial\varphi(x)}{\partial x_{t}}\Big|
\Big)
\le 1
.
\end{eqnarray}
%
%We will use this form in the following.
%We have all the ingredients to check if eq. (\ref{eq:OT_12_0}) is valid and deduce many more features of the KR adjoint-source.

\subsection{A piecewise linear function}
\label{sec:ot-as-2}

In this section, we formally study the ``shape'' of $\varphi_{max}$.
This will allow us to prove eq.~(\ref{eq:OT_12_0})
but also to draw interesting lessons like rigorously defining a ``texture'' for the KR adjoint-source.
We start by simplified cases before considering the ``full'' case.

\subsubsection{Simplified case (thresholding constraint only)}$\vspace{1mm}$

\noindent
We firstly consider only the thresholding constraint $||\varphi||_{\infty}^{ }\le \lambda$ in eqs. (\ref{eq:OT_7}) and (\ref{eq:OT_10}), i.e. removing the Lipschitz constraint.
\ref{app:proofs} demonstrates that the solution of the problem in eq. (\ref{eq:OT_10}) is then
\begin{eqnarray}
&&
x\in X \backslash X_{null}^{(\Delta f)} \mathrm{\hspace{1.5mm}with\hspace{1.5mm}} \Delta f(x)>0:
\quad
\varphi_{max}(x)=\lambda
\label{eq:proofs_5_text}\\
&&
{\hspace{3.05cm} \Delta f(x)<0:}
\quad
\varphi_{max}(x)=-\lambda
%\nonumber\\
%&&
%{$\forall x\in X_{null}^{(\Delta f)}$: $\varphi_{max}(x)$ values do not matter, remind \S \ref{sec:ot-as}}
.
\nonumber
\end{eqnarray}
This result defines $\varphi_{max}$ in $X \backslash X_{null}^{(\Delta f)}$,
where it clearly remains stable under infinitesimal perturbations of $\Delta f$.
%Only $\varphi_{max}(x\in X_{null}^{(\Delta f)})$ can be affected by such perturbations but this does not matter.
We thus can deduce eq. (\ref{eq:OT_12_0}) in the subset that matters $X\backslash X_{null}^{(\Delta f)}$.

Note that the thresholding constraint saturates everywhere in $X \backslash X_{null}^{(\Delta f)}$,
see Fig. \ref{fig:fig-adj-source-comparison} for an illustration on a simple synthetic trace (shifted Rickers).
%In $X_{null}^{(\Delta f)}$, the sign is not defined, i.e. problem ``hesitates'' between different constraints saturations.
Note also that the solution of eq. (\ref{eq:proofs_5_text}) can be rewritten
\begin{eqnarray}
x\in X \backslash X_{null}^{(\Delta f)}: \varphi_{max}[f](x)=\lambda{\hspace{1.5mm}\mathrm{sign}}(f(x)-f_{obs}(x))=\lambda\frac{\partial }{\partial f(x)} ||f-f_{obs}||^{ }_1
.
\end{eqnarray}
This simplified case is thus equivalent to the $L_1$ norm cost function case,
up to a multiplicative factor $\lambda$ that has no impact because of the line-search
that will recompute the global scaling.

\subsubsection{Intermediary case (1-Lipschitz constraint only)}
\label{sec:ot-as-2}$\vspace{1mm}$

\noindent
We now consider only the 1-Lipschitz constraint in eqs. (\ref{eq:OT_7}) and (\ref{eq:OT_10}), i.e. remove the thresholding constraint, which implies that the residual satisfies
\begin{eqnarray}
\label{eq:res_text}
\int_X\Delta f(x)d\mu(x)=0
\end{eqnarray}
to avoid singularities, remind \S \ref{sec:ot-rem}.

Considering firstly a 1D coordinate space $X=[0,T]$, in the time direction, eq. (\ref{eq:lip_1D_text}) implies 
$\forall x_t\in X:\sigma_t\big|{\frac{\partial\varphi(x_t)}{\partial x_t}}\big|\le 1$.
\ref{app:proofs} demonstrates that the solution of the problem in eq.~(\ref{eq:OT_10}) is then
\begin{eqnarray}
&&
{x_t\in X \backslash X_{null}^{(\Delta F_{\beta})} \mathrm{\hspace{1.5mm}with\hspace{1.5mm}} \Delta F_{\beta}(x_t)<0:}
\quad
\beta^+(x_t)=-\Delta F_{\beta}(x_t),\beta^-(x_t)=0
,\sigma_t\frac{\partial\varphi_{max}(x_t)}{\partial x_t}=1
\nonumber\\
%\label{eq:proofs_5b}\\
&&
{\hspace{3.4cm} \Delta F_{\beta}(x_t)>0:}
\quad
\beta^-(x_t)=\Delta F_{\beta}(x_t),\beta^+(x_t)=0
,\sigma_t\frac{\partial\varphi_{max}(x_t)}{\partial x_t}=-1
,
\nonumber\\
\label{eq:proofs_5b_text}
\end{eqnarray}
where
%
%\footnote{
%The effect of the constant $\beta^+(0)-\beta^-(T)$ is 
%to correct for the ``change of phase'' in $\Delta F_{\beta}$ due to the integral,
%i.e. to ``shift''
%$\varphi_{max}$ so that it ``follows'' the average phase of $\Delta f$.
%}
%
\begin{eqnarray}
\label{eq:proofs_4b_text}
\Delta F_{\beta}(x_t)=\int_{0}^{x_t}\Delta f(x)d\mu(x)
-\Big(\beta^+(0)-\beta^-(0)\Big)
\end{eqnarray}
defines the set $X_{null}^{(\Delta F_{\beta})}$ through eq. (\ref{eq:OT_15}).
Eq. (\ref{eq:proofs_5b_text}) contributes to define $\varphi_{max}$ in $X \backslash X_{null}^{(\Delta F_{\beta})}$ but implies the resolution of a self-consistent problem.
Indeed, $\beta^+$, $\beta^-$ and $\varphi_{max}$ are coupled through $\Delta F_{\beta}$,
but also through the necessary continuity of $\varphi_{max}$ in the whole set $X$,
i.e. also in $X_{null}^{(\Delta F_{\beta})}$.
%and the maximization of $\int_X\varphi_{max}(x_t)\Delta f(x_t)d\mu(x_t)$.

\begin{figure}[H]
\centering
\includegraphics[height=7.0cm,width=12.0cm]{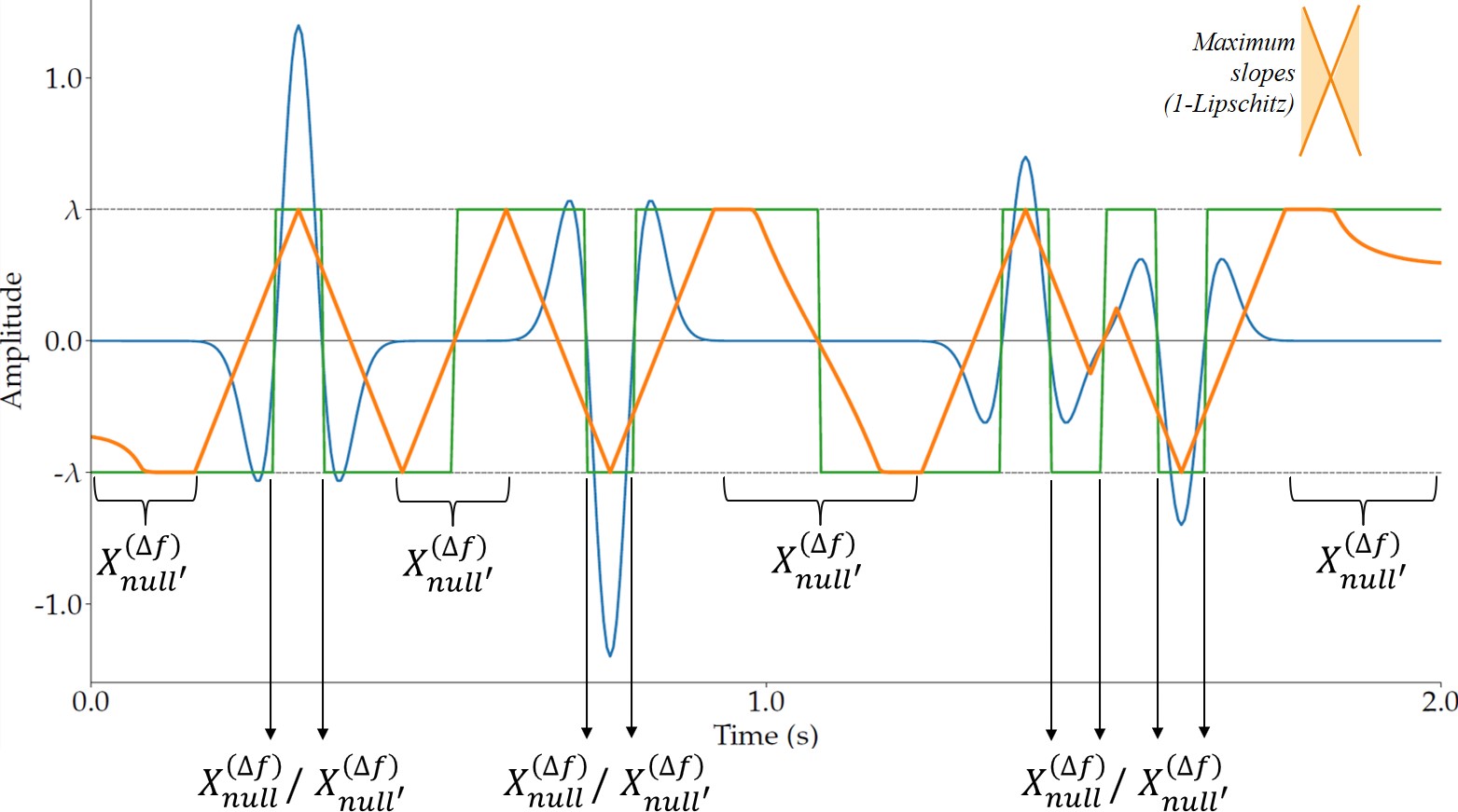}
\caption{
1D coordinate space $X=[0,T]$ (time direction).
Comparison of adjoint-sources for a simple synthetic trace, where $f_{obs}$ contains two Ricker wavelets (all with peak frequency at 6 Hz)
and $f$ contains the same Ricker wavelets but shifted. 
In blue, 
the LSQ residual.
%$X_{null}^{(\Delta f)}\backslash X_{null'}^{(\Delta f)}$ are point-wise contributions (shown by the four black arrows) that have a null Lebesgue measure in $X$.
In green, the $L_1$ adjoint-source.
In orange, the KR adjoint-source, which is computed using the default linear cone iterative solver of the open source sc{cvxopt} library \cite[]{andersen2013cvxopt}, ran until convergence.
At the bottom the various corresponding subsets of $X$.
The KR adjoint-source tends to saturate the (1-Lipschitz or thersholding) constraints in most of $X$ and especially in $X\backslash X_{null}^{(\Delta f)}$.
}
\label{fig:fig-adj-source-comparison}
\end{figure}

\begin{figure}[H]
\centering
\includegraphics[width=0.6\linewidth]{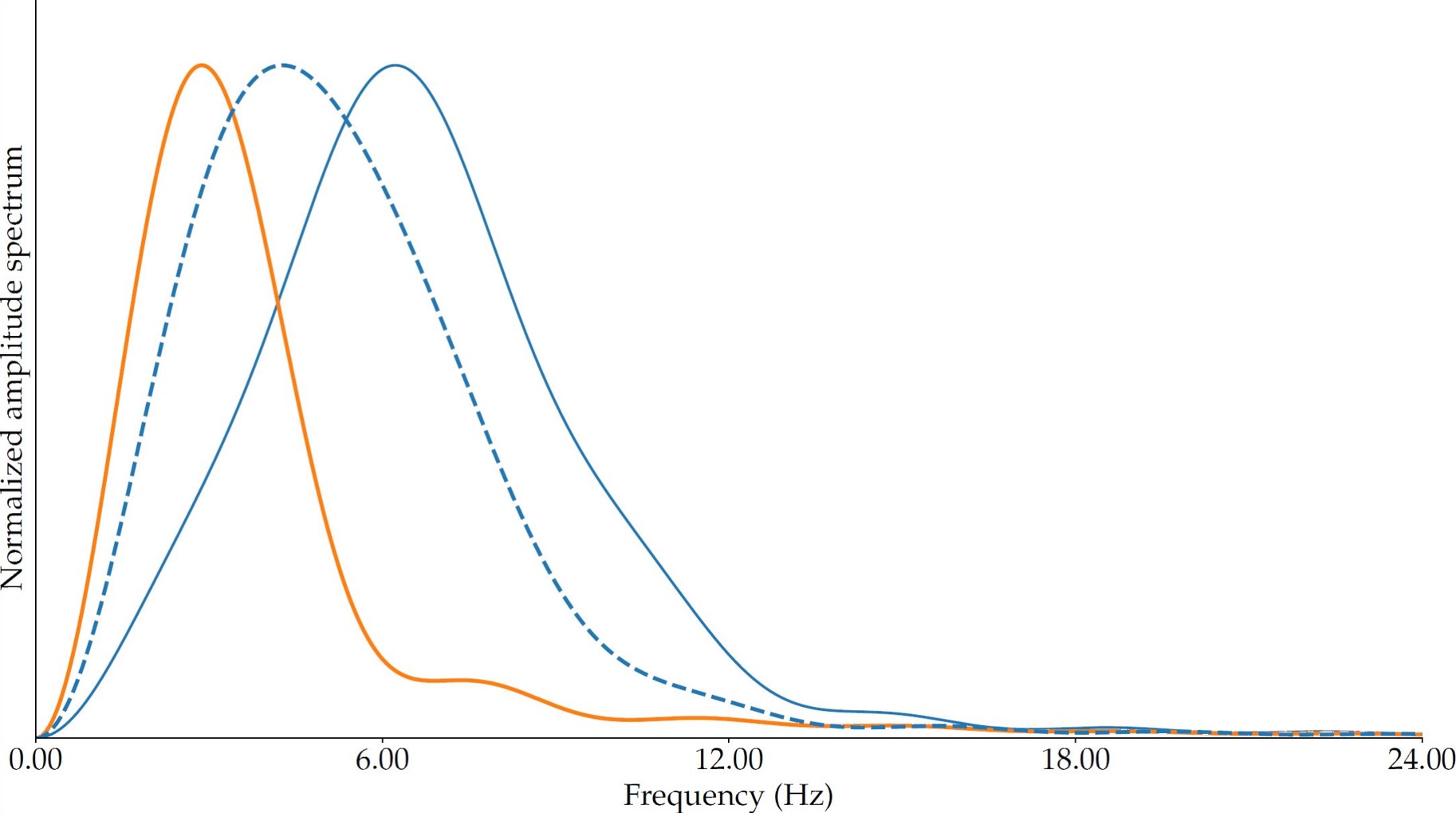}
\caption{
Comparison of the amplitude spectra of different adjoint-sources corresponding to the same setting as Fig.~\ref{fig:fig-adj-source-comparison}.
In blue, the LSQ residual
(same spectrum as the one of a Ricker wavelet whose peak frequency is at 6 Hz). 
In dashed blue, the spectrum of the integrated LSQ residual.
In orange, the KR adjoint-source. 
%In red the approximated KR adjoint-source, using SDMM (50 iterations). 
The KR adjoint-source
has its spectrum shifted towards the low frequencies, beyond the low frequencies of the integrated LSQ residual.
This is favorable to help overcoming cycle-skipping within FWI
(up to the extent of the low frequencies present in the forward propagated source wavelet).
}
\label{fig:fig-kr-freq}
\end{figure}

So, even if these equations would not lead to the most practical numerical scheme,
they allow us to draw interesting formal conclusions:
\begin{itemize}
\item
Eq. (\ref{eq:proofs_5b_text}) clearly remains stable under infinitesimal perturbations of $\Delta F_{\beta}$,
eq. (\ref{eq:proofs_4b_text}),
the latter remaining stable under infinitesimal perturbation of $\Delta f$.
%Only $\varphi_{max}(x\in X_{null}^{(\Delta F_{\beta})})$ can be affected by such perturbations but this does not matter.
This implies that eq. (\ref{eq:OT_12_0}) is satisfied in $X\backslash X_{null}^{(\Delta F_{\beta})}$,
and thus in $X \backslash X_{null}^{(\Delta f)}$
(as $X\backslash X_{null}^{(\Delta F_{\beta})}$ tends to be ``larger'' for seismic data).
\ref{app:proofs} gives further analysis.
\item
The derivative constraint saturates almost everywhere in $X \backslash X_{null}^{(\Delta F_{\beta})}$, thus in $X \backslash X_{null}^{(\Delta f)}$.
%This will produce a piecewise linear shape for $\varphi_{max}$,
%sometimes called ``triangular wavelet''.
But, in most cases, the derivative of $\varphi_{max}$ will also tend to saturate in $X_{null}^{(\Delta f)}$, which is illustrated in Fig. \ref{fig:fig-adj-source-comparison} on a simple synthetic trace (built from shifted Ricker wavelets).
Indeed, even if $X_{null}^{(\Delta f)}$ does not play an explicit role in eq. (\ref{eq:proofs_5b_text}),
it plays a role through the necessary continuity constraint on $\varphi_{max}$
and the maximization of $\int_X\varphi(x_t)\Delta f(x_t)d\mu(x_t)$.
This will tend to produce a piecewise linear shape for $\varphi_{max}$ in the whole $X$, as visible in Fig. \ref{fig:fig-adj-source-comparison}.
%(this figure contains just a little thresholding around 1 second, whose effect will be discussed in next section).
\item
Fig. \ref{fig:fig-adj-source-comparison} also shows 
how the Lischitz constraint saturation leads to a reduction in amplitude dynamics
in $\varphi_{max}$ compared to $\Delta f$.
\item
The ``oscillations'' of $\varphi_{max}$ are driven by $\Delta F_{\beta}$ through eq. (\ref{eq:proofs_5b_text}),
thus among others by the integral of $\Delta f$, eq. (\ref{eq:proofs_4b_text}).
This should give $\varphi_{max}$  a lower frequency content than $\Delta f$,
which can be deduced from Fig. \ref{fig:fig-adj-source-comparison}
and is quantified in Fig.~\ref{fig:fig-kr-freq}.
We observe that $\varphi_{max}$
has its spectrum shifted towards the low frequencies, beyond the low frequencies of the integrated LSQ residual.
\item
The argmax (i.e. $\varphi_{max}$) of the considered problem is not unique \cite[]{Villani2008}.
In particular, any
\begin{eqnarray}
\label{eq:phi_cte}
\varphi^{(c)}_{max}(x) = \varphi_{max}(x) + c,
\quad \forall c \in \mathbb{R},
\end{eqnarray}
clearly remains a solution of eqs. (\ref{eq:OT_10}) and (\ref{eq:OT_6}),
or of eqs.~(\ref{eq:proofs_5b_text}) and (\ref{eq:proofs_4b_text}),
because eq.~(\ref{eq:res_text}) must be satisfied.
The constant $c$ does not change the value of the max, i.e. of $\tilde{W}_d$, but only the argmax.
In schemes that use the argmax
(like FWI where the argmax is interpreted as the adjoint-source),
the constant may play a role,
possibly producing a low frequency smearing.
%We will come back to this point later.
%but we mention that choosing the $c$ that leads to a $\varphi^{(c)}_{max}(x)$ with no null frequency (or low frequency smearing) is often a pertinent way to make the solution unique.
\end{itemize}

\subsubsection{Full case and KR ``texture''}
\label{sec:ot-as-3}$\vspace{1mm}$

\begin{figure}[H]
\centering
\includegraphics[height=7.0cm,width=12.0cm]{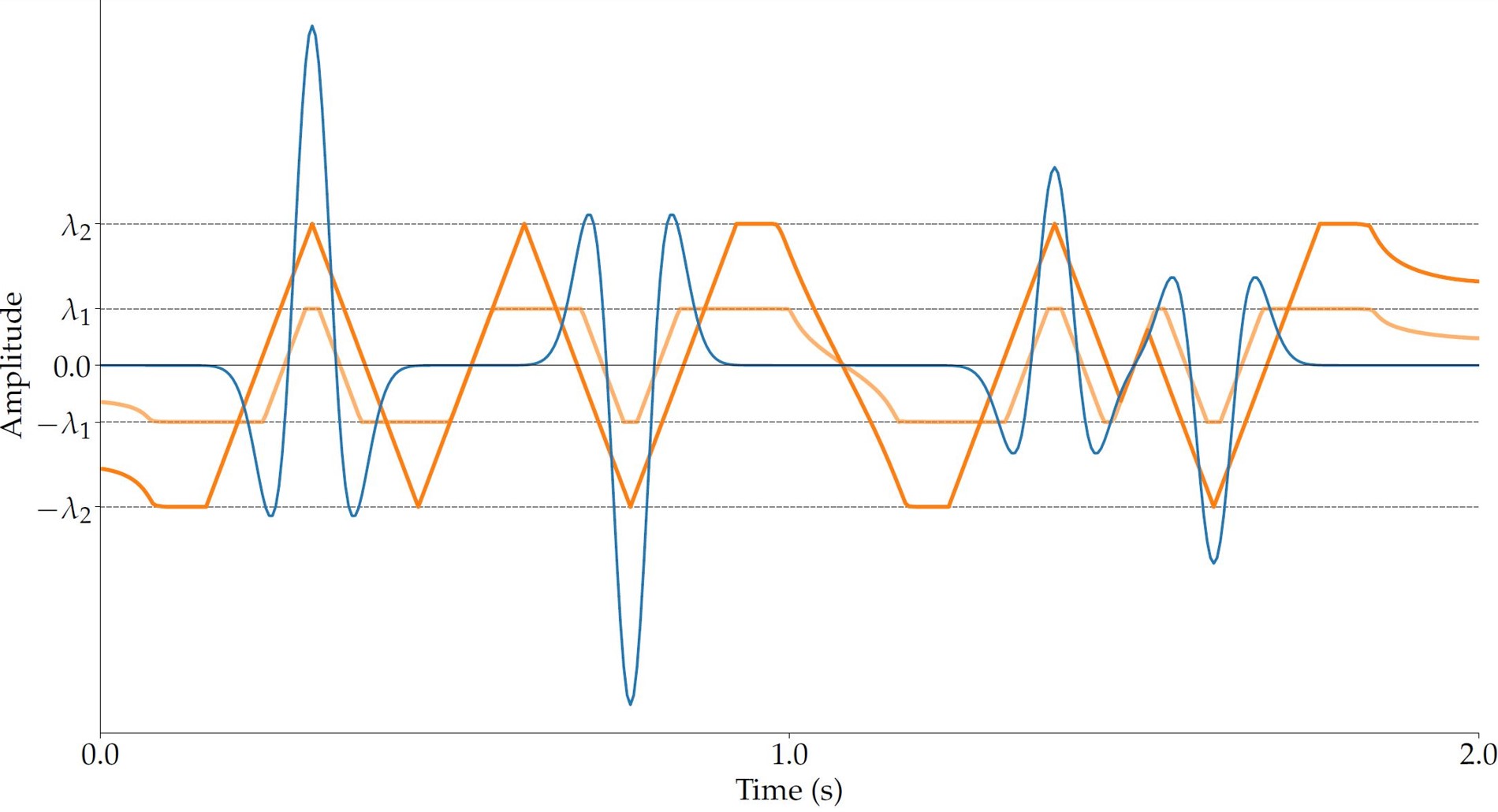}
\caption{
Comparison of adjoint-sources for the same setting as Fig.~\ref{fig:fig-adj-source-comparison}.
%Best viewed in colors. 
In blue, 
the LSQ residual.
In orange, the KR adjoint-sources corresponding to bound $\lambda_2$ (dark orange) and $\lambda_1$ ($<\lambda_2$, light orange).
Smaller $\lambda$ values make the thresholding constraint saturate more, producing more clipping of the adjoint-source.
%For a very small $\lambda$, the KR adjoint-source would become equivalent to the $L_1$ adjoint-source up to a global proportionality constant.
}
\label{fig:fig-kr-varying-bounds}
\end{figure}

\begin{figure}[H]
\centering
\includegraphics[height=7.0cm,width=12.0cm]{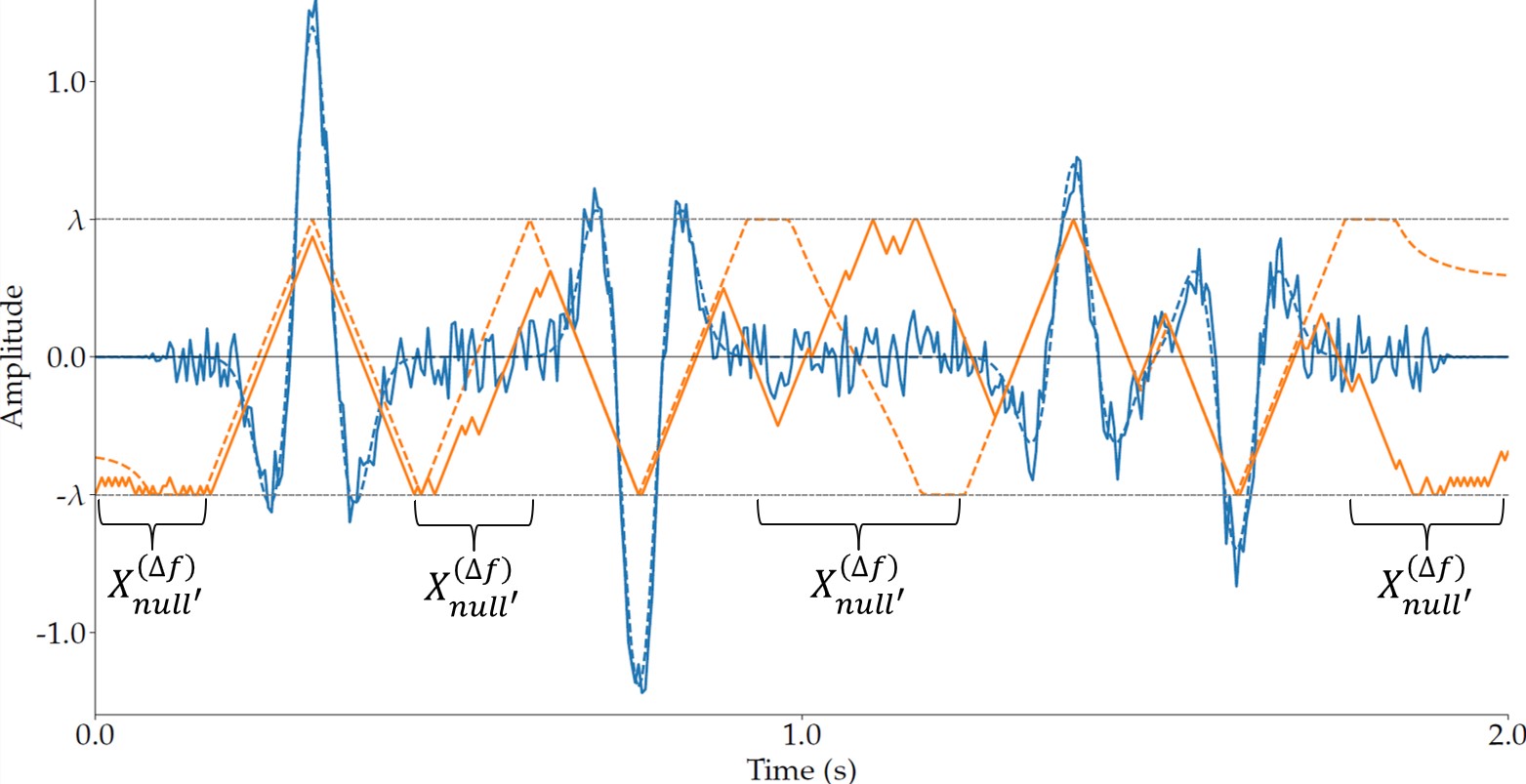}
\caption{
Comparison of adjoint-sources for a similar setting than in Fig.~\ref{fig:fig-adj-source-comparison},
with additional uniform noise.
%Best viewed in colors. 
In blue, 
the LSQ residual to which uniform noise has been added
(the LSQ residual without noise is blue dashed).
The level of noise corresponds to 30\% of the maximum amplitude of the LSQ residual.
% (or of the corresponding $f_{obs}$).
In orange, the KR adjoint-source inverted from noisy LSQ residual
(the KR adjoint-source  without noise is orange dashed).
The KR adjoint-source is perturbed by the presence of noise mostly in $X_{null}^{(\Delta f)}$, that does not really matter.
}
\label{fig:fig-kr-varying-noise}
\end{figure}

\noindent
As demonstrated in \ref{app:proofs},
considering both the thresholding and the 1-Lipschitz constraints in eqs. (\ref{eq:OT_7}) and (\ref{eq:OT_10})
and still a 1D coordinate space $X=[0,T]$,
we obtain for $\varphi_{max}$:
\begin{eqnarray}
\label{eq:proofs_5_2_text)}
&&
{x_t\in X \backslash X_{null}^{(\Delta F_{\beta,\alpha})} \mathrm{\hspace{1.5mm}with\hspace{1.5mm}} \Delta F_{\beta,\alpha}(x_t)<0:}
\quad
\beta^+(x_t)=-\Delta F_{\beta,\alpha}(x_t),\beta^-(x_t)=0
,\sigma_t\frac{\partial\varphi_{max}(x_t)}{\partial x_t}=1
\nonumber\\
%\label{eq:proofs_5b}\\
&&
{\hspace{3.6cm} \Delta F_{\beta,\alpha}(x_t)>0:}
\quad
\beta^-(x_t)=\Delta F_{\beta,\alpha}(x_t),\beta^+(x_t)=0
,\sigma_t\frac{\partial\varphi_{max}(x_t)}{\partial x_t}=-1
\nonumber\\
&&
{x_t\in X \backslash X_{null}^{(\Delta f_\beta)} \mathrm{\hspace{1.5mm}with\hspace{1.5mm}} \Delta f_\beta(x_t)>0:}
\quad
\alpha^+(x_t)=\Delta f_\beta(x_t),\alpha^-(x_t)=0
,\varphi_{max}(x_t)=\lambda
\\
&&
{\hspace{3.3cm} \Delta f_\beta(x_t)<0:}
\quad
\alpha^-(x_t)=-\Delta f_\beta(x_t),\alpha^+(x_t)=0
,\varphi_{max}(x_t)=-\lambda
,
\nonumber
\end{eqnarray}
where ($\Delta F_{\beta}$ being defined by eq.~(\ref{eq:proofs_4b_text}))
\begin{eqnarray}
&&
\Delta F_{\beta,\alpha}(x_t)=
\Delta F_{\beta}(x_t)
%\int_{0}^{x_t}\Delta f(x)d\mu(x)
%-\Big(\beta^+(0)-\beta^-(0)\Big)
-\int_{0}^{x_t} dx \Big(\alpha^+(x)-\alpha^-(x)\Big)d\mu(x)
\nonumber
\\
&&
\Delta f_\beta(x_t)
=
\Delta f(x_t)
+\frac{\partial\beta^+(x_t)}{\partial x_t}-\frac{\partial\beta^-(x_t)}{\partial x_t}
.
\label{eq:proofs_5b_2_text}
\end{eqnarray}
The two group of equations in eq.~(\ref{eq:proofs_5_2_text)}) define $\varphi_{max}$ quite similarly than in eq. (\ref{eq:proofs_5b_text}),
with an additional explicit coupling.
An implicit coupling is also imposed through the necessary (strong) continuity of $\varphi_{max}$ and the fact that none of the constraints can saturate simultaneously, i.e. (see \ref{app:proofs})
\begin{eqnarray}
X_{null}=X_{null}^{(\Delta f_\beta)}\cup  X_{null}^{(\Delta F_{\beta,\alpha})}
\quad\mathrm{and}\quad
\varnothing=X_{null}^{(\Delta f_\beta)}\cap   X_{null}^{(\Delta F_{\beta,\alpha})}
.
\end{eqnarray}
All this allows us to draw similar lessons than these underlined above:
\begin{itemize}
\item
We deduce that eq. (\ref{eq:OT_12_0}) is satisfied in the subset that matters $X \backslash X_{null}^{(\Delta f)}$, concluding the proof that
the KR adjoint-source is given by eq. (\ref{eq:OT_12bis}).
\item
The 1-Lipschitz constraint or the thresholding constraint tend to saturate  (not simultaneously) almost everywhere in $X \backslash X_{null}^{(\Delta f)}$
and also in most of $X_{null}^{(\Delta f)}$.
This will tend to produce a piecewise linear shape for $\varphi_{max}$ in the 1D case, as illustrated in Fig.~\ref{fig:fig-kr-varying-bounds}.
Note that, in the multiD case, the piecewise linear shape becomes approximate. Indeed, at least one component of the 1-Lipschitz constraint then needs to saturate at a given position (mostly the time direction in our implementation), but nothing constrains many components to saturate simultaneously in the general case. This will be further discussed in the next section.
\item
Visible in Fig.~\ref{fig:fig-kr-varying-bounds} and for the reason mentioned in \S \ref{sec:ot-as-2}, the KR adjoint-source $\varphi_{max}$ has a lower frequency content than the LSQ residual.
\item
Fig. \ref{fig:fig-kr-varying-bounds} also shows 
how the constraints saturation lead to a reduced amplitude dynamics
in $\varphi_{max}$ compared to $\Delta f$,
the thresholding constraint parameter $\lambda$ allowing further control of these dynamics
(in the extreme case of a very small $\lambda$, the KR adjoint-source would become equivalent to the $L_1$ adjoint-source up to a global proportionality constant).
\item
These three previous items (piecewise linearity, lower frequency content, reduced amplitude dynamics) represent what we call KR ``texture'' for the adjoint-source in the following.
It is favorable for reducing the sensitivity of FWI to cycle-skipping and amplitudes.
\item
The thresholding constraint, making eq. (\ref{eq:res_text}) useless,
limits the additive constant (or null frequency) issue discussed in \S \ref{sec:ot-as-2}.
However, in a situation where the chosen $\lambda$ value does not provide enough constraint  
and eq. (\ref{eq:res_text}) approximately holds,
the additive constant issue can still exist,
that could produce a low frequency smearing in the FWI adjoint-source.
A possible effect will be discussed later in \S \ref{sec:ot-int-fwi}.
\item
Eq. (\ref{eq:OT_12_0}) can remain valid for non-infinitesimal perturbations $h$,
hence the often mentioned robustness of the KR adjoint-source to noise.
Fig. \ref{fig:fig-kr-varying-noise} gives an illustration for a quite strong uniform noise.
We observe that the KR adjoint-source is perturbed by the presence of noise mostly in $X_{null}^{(\Delta f)}$, that does not really matter, and tends to remain very stable in the subset that matters $X \backslash X_{null}^{(\Delta f)}$.
\item
Finally, we mention that the approximate piecewise linear shape for $\varphi_{max}$
in the multiD case is due to the use of a distance $d$ induced by the $L_1^{(X)}$ norm
in the 1-Lipschitz constraint of eq. (\ref{eq:OT_7}),
that leads to the constraint in eq. (\ref{eq:lip_1D_text}).
If a $L_p^{(X)}$ norm with $p\in]1,\infty[$ was used instead,
the constraint in eq. (\ref{eq:lip_1D_text}) would have to be replaced by
the following constraint (as demonstrated in \ref{app:proofs_0})
\begin{eqnarray}
\Big(
\sigma_{xl}^q\Big|\frac{\partial \varphi(x)}{\partial x_{xl}}\Big|^q
+
\sigma_{inl}^q\Big|\frac{\partial \varphi(x)}{\partial x_{inl}}\Big|^q
+
\sigma_t^q\Big|\frac{\partial \varphi(x)}{\partial x_t}\Big|^q
\Big)^{1/q}
\le 1
\quad\mathrm{with}\quad
1/p+1/q=1
.
\nonumber\\
\end{eqnarray}
The saturation of this constraint would not lead to approximate piecewise linearity in the general multiD case, as the derivatives in the various coordinate space directions would be strongly coupled.
\end{itemize}

%Note also the ``triangular'' shape of the KR adjoint-source wavelet,
%due to the 1-Lipschitz constraint saturation
%within the 1-Wasserstein formulation.

%Also, let us consider perturbations of the form $h(x)=\alpha(x)(f(x)-f_{obs}(x))$.
%Because of the considerations of Appendix \ref{app:proofs},
%the inverted $\varphi_{max}[f]$ will tend to remain invariant
%for any $\alpha$ so that $(1+\alpha)(f-f_{obs})$ and $(f-f_{obs})$ keep the same sign,
%and $|\int_{x_{min}}^x (f(x)-f_{obs}(x))d\mu(x)| >> |\int_{x_{min}}^x \alpha(x)(f(x)-f_{obs}(x))d\mu(x)|$.
%This can be achieved with non-infinitesimal $\alpha$, for instance rapidly and randomly varying ones that remain strictly greater than $-1$. 
%\JMcomm{(Raphael: an illustration?)}.

\begin{figure}[H]
\centering
\includegraphics[height=7.0cm,width=12.0cm]{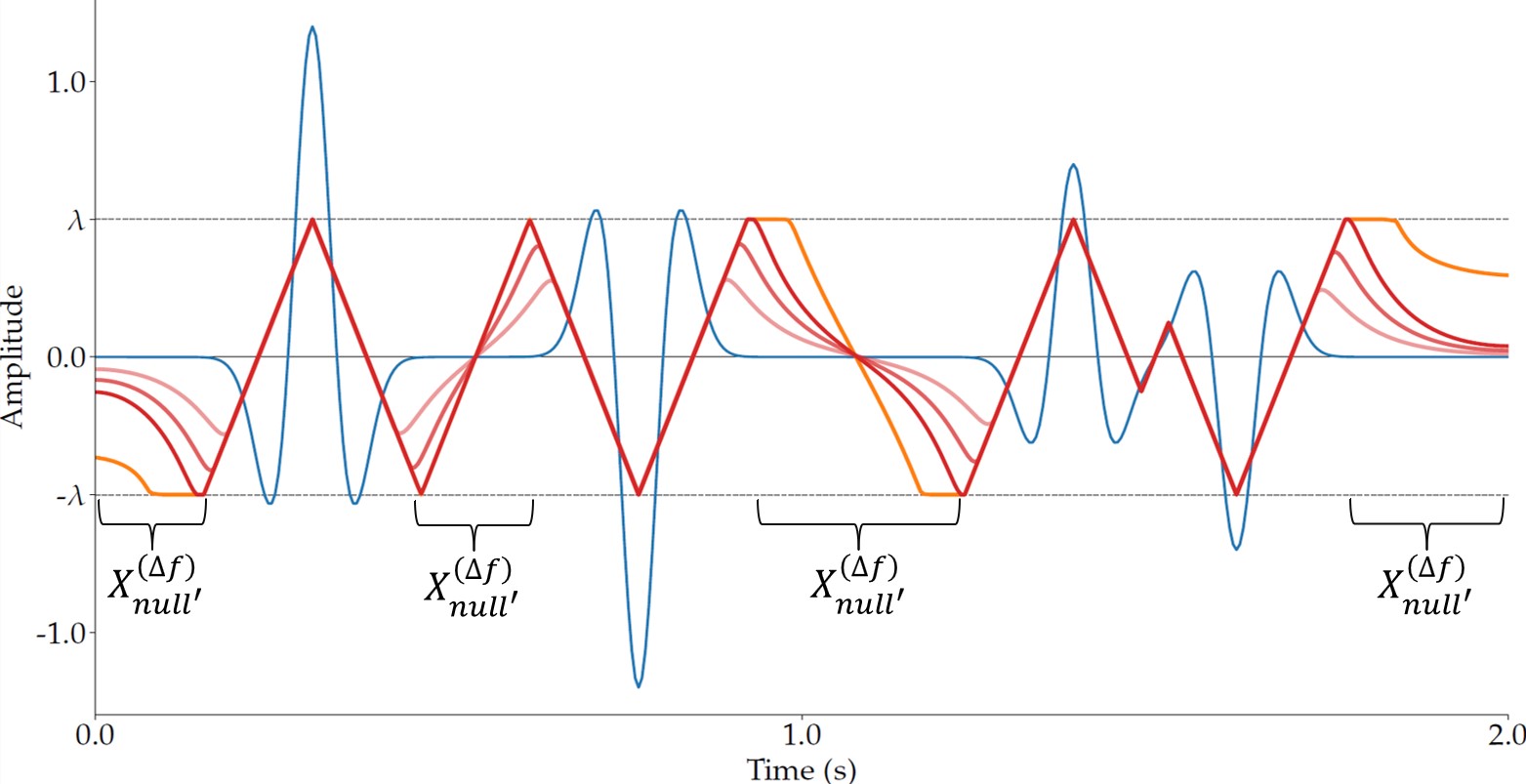}
\caption{
Comparison of adjoint-sources for a similar setting than in Fig.~\ref{fig:fig-adj-source-comparison}.
In blue, 
the LSQ residual.
In orange, the exact KR adjoint-source.
In various reds, the KR adjoint-source approximated using SDMM with $50$ (light red), $500$ and $5000$ (dark red) iterations.
The various SDMM KR adjoint-sources differ from the exact KR adjoint-source mostly on $X_{null}^{(\Delta f)}$, that does not really matter.
}
\label{fig:fig-adj-source-SDMM-cv}
\end{figure}

\begin{figure}[H]
\centering
\includegraphics[height=7.0cm,width=12.0cm]{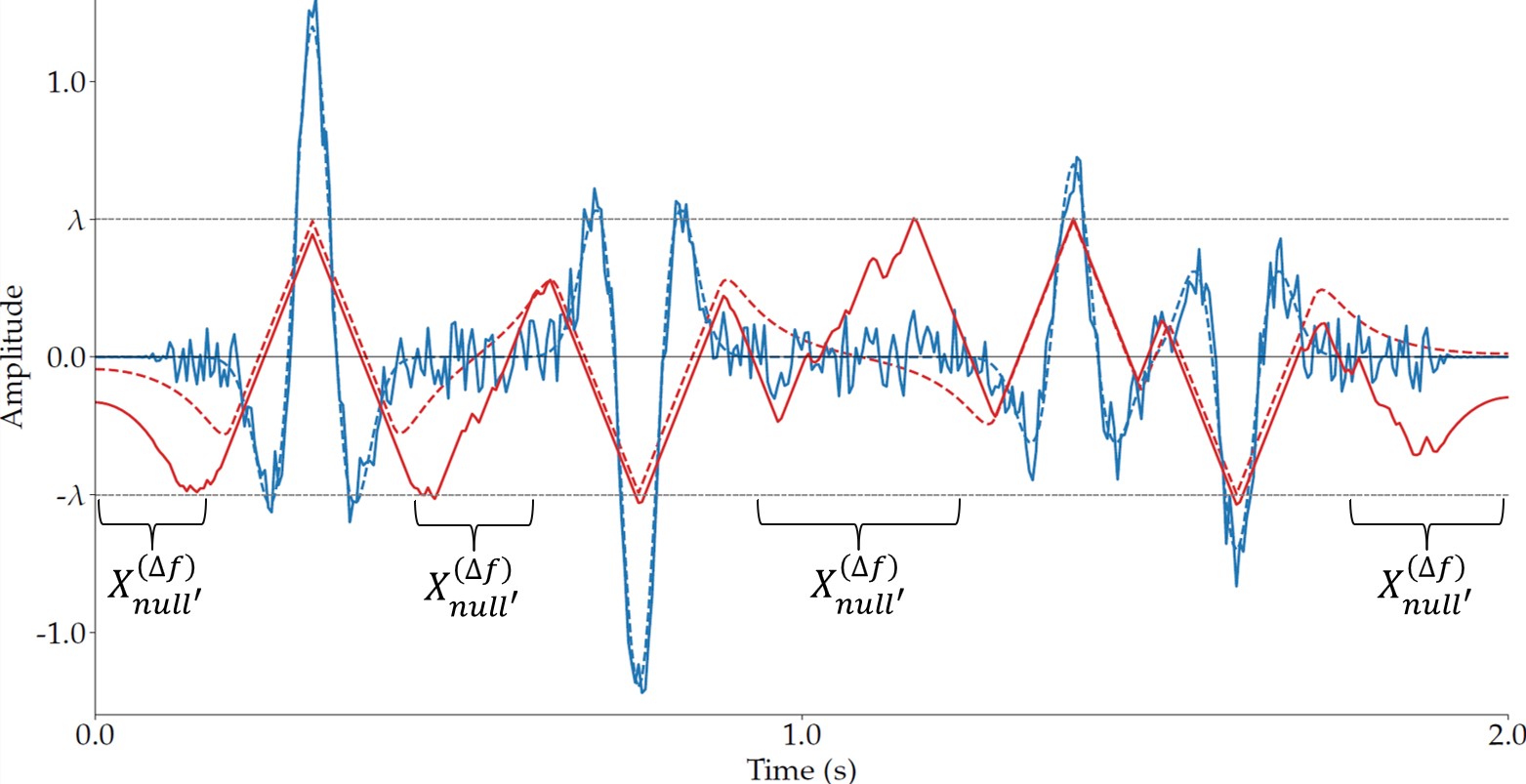}
\caption{
Comparison of adjoint-sources for the same setting as Fig.~\ref{fig:fig-kr-varying-noise}.
In blue, 
the noisy LSQ residual
(the LSQ residual without noise is dashed).
In red, the SDMM ($50$ iterations) KR adjoint-source inverted from the inverted the noisy LSQ residual
(the corresponding SDMM KR adjoint-source without noise is dashed).
The SDMM KR adjoint-source is perturbed by the presence of noise mostly in $X_{null}^{(\Delta f)}$, that does not really matter.
Note that it seems a little less perturbed by the noise (in $X_{null}^{(\Delta f)}$) than the exact KR adjoint-source of Fig. \ref{fig:fig-kr-varying-noise}.
}
\label{fig:fig-kr-varying-noise_2}
\end{figure}

\section{Practical aspects}

\subsection{SDMM approach and illustration on a simple synthetic trace}

%Having discussed theoretical aspects about the KR adjoint-source,
%we now discuss practical aspects still focusing on the adjoint-source.

Up to now, we have considered an exact resolution of the KR problem.
% eq. (\ref{eq:OT_10}) with a $L_1^{(X)}$-norm in the coordinate space.
In this section, we underline the advantages of using an approximate resolution method, the SDMM convex optimization with a Laplace solver \cite[]{Metivier2016}.

$N\ge 0$ denotes the number of SDMM inner iterations.
In our implementation, we start iterating taking the LSQ residual $\Delta f$ as the initial
function for $\varphi$. Thus, for $N=0$, 
the scheme is equivalent to LSQ.
Increasing $N$ will ``add'' to the adjoint-source more and more of the KR texture defined in \S \ref{sec:ot-as-3}, linearly increasing the computational cost.
An exact resolution of the KR problem would be obtained using a very large $N$,
which would be too costly for large scale applications and, mostly, unecessary.
Fig. \ref{fig:fig-adj-source-SDMM-cv} shows the exact KR adjoint-source
and the ones approximated by SDMM ($N=50$, $500$ and $5000$)
on a simple synthetic trace (shifted Ricker wavelets).
Interestingly, all results are very similar in the subset that matters $X\backslash X_{null}^{(\Delta f)}$
and differ mostly on $X_{null'}^{(\Delta f)}$ where SDMM takes more iterations to converge.
As what happens in $X_{null'}^{(\Delta f)}$ does not really matter in practice,
the adjoint-source obtained with the smaller $N$ value, i.e. $N=50$, seems sufficient.
We observe that the KR texture (piecewise linearity, lower frequency content, reduced amplitude dynamics) is mostly preserved even for $N=50$.

The $N$ parameter thus has a clear meaning, controlling the desired amount of KR texture added to the LSQ residual, allowing us to consider the KR norm as a hybrid misfit that mixes OT with the LSQ cost function.
This flexibility is a strong point for industrial applications.
For instance, at the first FWI iteration, more KR texture can be added by increasing $N$ and,
as FWI converges, $N$ may be reduced to tend to LSQ.

What about SDMM  robustness to noise?
Fig. \ref{fig:fig-kr-varying-noise_2} illustrates
the KR adjoint-source inverted by SDMM with $N=50$, for a noisy LSQ residual.
We observe that the KR adjoint-source is perturbed by the presence of noise mostly in $X_{null}^{(\Delta f)}$, that does not really matter, and tends to remain very stable in $X \backslash X_{null}^{(\Delta f)}$, which is satisfying.
Interestingly, the SDMM KR adjoint-source seems a little less perturbed by the noise in $X_{null'}^{(\Delta f)}$ than the exact KR adjoint-source, comparing Figs.~\ref{fig:fig-kr-varying-noise}~and~\ref{fig:fig-kr-varying-noise_2}.

This section justifies the use of SDMM to compute the KR adjoint-source
far beyond computational cost considerations only.
In practice, our optimizations and tuning of SDMM allow ta satisfying result to be reached for $N=20$ to $60$.

\subsection{Data space ``splitting'', and 1D and 2D KR}

\begin{figure}[H]
\centering
\includegraphics[width=0.5\linewidth]{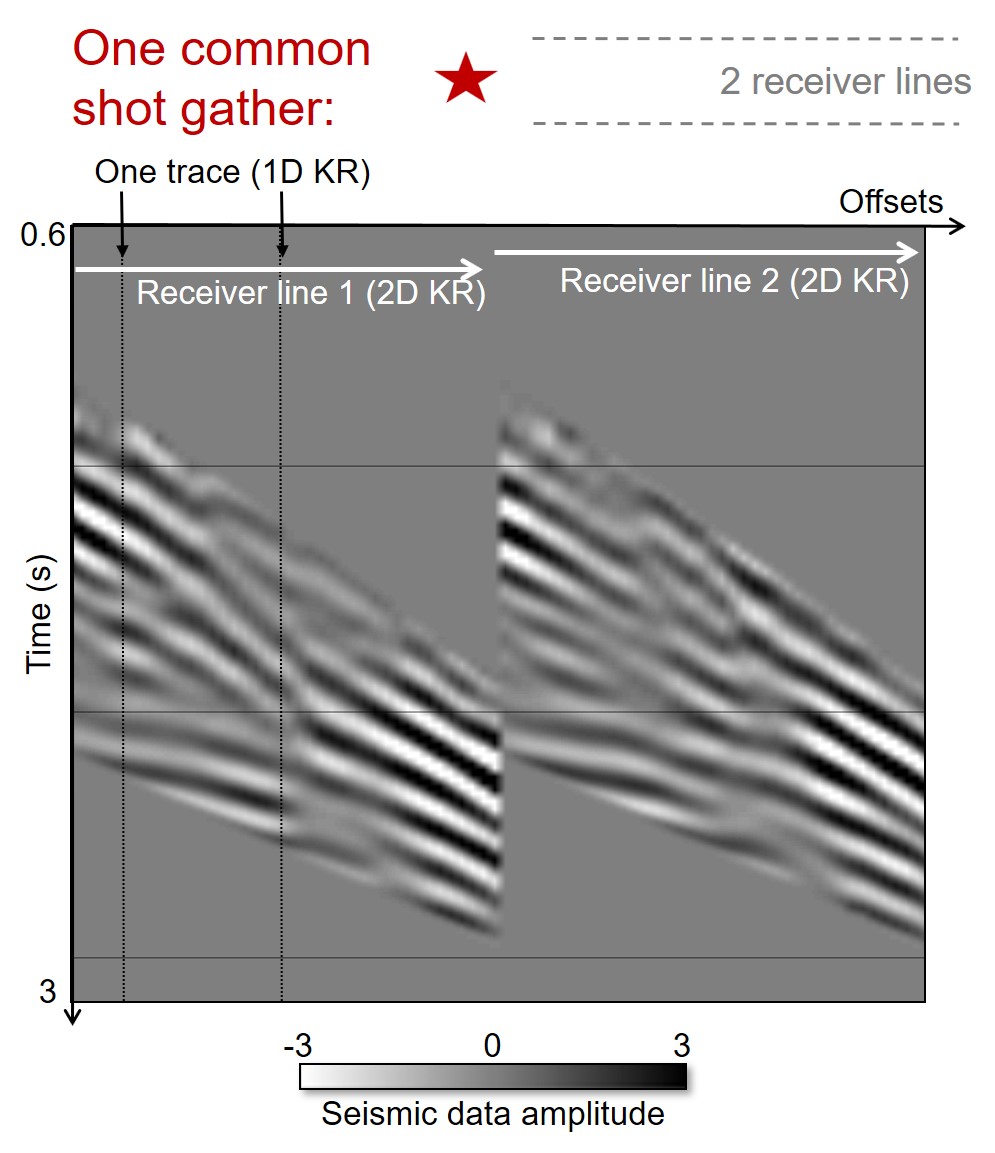}
\caption{
Marine field data set corresponding to two receiver lines within a common shot gather.
LSQ residual at 6 Hz, muted for the FWI,
that represents the initial adjoint-source for SDMM iterations.
2D KR considers each receiver line independently and 1D KR considers each trace independently.
%Left: LSQ adjoint-source (or residual). Middle: KR adjoint-source. Right: Corresponding amplitude spectra comparison.
Modified from \cite{Poncet2018} (see the article for more details).
}
\label{fig:fig-kr-lines}
\end{figure}

We mentioned that eq. (\ref{eq:OT_10}) can be resolved considering different effective dimensionalities for the coordinate space $X$.
In our implementation, we considered two cases:
\begin{enumerate}
%\item
%The largest effective dimensionality is $3$, when eq. (\ref{eq:OT_10}) is resolved for a full shot. Then $X=]0,H_{xl}]\times]0,H_{inl}]\times]0,T]$.
%Any correlation in the data may be accounted for.
\item
Each trace  within a shot independently.
The coordinate space is not anymore multiD but $1D$
(that is still ``much more'' than $L_p$ norms whose coordinate space dimensionality is $0$ as they are computed point-wise), the effective $X$ being $[0,T]$.
Only correlations in the time direction can be accounted for
and we call this scheme 1D KR in the following.
\item
Group of traces in one spatial direction, the best sampled or less aliased one,
thus an effective dimensionality of $2$.
Correlations in the chosen spatial and time directions can be accounted for
and we call this scheme 2D KR in the following.
For instance in the marine case, we consider each receiver line within a shot independently,
the effective $X$ being $[H_{inl}^{min},H_{inl}^{max}]\times[0,T]$.
\end{enumerate}
Fig. \ref{fig:fig-kr-lines} illustrates this data space ``splitting'' for 1D and 2D KR.

\subsection{SDMM KR illustrated on common shot data}
\label{sec:ot-num}

\begin{figure}[H]
\centering
\includegraphics[width=1.1\linewidth]{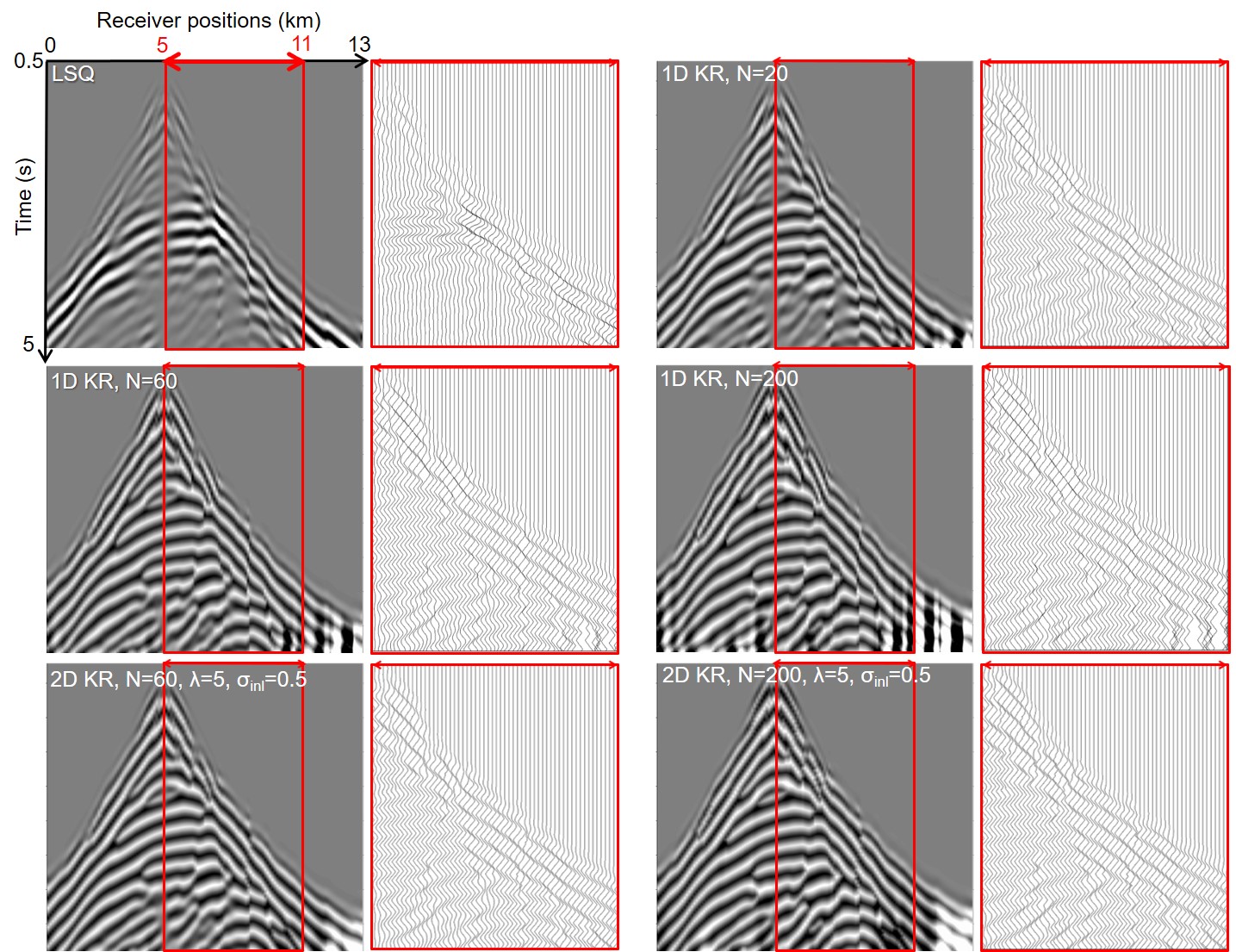}
\caption{
Marmousi 2 data set \cite[]{Martin2006} (Ricker wavelet with peack frequency at 6 Hz and frequencies below 3 Hz muted).
A smooth high-cut filter above 4 Hz has been applied to make it easier to see the wiggle traces.
LSQ residual, and 1D and 2D KR adjoint-sources (effect of the number of iterations $N$) are shown, for one common shot gather.
}
\label{fig:figure3_N}
\end{figure}

We firstly consider a 2D common shot data modelled in the Marmousi 2 model 
\cite[]{Martin2006}.
The source is a Ricker wavelet with peak frequency at 6 Hz and the data has been low-pass filtered below 3 Hz to be more realistic.
Fig. \ref{fig:figure3_N} illustrates the effect of the SDMM number of iterations $N$.
The results are shown with a smooth high-cut filter above 4 Hz to make it clearer to see the wiggle traces.
We can observe how more and more KR texture is added to the KR adjoint-source as $N$ increases. 
The wiggles highlight how each KR adjoint-source trace tends to becomes more piecewise linear,
especially in the time direction that matters the most to overcome cycle-skipping.
%The theoretical reason has been discussed in \S \ref{sec:ot-as-2}
%for 1D KR and 2D KR case
%(with the subtlety that in the latter case at least one component of the Lipschitz constraint needs to saturate at a given position, but nothing constrains all the components to saturate simultaneously).

Some artefacts can be observed in Fig. \ref{fig:figure3_N} at large offsets as the amplitudes tend to be larger
and the traces are further from null average.
However, these artefacts remain very localized.
We can observe that using 2D KR diminishes
the artefacts and improves the adjoint-source continuity compared to using 1D KR.
We will detail the corresponding specifics of 2D KR further in \S \ref{sec:ot-int}.

\begin{figure}[H]
\centering
\includegraphics[width=0.8\linewidth]{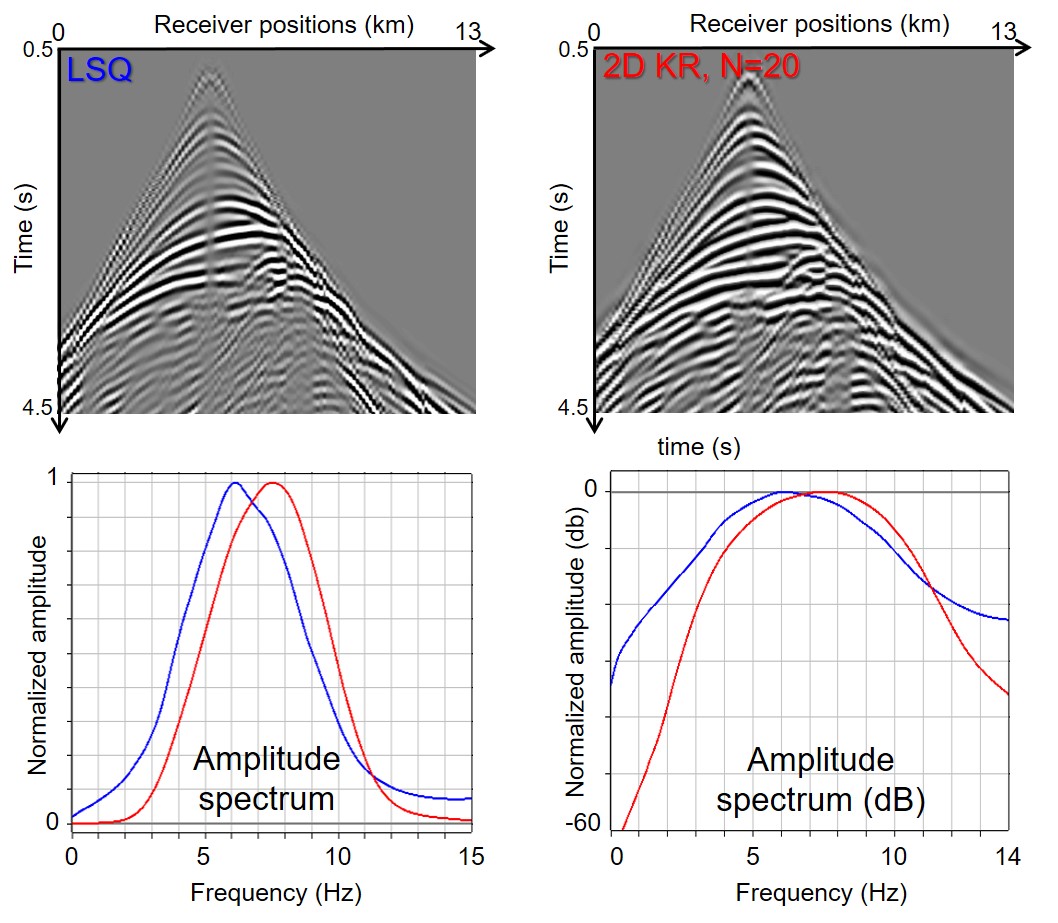}
\caption{
Same data as Fig. \ref{fig:figure3_N}, but with the high-cut filter at 10 Hz (instead of the smooth high-cut filter above 4 Hz).
%, related to the following Marmousi 2 10 Hz FWI results.
LSQ residual and 2D KR ajoint-source, together with their frequency spectra.
}
\label{fig:figure3_0}
\end{figure}

Fig. \ref{fig:figure3_N} also illustrates how the KR adjoint-source amplitude dynamics are reduced as $N$ increases, and lets guess that the low frequency content is reinforced.
This is confirmed by Fig. \ref{fig:figure3_0} (high-cut at 10 Hz instead of 4 Hz to make frequency spectra visualization clearer), even for the smaller $N=20$ value.

\begin{figure}[H]
\centering
\includegraphics[width=0.95\linewidth]{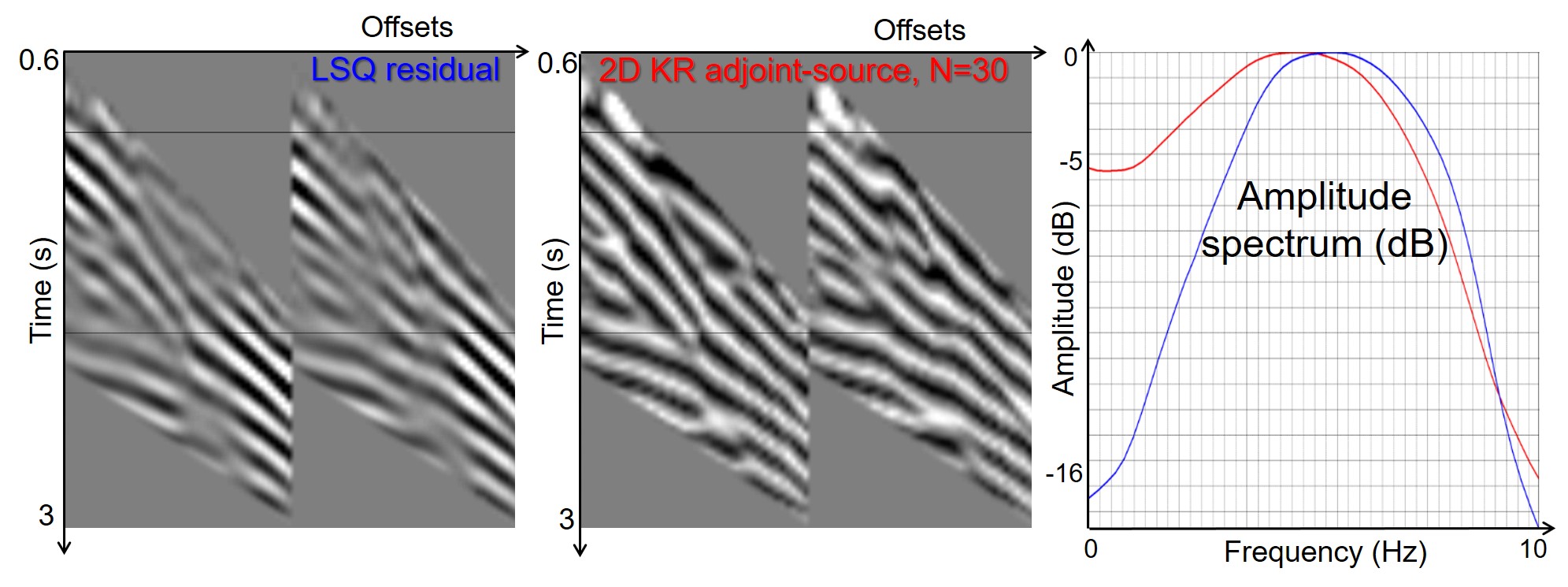}
\caption{
3D field data set with a mute for a 6 Hz inversion.
The two left figures represent the LSQ residual and the KR adjoint-source, muted as usual for the FWI.
The right figure represents the corresponding amplitude spectra.
%Left: LSQ adjoint-source (or residual). Middle: KR adjoint-source. Right: Corresponding amplitude spectra comparison.
Modified from \cite{Poncet2018} (see the article for more details).
}
\label{fig:figure3}
\end{figure}

Secondly, we consider a field 3D common shot data.
Fig. \ref{fig:figure3} compares a LSQ residual with the corresponding 2D KR adjoint-source computed by SDMM, with a mute applied as usually done for FWI.
Again, the 2D KR adjoint-source has more low frequencies and a reduced amplitude dynamics.

In this section, the KR texture for the adjoint-source computed by SDMM has been illustrated on shot data and in the 1D and multiD cases.
This texture represents a key to mitigate the cycle-skipping and reduce the sensitivity to the amplitudes within FWI, enhancing the kinematic information present in the adjoint-source
(like a smart processing of the LSQ residual).
To make this point explicit, the KR texture is sometimes called ``skeleton-like'' texture.
 
%We discuss further the specific properties of the multiD case, i.e. to enhance the adjoint-source continuity in the move-out direction.


\subsection{Thresholding and link with standard cost functions}
\label{sec:ot-num-2}

We now illustrate the effect of the thresholding constraint on common shot data.
Choosing a good $\lambda$ value is important for the success of the scheme. 
Using a value that is too large would lead to instabilities (possible ``singularities''), as explained in \S \ref{sec:ot-rem}.
Using a value that is too small would be equivalent to using a $L_1$ norm cost function.
A good $\lambda$ value must let the 1-Lipschitz constraint saturate at most positions in $X \backslash X_{null}^{(\Delta f)}$ and limit the additive constant (or null frequency) issue discussed in \S \ref{sec:ot-as-3}.
Within reasonable limits, slightly smaller $\lambda$ values would make the thresholding constraint saturate slightly more, producing a clipping of the adjoint-source and thus slightly reducing the sensitivity of the FWI to amplitudes.

\begin{figure}[H]
\centering
\includegraphics[width=1.1\linewidth]{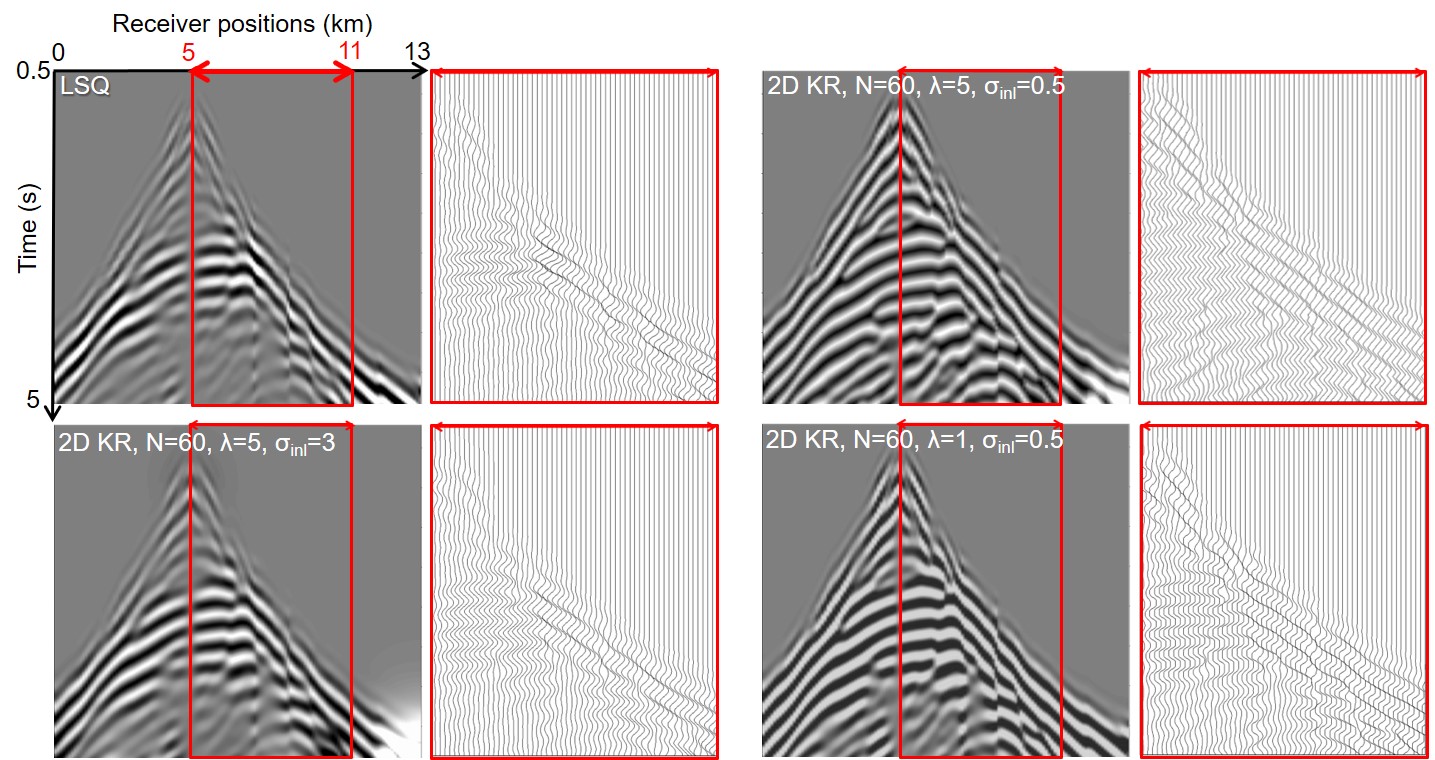}
\caption{
Same data as Fig. \ref{fig:figure3_N}.
LSQ residual and 2D KR adjoint-sources (effect of $\lambda$ and $\sigma_{inl}$) are shown.
}
\label{fig:figure3_lamsig}
\end{figure}

Fig.~\ref{fig:figure3_lamsig} (bottom-right) illustrates the thresholding that occurs on Marmousi data when $\lambda$ is reduced;
we can observe that the clipping remains slight (the amplitude dynamics are not strong with this data)
and that larger $\lambda$ values may lead to a better amplitude equalization for a finite number of SDMM iterations (top-right).
%So, $\lambda$ may in general be considered more as a regularizer than an efficient amplitude equalizer.

We do not observe an additive constant or null frequency in our Marmousi KR adjoint-source results,
contrary to \cite{Metivier2016,Metivier2016a}.
It is probable that this is due to our $\lambda$ tuning, that contribute to limit the additive constant effect as discussed in \S \ref{sec:ot-as-3}.
A possible effect will be discussed further on the Marmousi FWI example in \S \ref{sec:ot-int-fwi}.

Note that the $N$ and $\lambda$ parameters allow us to consider KR as a hybrid cost function that mixes OT with the standard LSQ and $L_1$ cost functions,
being able to pass continuously from one to the other.

\subsection{Improvement of the rate of convergence with one of the standard-deviation-like weights}

Let us come back to the chosen norm for the data coordinate space, i.e. eq. (\ref{eq:lq_omega}) with $p=1$.
%For efficiency, we mentioned that  derives from a $L_1$ norm,
%i.e. eq. (\ref{eq:lq_omega}) with $p=1$.
We have, in the 2D case
\begin{eqnarray}
|| x||_1^{(X)}
=
\frac{1}{\sigma_{inl}}
\Big(
|x_{inl}|
+
v
|x_t|
\Big)
=
\frac{1}{\sigma_{t}}
\Big(
\frac{1}{v}|x_{inl}|
+
|x_t|
\Big)
\quad\quad{with}\quad\quad
v
=
\frac{\sigma_{inl}}{\sigma_t}
,
\label{eq:l1_omega_2}
\end{eqnarray}
where $\sigma_{inl}$  (or $\sigma_{t}$) represents a standard-deviation-like weight in the inline (or time) direction, thus has the dimension of a distance (or a time), and $v$ represents an apparent velocity parameter. 

It can be demonstrated that, at full convergence (i.e. a sufficiently large number $N$ of SDMM iterations), any choice for $\sigma_{inl}$ (or $\sigma_{t}$) is equivalent when an optimal threshold ($\lambda$ value) is considered,
as the corresponding adjoint-sources would almost only differ by a global proportionality constant compensated by the line search. But, in practice for large scale applications, we can only run a small number of iterations $N$ (usually around 30-60). Hence, $\sigma_{inl}$ (or $\sigma_{t}$) can play a role, mostly related to a change of the rate of convergence.
Fig. \ref{fig:figure3_lamsig} illustrates the much slower convergence that occurs with a too large $\sigma_{inl}$ value (the KR adjoint-source remains much closer to the LSQ adjoint-source for a given number of iterations $N=60$).
This is thus not directly related to the 2D (or multiD) behavior in the data coordinate space.
In 1D, only the parameter $\sigma_t$ can play this role (the parameter $\sigma_{inl}$ then does not exist).

\subsection{Enhancement of the KR adjoint-source continuity with the apparent velocity (multidimensional feature)}
\label{sec:ot-int}

Now, let us define which parameter controls the multiD (here 2D) KR feature.
For a fixed $\sigma_{inl}$ (or $\sigma_{t}$),
the velocity $v$ in eq. (\ref{eq:l1_omega_2}) plays a crucial role in the success of 2D KR FWI.
It defines the average direction along which most correlations between inline traces events occur, thus must be parameterized to characterize the average moveout direction.

\begin{figure}[H]
\centering
\includegraphics[width=1.\linewidth]{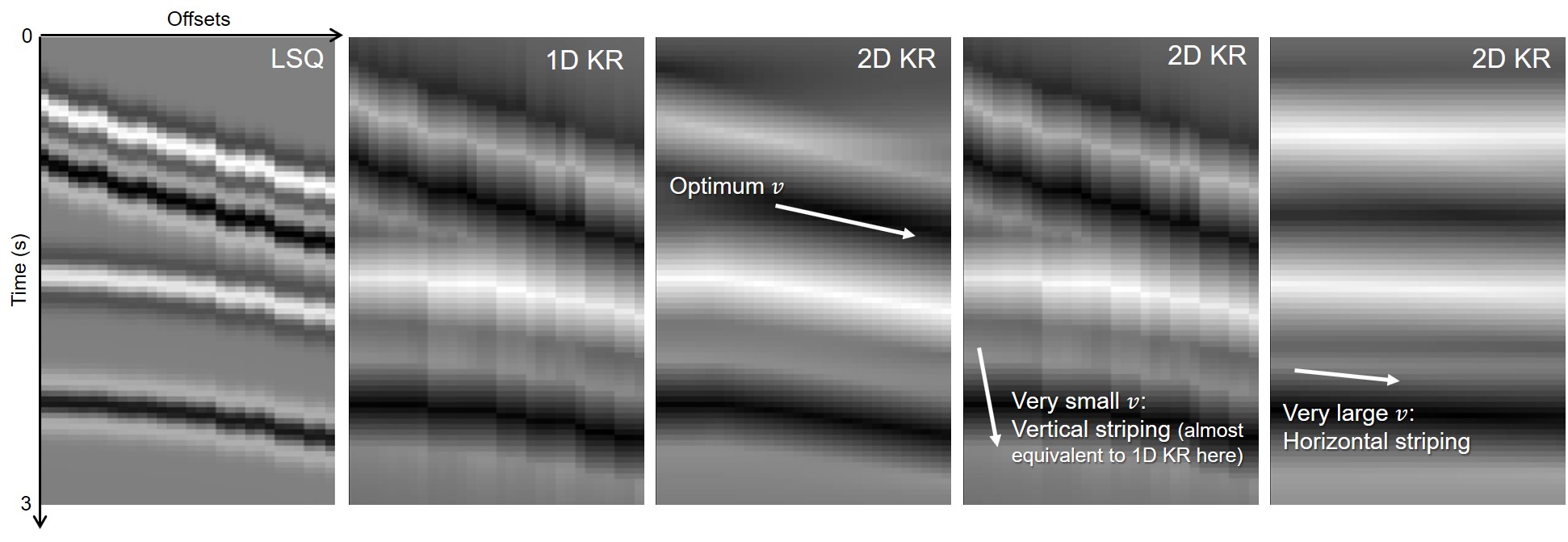}
\caption{
Comparison of adjoint-sources for a simple example, where $f_{obs}$ contains events built from Ricker wavelets (peak frequency at 6 Hz)
and $f$ contains the same events but shifted.
The setting is similar than in Fig.~\ref{fig:fig-subest}b but the events are extended in a second (inline) dimension applying linear and hyperbolic moveouts,
and then the traces are randomly shifted in time to degrade the continuity.
LSQ residual, 1D KR adjoint-source, and 2D KR adjoint-sources (effect of $v$) are shown.
}
\label{fig:figure4_Raph}
\end{figure}

\begin{figure}[H]
\centering
\includegraphics[width=0.8\linewidth]{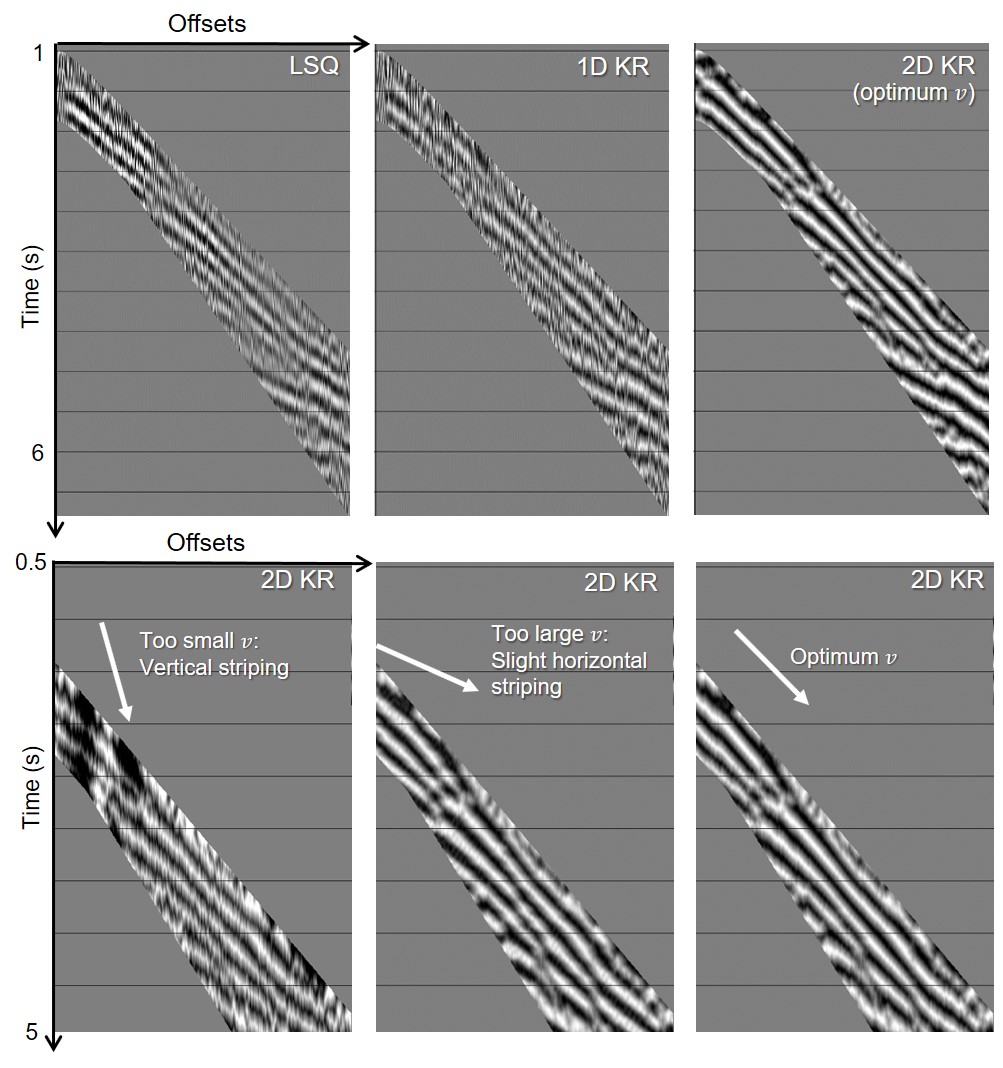}
\caption{
Marine field data at 4 Hz. LSQ residual, 1D KR adjoint-source, and 2D KR adjoint-sources (effect of $v$) are shown.
%LSQ FWI (left), 1D KR norm FWI (middle) and multiD KR norm FWI (right).
%From \cite{Messud2019} (see the article for more details).
}
\label{fig:figure4}
\end{figure}

\begin{figure}[H]
\centering
\includegraphics[width=1.1\linewidth]{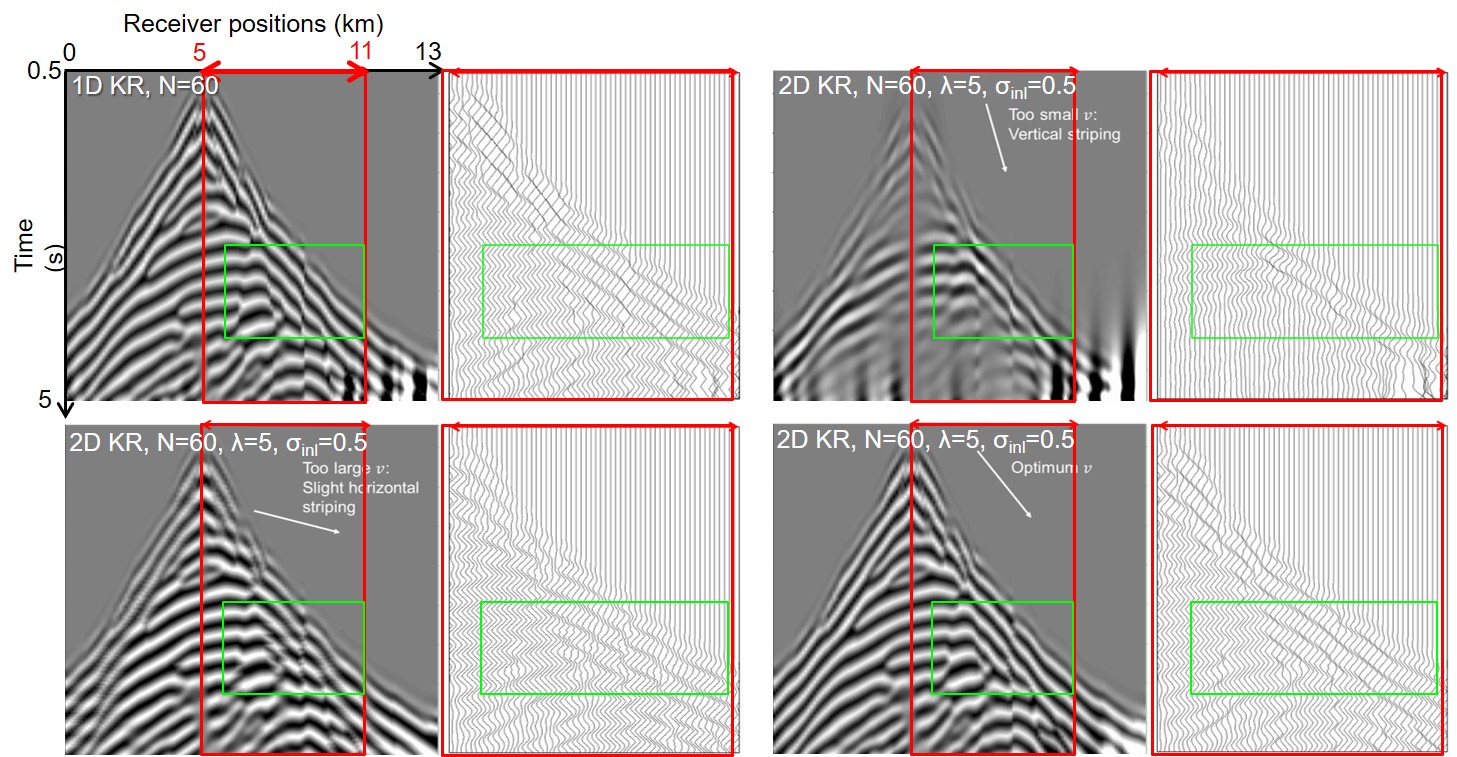}
\caption{
Same data as Fig. \ref{fig:figure3_N}.
1D KR adjoint-source and 2D KR adjoint-sources (effect of $v$) are shown.
}
\label{fig:figure4_marm}
\end{figure}

To better understand this point we note that, according to eq. (\ref{eq:lip_1D_text}), the 1-Lipschitz constraint corresponding to eq. (\ref{eq:l1_omega_2}) implies almost everywhere in $X$
\begin{eqnarray}
\Big|{\frac{\partial\varphi_{max}(x)}{\partial x_{inl}}}\Big|\le \frac{1}{\sigma_{inl}}
\quad\quad\mathrm{ and }\quad\quad
\Big|{\frac{\partial\varphi_{max}(x)}{\partial x_{t}}}\Big|\le \frac{v}{\sigma_{inl}}
.
\end{eqnarray}
Taking a very small $v$ value would constrain $\varphi_{max}$ relative variations to be slower in the time direction than in the inline direction, possibly leading possible to vertical artifacts in the 2D KR adjoint-source; for sufficiently small $v$ and $\sigma_{inl}$, the result becomes equivalent to the 1D KR adjoint-source (up to an irrelevant global proportionality constant).
Taking a large $v$ value would constrain $\varphi_{max}$  relative variations to be slower in the inline direction than in the time direction, possibly leading to some horizontal artifacts.
Taking an optimum $v$ value would tend to:
\begin{itemize}
\item
Favor piecewise linearity of the KR adjoint-source in the time direction, the latter contributing the most to reduce the cycle-skipping issue.
\item
Within these bounds, favor the balance between the $\varphi_{max}$ derivatives in the time and inline directions,
i.e. favor the average moveout direction.
\end{itemize}
This is illustrated in Figs.~\ref{fig:figure4_Raph}-\ref{fig:figure4_marm}.
%, together with the avantage of 2D KR with a good $v$ choice compared to 1D KR and LSQ.

Figs.~\ref{fig:figure4_Raph} considers a simple synthetic 2D LSQ adjoint-source with events built from Ricker wavelets that are similar than these in Fig.~\ref{fig:fig-subest}b.
However, theses events are here extended in a second (inline) dimension applying linear (first two events) and hyperbolic  (last two events) moveouts.
Then, each trace is shifted randomly in the time direction to degrades the continuity of the LSQ adjoint-source's events as visible in Fig.~\ref{fig:figure4_Raph} (left).
The 1D KR adjoint-source, while producing the KR texture for each trace, does not manage to recover the continuity in the moveout direction.
In contrast, the 2D KR adjoint-source corresponding to an optimum $v$ manages to also recover a much better continuity, producing a denoising effect that enhances even more the kinematic information present in the data, which is beneficial to FWI. However, without a suitable choice for $v$, the 2D KR adjoint-source can exhibit striping artefacts to the point of becoming unusable (Fig.~\ref{fig:figure4_Raph}, right).

Fig. \ref{fig:figure4} illustrates the avantage of 2D KR on a field marine data.
This data is polluted by a noise that is different from one trace to the other.
% degrading the amplitude continuity between traces.
Note that a mute is applied to the data, common in industrial FWI to help the inversion concentrate on the most relevant data, i.e. the diving waves.
Fig. \ref{fig:figure4} (top) shows that
the noise in the data degrades the continuity of LSQ adjoint-source and that the 1D KR adjoint-source does not recover any better continuity.
In contrast, the 2D KR adjoint-source with an optimum $v$ is strongly denoised, with an increased continuity in the moveout direction and better amplitudes balancing.
This may be useful to start FWI at even lower frequency and to relax sensitivity to amplitudes.
Fig.~\ref{fig:figure4} (bottom) shows how $v$ affects the 2D KR adjoint-source and how a good choice enhances the continuity of events.

Finally, Fig.~\ref{fig:figure4_marm} illustrates on a synthetic Marmousi adjoint-source
that 2D KR with a good $v$ choice allows
to produce an adjoint-source with an enhanced continuity of events in the moveout direction,
despite the absence of noise in the data.
The effect is then more subtle but nevertheless allows to improve the Marmousi inverted model as further in this section.

We thus observe that accounting for correlations between group of traces 
in a well sampled or non aliased spatial direction
allows to exploit more information than only the time related the events,
like here information related to events continuity.


\subsection{FWI adaptations and interest of KR for the inversion}
\label{sec:ot-int-fwi}

Up to now, the analysis has remained at the adjoint-source level, which has allowed us to understand many features of the use of the KR norm within FWI.
In the rest of this article, we illustrate how these features translate into a better velocity update.
A SDMM number of iterations $N=30$ is considered in all the following results.

Once the gradient in the model space has been computed using the adjoint-state method described in \S \ref{sec:FWI-prob},
it is common to apply to it a diagonal preconditioning followed by a L-BFGS preconditioning and a line search
\cite[]{plessix-2006-adjoint,Virieux2009}.
The cost function choice directly affects the adjoint-source only, but it indirectly affects the FWI through the preconditioning and the line search.
For instance, the derivation of the classical FWI gradient diagonal preconditioning of \cite{Warner2013}
is only valid for the LSQ cost function. However, we have not found any pratical downsides of using it for KR FWI.
Similarly, a single line-search iteration is valid only for the LSQ cost function, with a Hessian equal to the identity; but the line-search can be iterated when using non LSQ cost functions.

\begin{figure}[H]
\centering
\includegraphics[width=1.\linewidth]{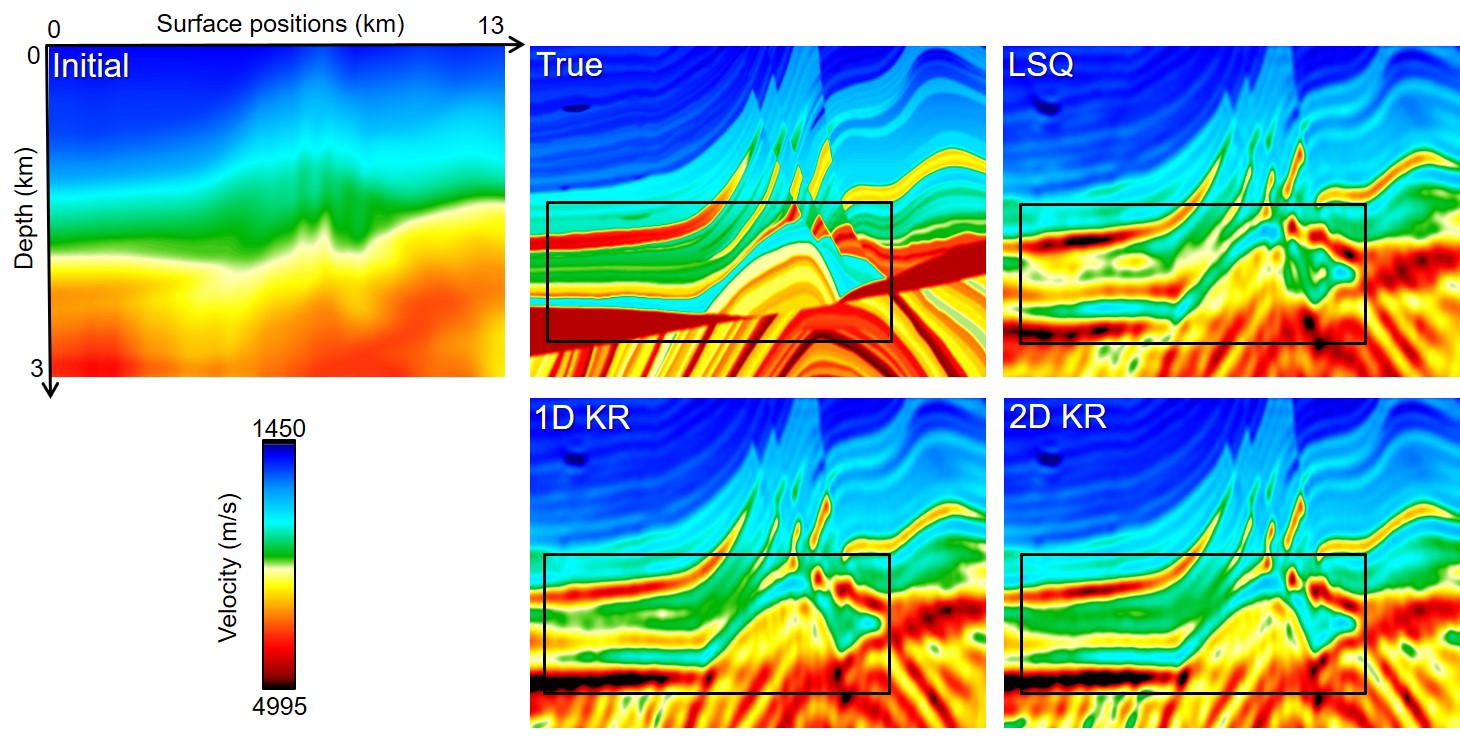}
\caption{
Marmousi 2 model  \cite[]{Martin2006}.
FWI inversion directly at 10 Hz (adjoint-sources in Fig. \ref{fig:figure3_0} are related to this case).
}
\label{fig:figure11_marm}
\end{figure}

%We now study the interest of 2D KR over 1D KR on models inverted by FWI.
We firstly consider the Marmousi 2 data set \cite[]{Martin2006}.
Like above, a Ricker wavelet with a peak frequency at 6 Hz and all frequencies below 3 Hz muted
is used (the adjoint-source examples in Fig. \ref{fig:figure3_0} are related to this case).
The constant-density acoustic approximation has been used for the modelling and only the velocity is inverted by the FWI.
20 FWI iterations were performed directly at up to 10 Hz, keeping all the data (first break, reflections, multiples...) and starting with a smooth initial velocity model where LSQ is cycle-skipped (obtained by Gaussian filtering of the true model), see Fig. \ref{fig:figure11_marm}. 
The model estimated by 1D KR matches the true model better than LSQ, especially in the highlighted zones, mitigating cycle-skipping and producing more continuity along structures, see Fig.~\ref{fig:figure11_marm}.
The texture of the KR adjoint-source (including the enhanced low frequency content) helps overcome cycle-skipping.
(The strength of this effect may depend on the extent of the low frequencies present in the forward propagated source wavelet; sufficient coupling occured in our experiments.)

Note that our Marmousi 1D KR FWI result in Fig. \ref{fig:figure11_marm} looks improved compared to Fig.~27d in \cite{Metivier2016}.
This may be due to the additive constant or null frequency observed in the Marmousi adjoint-sources of \cite{Metivier2016}  (see Fig.~26b of the latter article for a 2D KR adjoint-source).
In practice, this could produce a low frequency smearing that may somewhat overlap the low frequencies present in the forward propagated source wavelet;
then, if the low frequency smearing is different from one trace to the other, and as 1D KR treats each trace independently, this may degrade the 1D KR FWI update.
As discussed in \S \ref{sec:ot-num-2}, we do not observe such an additive constant in our KR adjoint-source results, that may explain our improved Marmousi 2 model update.
However, the conclusion remains the same:
the 2D KR FWI result is better than the 1D KR FWI result, producing a better structural coherency.
The 2D KR adjoint-source continuity enhancement (achieved with the $v$ parameter) maps into a FWI model update structural coherency enhancement.
This remains true even if there is not a lot of noise in the data, i.e. when the adjoint-source improvement brought by 2D KR compared to 1D KR is more subtle than in the case of Fig. \ref{fig:figure4} for instance.

\section{Application to field data}
\label{sec:field}

%Having demonstrated the advantages of 2D KR from the adjoint-source texture and quality of the FWI model update,
%we now illustrate how 2D KR FWI outperforms LSQ FWI on field data examples.

Successful applications of KR FWI on large scale field data have been published in
\cite[]{Poncet2018,Messud2019,Sedova2019,Hermant2019,Carotti2020,Hermant2020}.
In this section, we gather few illustrations from \cite{Messud2019} and \cite{Carotti2020}.
For further details or more illustrations, we invite the reader to refer to the aforementioned articles.

\subsection{Land data}

\begin{figure}[H]
\centering
\includegraphics[width=0.85\linewidth]{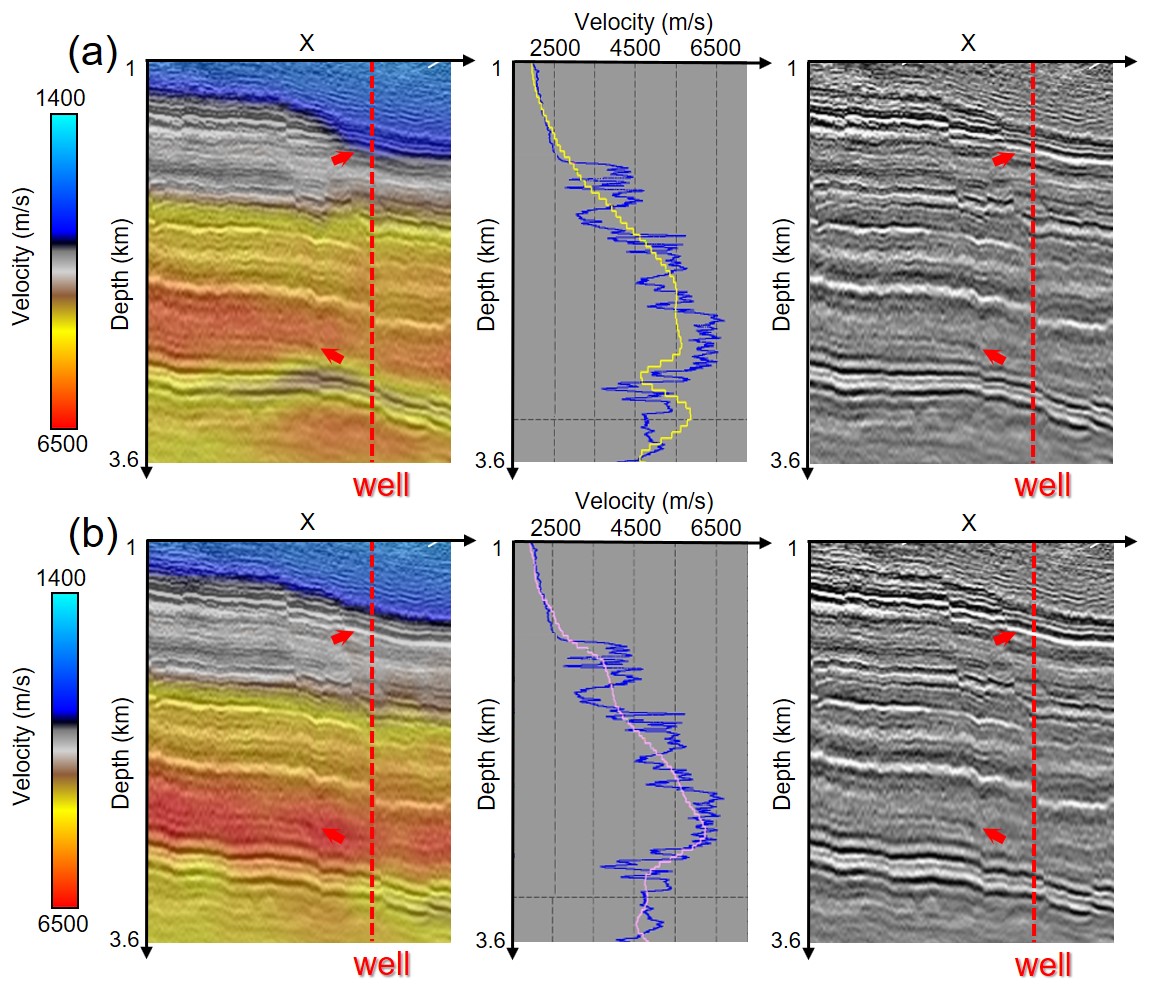}
\caption{
FWI results at 9 Hz on Oman land data. 
%Results of velocity model building using LSQ FWI (left) and multiD KR norm FWI (right) at 9 Hz.
(a) LSQ FWI result and (b) 2D KR FWI result
(same input data, initial model, mute and number of iterations).
Three images are presented for each case.
Left: velocity model superimposed on the corresponding Kirchhoff depth migrated stack.
Middle: 
well measurement (blue) and inverted model trace at the position schematized by the dashed red line.
Right: Kirchhoff depth migrated stack. 
Modified from \cite{Messud2019} (see the article for more details).
}
\begin{figure}[H]
\centering
\includegraphics[width=0.85\linewidth]{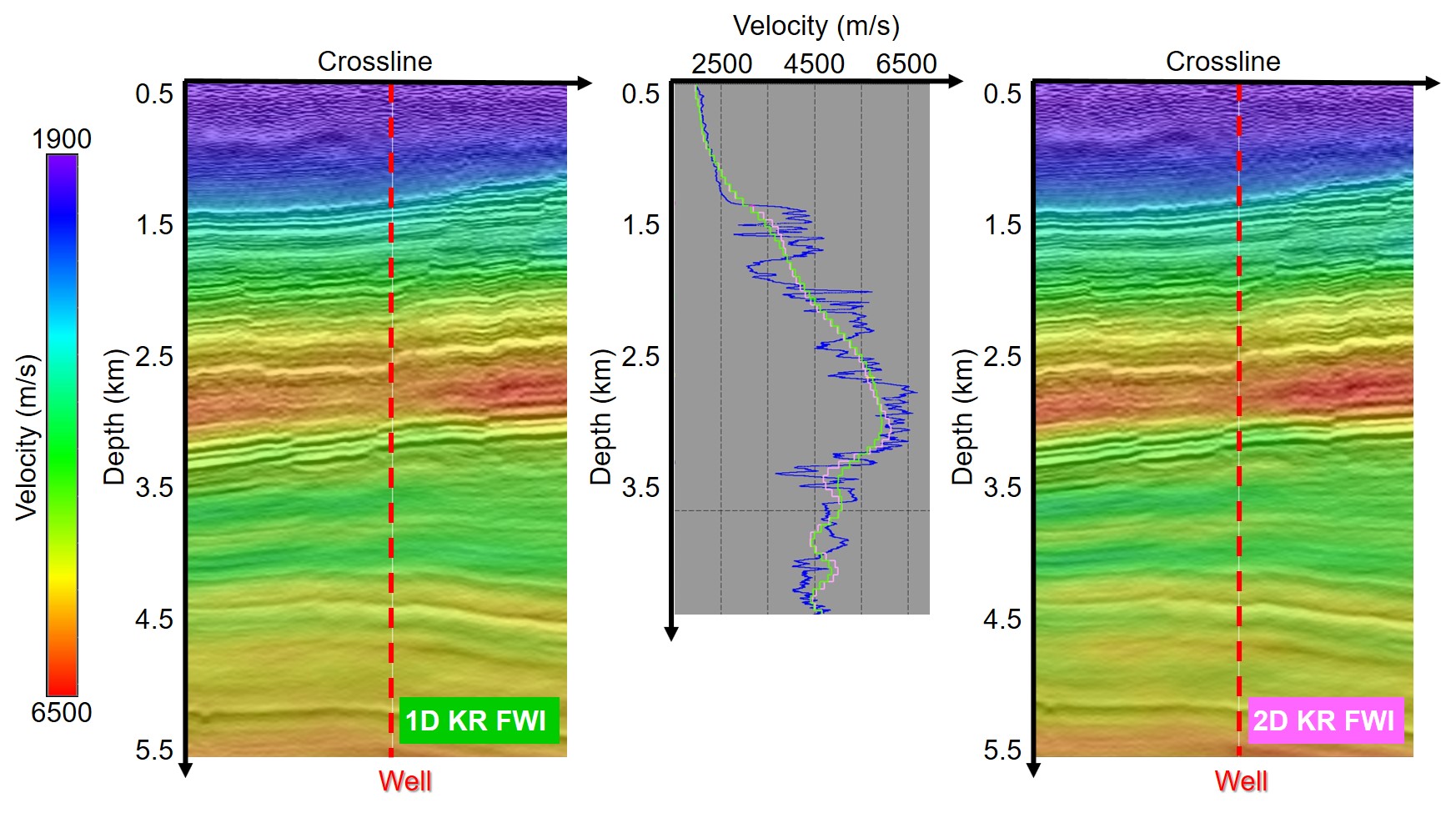}
\caption{
FWI results at 9 Hz on Oman land data. 
Left, right: models inverted at 9 Hz superimposed on the migrated stacks
(same input data, initial model, mute and number of iterations).
Middle: well measurement (blue) and inverted model traces (green is 1D KR and magenta is 2D KR) at the position schematized by the dashed red line.
%Benefit of multiD KR norm FWI (right) compared to 1D KR norm FWI (left), .
Modified from \cite{Messud2019} (see the article for more details).
}
\label{fig:figure5}
\end{figure}
\label{fig:figure6}
\end{figure}

%We used $N\le 30$ inner iterations for each KR problem and a preconditioned L-BFGS optimization scheme for the FWI velocity optimization process.
The first example refers to an Oman land broadband full azimuth data set consisting of three separate acquisitions that have been merged, with challenges associated with irregular offset distributions among the merged surveys.
%As careful denoising of the data has been performed here before FWI, the 1D KR norm adjoint-source is quite similar to the multiD KR norm one. But multiD KR norm nevertheless leads to an improved velocity inversion result as illustrated in
%Fig. \ref{fig:figure5},
%with better structural consistency and well matching.
Fig. \ref{fig:figure6}
shows various FWI results obtained at 9 Hz.
% using the same input data and starting model.
LSQ FWI leads to poor structural consistency and poor fit to the sonic log at the location of the well.
2D KR FWI gives a better fit to the sonic log  and a better consistency with the geology compared to LSQ FWI. This can be related to a reduced sensitivity to cycle-skipping
and to the ability of 2D KR to enhance the continuity along structures.
%Note that 1D KR norm FWI also performs better than LSQ FWI (but less well than multiD KR norm FWI as alread\mu(y) illustrated).
Fig.~\ref{fig:figure5} 
compares 2D KR FWI to 1D KR FWI on the same survey.
As a careful preconditioning of the data has been performed before FWI
(to enhance events coherency, signal to noise ratio and thus convergence of FWI), following the workflow proposed by 
\cite{Sedova2019}, the 1D KR adjoint-source looks quite similar to the 2D KR adjoint-source (not shown here).
1D KR FWI is capable of resolving the cycle-skipping present in LSQ FWI.
However, 2D KR FWI leads to a better model update than 1D KR FWI,
with better structural consistency and well matching.
We underline that the level of improvement of 2D KR FWI over 1D KR FWI needs to be considered in the frame of the preconditioning steps applied to the data.

\begin{figure}[H]
\centering
\includegraphics[width=0.9\linewidth]{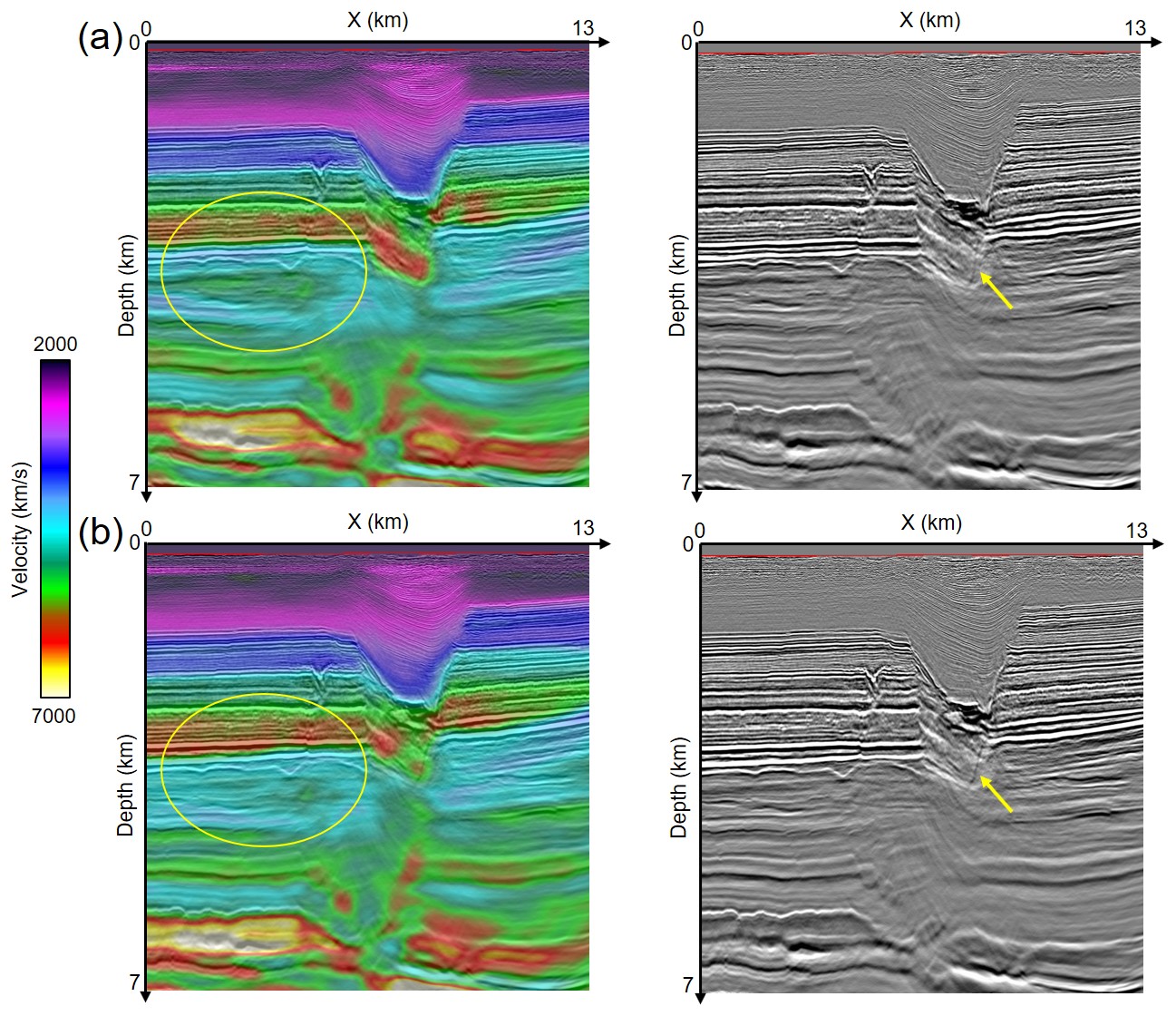}
\caption{
FWI results at 16Hz on North of Oman land data. 
(a) LSQ FWI result and (b) 2D KR FWI result
(same input data, initial model, mute and number of iterations).
Two images are presented for each case.
Left: velocity model superimposed on corresponding Kirchhoff depth migrated stack
(the yellow ovals highlight the improved delineation of the velocity contrast 
and the correction of the velocity increase achieved by 2D KR FWI).
Right: Kirchhoff depth migrated stack
(the yellow arrows highlight the improved focusing of the major fault achieved by 2D KR FWI).
Modified from \cite{Carotti2020} (see the article for more details).
}
\label{fig:figure9}
\end{figure}

The second example refers to a 3D land broadband data set with full-azimuth and offsets of up to 13 km,
again acquired in Oman.
This data set was processed to enhance diving and post-critical waves.
The FWI has been run with the frequency increasing from 2 Hz to 16 Hz, following the workflow proposed by 
\cite{Sedova2019}.
%Starting from a heavily smoothed VTI initial model obtained using a previous FWI result. 
Fig. \ref{fig:figure9}
compares LSQ FWI to 2D KR FWI.
%using the same input data and starting model.
The yellow oval in Fig. \ref{fig:figure9} highlights the improved delineation of the velocity contrast achieved by 2D KR FWI, and the unexpected velocity increase obtained with LSQ FWI that is corrected by 2D KR FWI. This improved velocity provides imaging uplifts of the deep reflectors observable in
Fig.~\ref{fig:figure9}b (right).
Moreover, the focusing of the major fault, a difficult challenge for the area, is enhanced, as highlighted by the yellow arrow in 
Fig.~\ref{fig:figure9}b (right).

\subsection{Marine data}

\begin{figure}[H]
\centering
\includegraphics[width=1.12\linewidth]{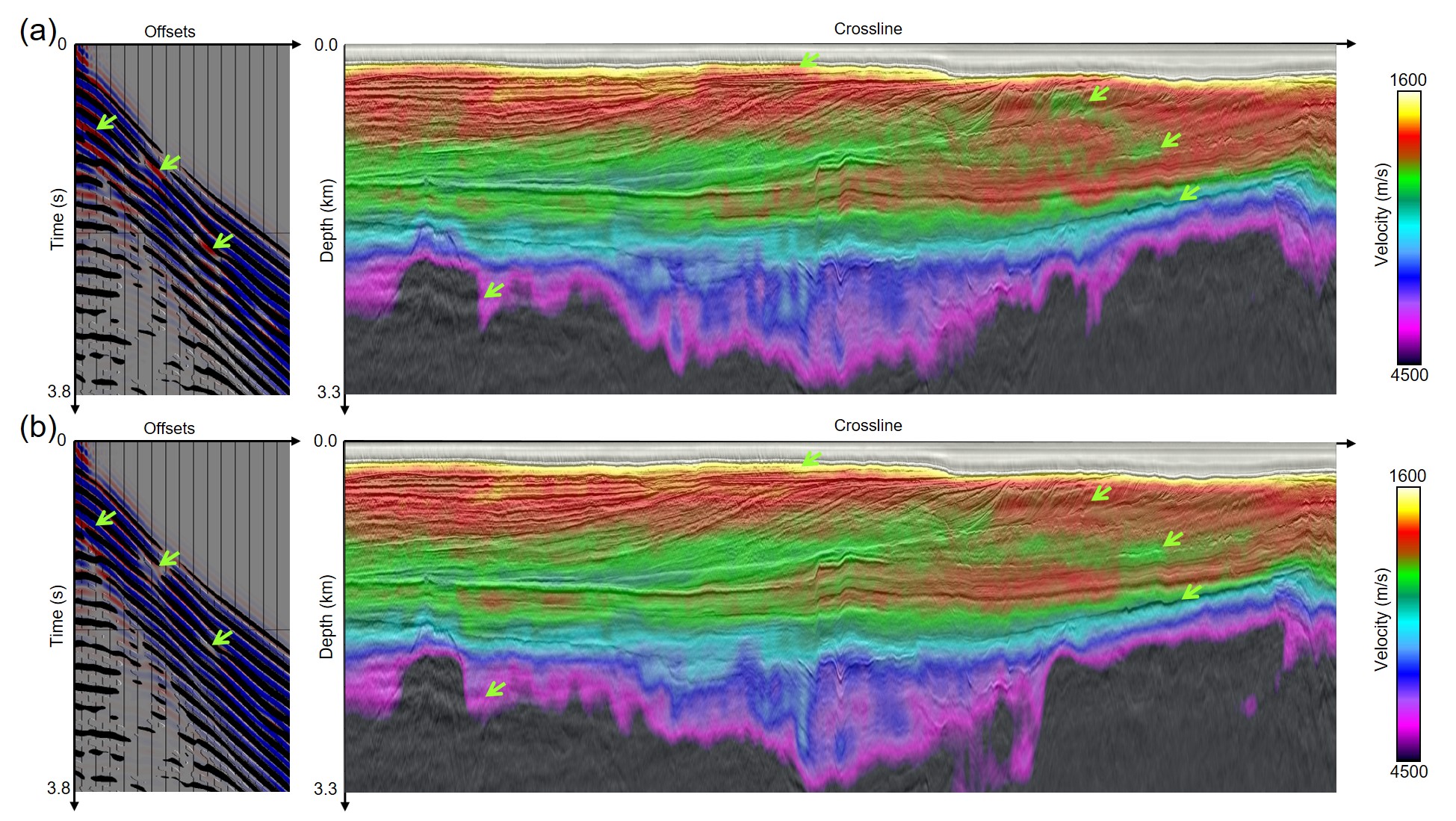}
\caption{
North sea data.
(a) LSQ FWI result and (b) 2D KR FWI result
(same input data, initial model, mute and number of iterations).
Two images are presented for each case.
Left: observed data (in black-grey-white) superimposed to modelled data (in red-blue)
(the green arrows highlight where the modelled data suddently jumps from one ``cycle'' to another with LSQ).
Right: FWI updated models at 7 Hz
(the green arrows highlight some areas where 2D KR FWI gives an improved and more structurally consistent velocity model).
Modified from \cite{Messud2019} (see the article for more details).
}
\label{fig:figure10}
\end{figure}

The third example refers to a North Sea marine data set.
Fig. \ref{fig:figure10}
shows results obtained with a 7 Hz FWI inversion.
%  using the same input data and starting model.
LSQ FWI is cycle-skipped, as indicated by the ``red spots'' in the observed data overlaid on top of the modelled data (highlighted by green arrows).
These red spots are due to events that suddently jump from one ``cycle'' to another in the modelled data, 
resulting in a lack of structural consistency and continuity in the inverted velocity model.
Fig.~\ref{fig:figure10}
shows how 2D KR FWI solves for these issues and inverts for an improved velocity model
(especially in the areas highlighted by green arrows).

\begin{figure}[H]
\centering
\includegraphics[width=1.\linewidth]{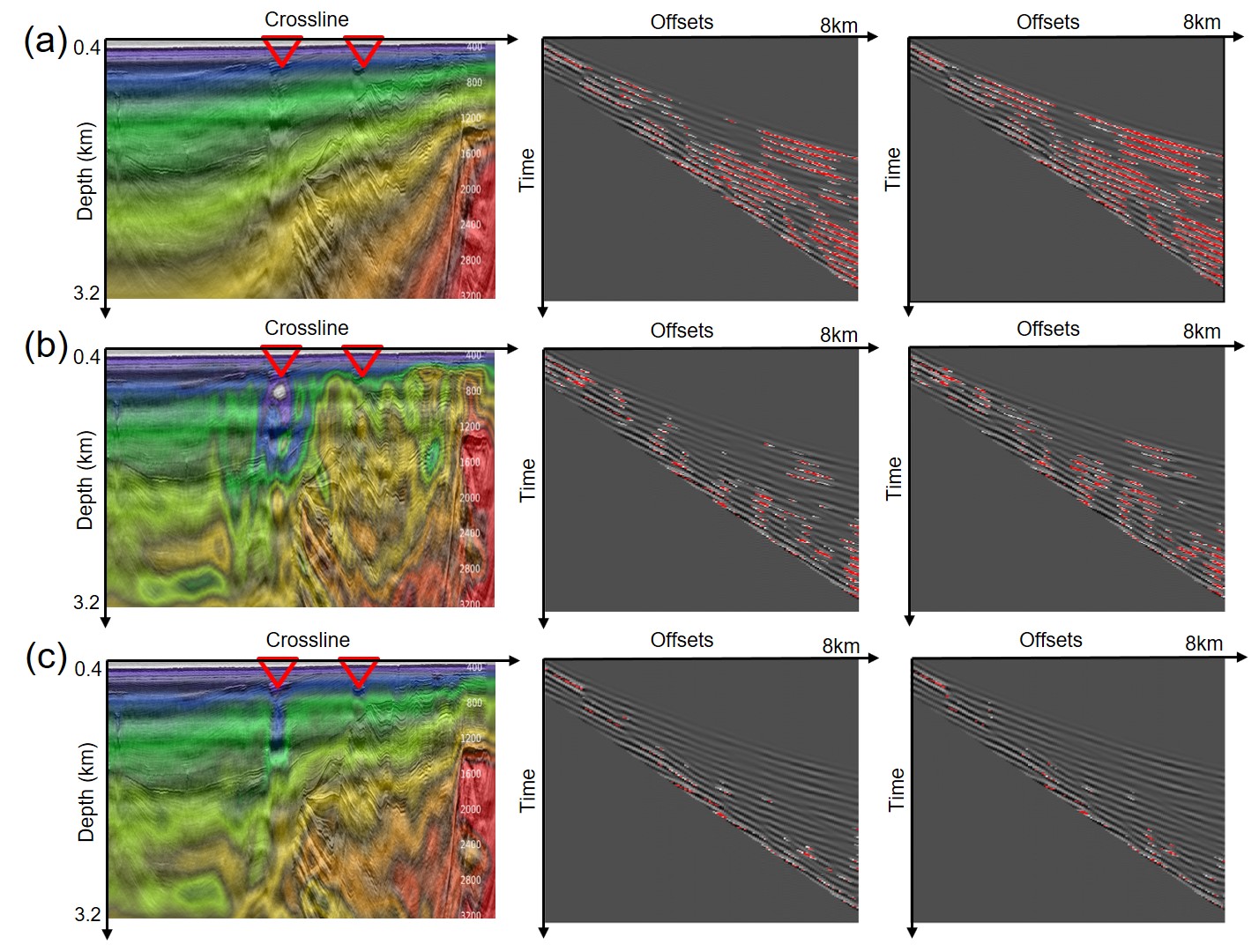}
\caption{
Barents Sea data.
(a) Initial model for FWI, (b) LSQ FWI result and (c) 2D KR FWI result
(same input data, initial model, mute and number of iterations).
%Results of velocity model building using LSQ FWI (left) and multiD KR FWI (right) at 6 Hz.
Three images are presented for each case.
Left: velocity model and migrated stack.
Middle and right: normalized absolute values of the difference between observed and modelled data
(red means large values thus poor matching),
at the two positions highlighted by the red triangles.
Modified from \cite{Carotti2020} (see the article for more details).
}
\label{fig:figure7}
\end{figure}

The last example refers to a Barents Sea marine data set, that is challenging because of gas accumulations of varying size and depth location.
Fig. \ref{fig:figure7}
illustrates the stability of 2D KR FWI over LSQ FWI, the latter being cycle-skipped.
% using the same input data (preprocessed marine data, 8km offset maximum, same mute) and starting model. The cycle-skipping QC on the right (normalised absolute value of the difference between observed and calculated data) allows to analyze the model quality. Both LSQ FWI and 2D KR FWI improve the matching between observed and modelled data. However, 2D KR FWI shows a much better matching than LSQ FWI (less red in Fig. \ref{fig:figure7}, right).
Even with a poor initial model, 2D KR FWI converges to a more structurally consistent model update, while mitigating cycle-skipping
and matching better the observed data.


\section{Perspectives}
\label{sec:perspect}

We end up by providing some perspectives and come-back to adjoint-source considerations.
Examining previous KR adjoint-sources, e.g. Figs. \ref{fig:fig-adj-source-comparison}-\ref{fig:figure4}, 
we notice that:
\begin{itemize}
\item
The kinematics present in the adjoint-source is enhanced by the
KR texture (enhanced low frequencies, more balanced amplitudes and events continuity), 
explaining why the KR norm is more robust to cycle-skipping and amplitudes.
\item
But the kinematics present in the adjoint-source is not modified,
unlike other methods that help to reduce the cycle-skipping issue by shifting events in the data 
and using the corresponding kinematically-modified residual.
See e.g. the method of~\cite{Baek2014} and \cite{Wang2016} that we call ``registration-guided'' FWI in the following (sometimes also called ``dynamic warping'' FWI),
%\cite[]{hale-2013,hale-ma-2013}
or the method of~\cite{Metivier2018,Metivier2019} called ``graph-space'' (GS) OT FWI.
\end{itemize}
Ultimately, we would like to bring both features into a common framework
to combine their advantages (texture change together with kinematics change) and become even more robust to cycle-skipping.
This is possible by introducing the concept of ``kinematic transformation'' (KT) into the KR norm.

We call KT any transformation
that makes the kinematics of the events present in the observed data $f_{obs}$ closer to the kinematics of another (relatively close) data $f$, usually the modelled data.
We consider in this section the example of a KT that works
in the time-direction only, with generalization being straightforward.
$f_{obs}(x_{xl}, x_{inl}, x_t)$
is transformed into
$f_{obs}(x_{xl}, x_{inl}, \sigma[f](x_t) )=f_{obs}^{(\sigma)}(x_{xl}, x_{inl},x_t)$ to become kinematically closer to a data $f(x_{xl}, x_{inl}, x_t)$ from a certain similarity measure point of view.
$\sigma[f]:]0,T]\rightarrow ]0,T]$ represents the KT,
a functional of the data whose kinematics should be (partially) matched.
The registration-guide~\cite[]{Baek2014,Wang2016,Hale2013}
and the permutation resulting from GS \cite[]{Metivier2019} are examples of KT.
Used within FWI, the goal is to change the kinematics of the observed data 
to make it closer to the modelled data at each FWI iteration, to improve robustness to cycle-skipping.
%$\sigma[f]$ tends to the identity as the FWI converges.

%\JMcomm{(a developper bien plus??? Pas sur)}.
Our goal here is not to detail the differences between KT but to give a framework, so we remain general.
To embed a KT into KR, we can simply consider instead of eq. (\ref{eq:OT_10}):
%(remind $x=[x_{xl},x_{inl},x_t]^t$):
%
\begin{eqnarray}
%&&
%{For a given $f_{obs}$}:
&&
\tilde{W}_d(f,f_{obs}^{(\sigma)})
=
\max_{\varphi\in \mathrm{BLip}(||.||_1^{(X)},1,\lambda)}
\int_{X}
\varphi(x)
\Big(
f(x_{xl}, x_{inl}, x_t ) -f_{obs}( x_{xl}, x_{inl}, \sigma[f](x_t) )
\Big)
d\mu(x),
\nonumber
\\
\label{eq:KR_kt_1}
\end{eqnarray}
%
%As any pertinent $\sigma[f]$ should preserve the distinctive marks of $f_{obs}$,
where $\sigma[f]$ is defined by any KT method.
Applying the considerations of \S \ref{sec:ot-as} to eq.~(\ref{eq:KR_kt_1}), we can
consider the following expression as the adjoint-source
%for $x\in X \backslash X_{null'}$
%
\begin{eqnarray}
\frac{\partial \tilde{W}_d(f,f_{obs}^{(\sigma)})}{\partial f(x)}
&=&
\varphi_{max}[f](x)
-
\int_{X}
\varphi_{max}[f](x')
\frac{\partial f_{obs}(x_{xl}', x_{inl}', \sigma[f](x_t') )}{\partial f(x)}
d\mu(x')
%\\
%&=&
%\varphi_{max}[f](x)
%-
%\int_{X}
%\varphi_{max}[f](x')
%\frac{\partial f_{obs}(x_{xl}', x_{inl}', x_t'')}{\partial x_t''}\Big|_{x_t''=\sigma[f](x_t')}
%\frac{\partial \sigma[f](x_t')}{\partial f(x)}
%d\mu(x)'
.
\label{eq:KR_kt_2}
\end{eqnarray}
If
\begin{eqnarray}
\label{eq:KR_kt_3}
\exists \gamma >0,
\forall h \in L^1(X) \mathrm{\hspace{1.5mm}such\hspace{1.5mm}as\hspace{1mm}} ||h||_1 < \gamma:
\sigma[f+h](x)
=
\sigma[f](x)
%\quad\Rightarrow
%\frac{\partial \sigma[f](x_t')}{\partial f(x)}=0
%\quad\Rightarrow
%\frac{\partial \tilde{W}_d(f,f_{obs}^{(\sigma)}))}{\partial f(x)}
%=
%\varphi_{max}[f](x)
,
\end{eqnarray}
i.e. if $\sigma[f]$ is insensitive to infinitesimal perturbations of $f$,
the second term in eq. (\ref{eq:KR_kt_2}) can be neglected and
the adjoint-source can keep the simple form $\varphi_{max}[f]$ used in KR FWI
without KT (the adjoint-source form then would remain the same, but not the adjoint-source itself as the KT would affect the $\varphi_{max}$ values through eq. (\ref{eq:KR_kt_1})).
So, do some existing KT fulfill eq. (\ref{eq:KR_kt_3})?
\begin{itemize}
\item
It seems not strictly the case for the registration-guide~\cite[]{Baek2014,Wang2016},
even if in practice the scheme neglects $\partial f_{obs}(x_{xl}', x_{inl}', \sigma[f](x_t') )/\partial f(x)
$ ``by hand'', arguing it represents a second order term, non-crucial because of the non-linearity of the FWI.
\item%In practice, any dynamic warping scheme neglect this term~\cite[]{Baek2014,Wang2016,Hale2013},
%even if not strictly correct.
But it is strictly the case for GS as demonstrated formally in \cite{Metivier2019}
(in a discrete $X$ space case as GS only deals with discrete data coordinate spaces).
This is a result in favor
of the use of GS as a KT within KR FWI, at least mathematically.
Physically, the pertinence of GS OT FWI compared to registration-guided FWI still needs to studied.
\end{itemize}

\begin{figure}[H]
\centering
\includegraphics[width=1.\linewidth]{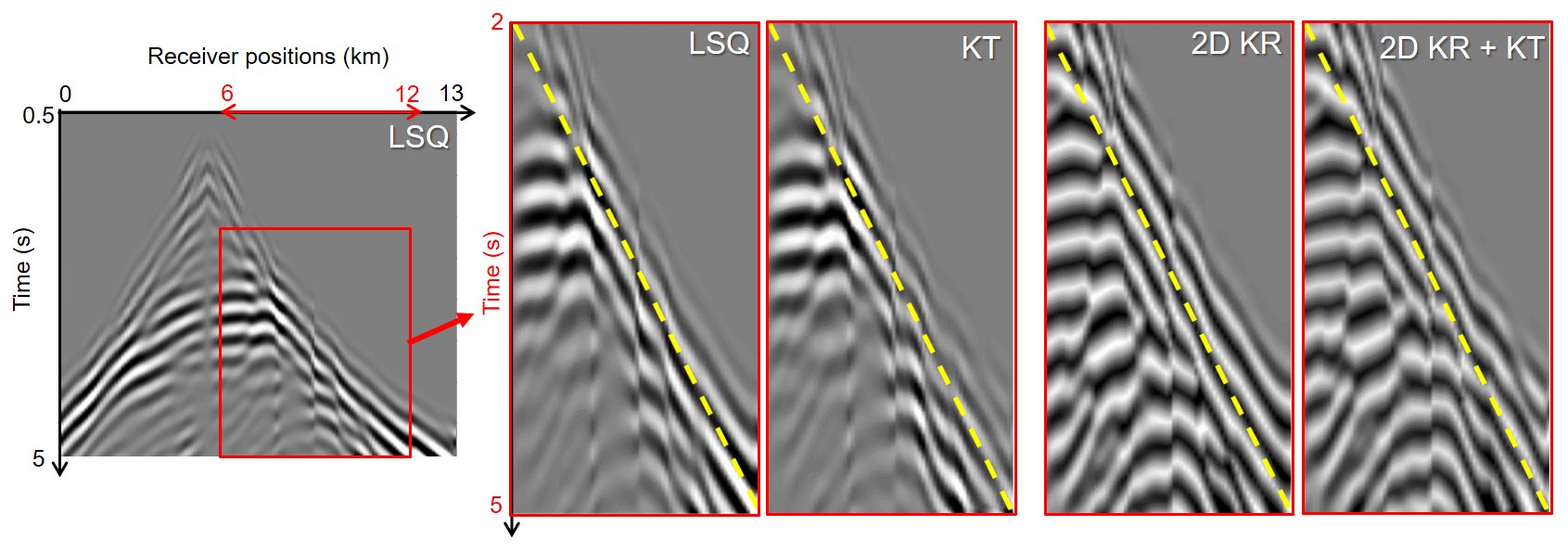}
\caption{
Same data as Fig. \ref{fig:figure3_N}, zoom.
Various adjoint-sources compared: LSQ, registration-guided KT, 2D KR and 2D KR-embedded registration-guided KT.
The yellow dashed line helps comparing the kinematic change brought by the KT in the diving waves area.
}
\label{fig:figure12_marm}
\end{figure}

To conclude with an illustration,
Fig. \ref{fig:figure12_marm} (left) shows how the registration-guided KT can affect the LSQ residual.
We notice that the texture remains quite the same
but the kinematics of some events in the residual change, especially in the diving wave area.
On the contrary, the texture of the 2D KR adjoint-source is quite different from the LSQ residual,
but the kinematics of the events does not change.
Fig. \ref{fig:figure12_marm} (right) shows the KR adjoint-source with a KT embedded;
we can observe the KR texture together together with some events kinematics change.
Embedding a KT can thus naturally ``augment'' the KR scheme, possibly for even more robustness to cycle-skipping.
Comparing  various possible KTs goes beyond the scope of this article.
We mention this point as a perspective,
that has started to be explored in \cite{Kpadonou2021}.

\section{Conclusion}
\label{sec:concl}

The use of the KR norm in FWI is now well established and has proven its benefits for practical large-scale applications. 
Thanks to a significantly wider global minimum valley, several field data case studies have shown that KR FWI clearly outperformed LSQ FWI in overcoming moderate cycle-skipping. 
In this article, 
we have developed an analysis to better understand the benefits of the KR norm for FWI, mainly focusing on the adjoint-source. 
We have addressed both theoretical aspects, fundamentally justifying the KR adjoint-source expression and qualifying its ``texture'', and practical points that are critical for an efficient use within FWI. 
The KR adjoint-source can be conceptualized as the result of smart processing applied to the LSQ residual,
that enhances the low frequency content and reduces the dynamics of amplitudes. 
When correctly tuned, the 2D (or multiD) KR adjoint-source appears to also enhance the lateral continuity of events (in a well sampled or non aliased spatial direction) within a common shot gather. All these components, reinforcing the kinematic content in the adjoint-source, are beneficial to a successful FWI implementation.

If the use of the KR norm within FWI significantly widens the global minimum valley compared to the use of LSQ, it does not fully solve the cycle-skipping issue or the challenge of FWI with reflected waves only.
Further investigations in these directions are ongoing
\cite[]{Metivier2019,Sun2019,Tang2020,Kpadonou2021}.

\ack{
We are grateful to CGG for granting permission to publish this work and to CGG Multi-Client, TGS, Occidental Petroleum, PDO, the Ministry of Oil and Gas of the Sultanate of Oman and INEOS for providing the data and permission to present the field data results.
We are grateful to Klaas Koster for help and support and to Ludovic M\'etivier for enlightening discussions.
We are indebted to F\'elix Kpadonou, Anna Sedova, Nabil Masmoudi, Adel Khalil and Roger Taylor for collaboration and enlightening discussions.
We are indebted to Thibault Lesieur and Thibaut Allemand for collaboration and enlightening advice regarding \ref{app:proofs_0}.
}

\newpage
\clearpage

\appendix
\section{Dual representation of the Lipschitz constraint (proof)}
\label{app:proofs_0}

We consider $d\rightarrow ||.||_p^{(X)}$ in the Lipschitz constraint, eq. (\ref{eq:OT_7}),
and $(x,y)\in X\times X$.
The Lipschitz semi-norm is defined by
\begin{eqnarray}
\label{eq:W0_app00}
||\varphi||^{ }_{Lip_p}
=
\max_{x\ne y}\frac{|\varphi(x)-\varphi(y)|}{||x-y||_p^{(X)}}
.
\end{eqnarray}
We obtain a norm identifying functions $\varphi$ defined up to an additive constant.

Taking $x=y+z$ with $z\in X$
and choosing a $\varphi(y)$ that is differentiable, the Lipschitz norm can be rewritten ($p\ge 1$)
\begin{eqnarray}
\label{eq:W0_app2}
||\varphi||^{ }_{Lip_p}
&=&
\max_{y}\max_{z \ne 0}
\frac{|\varphi(y + z)-\varphi(y)|}{||z||_p^{(X)}}
=
\max_{y}
\max_{z\ne 0}
\frac{
\Big|\frac{\partial \varphi(y)}{\partial y}.z\Big|
}{
||z||_p^{(X)}
}
=
\max_{y}
\max_{||z||_p^{(X)}\le 1}
\Big|\frac{\partial \varphi(y)}{\partial y}.z\Big|
\nonumber\\
&=&
\max_{y}
\Big|\Big|\frac{\partial \varphi(y)}{\partial y}\Big|\Big|_q^{(X)}
\quad\mathrm{with}\quad
1/p+1/q=1
.
\end{eqnarray}
The last equality is obtained by definition of the dual norm \cite[]{Rudin1991,Brezis1983}.
The Lipschitz norm can thus be ``represented'' by a $L_q$ norm $||.||^{(X)}_{q}$ on $\partial \varphi(y)/\partial y$.

Let us consider standard-deviation-like weights that satisfy $\infty>1/\sigma>\epsilon$ where $\epsilon >0$.
When the inverses of the $\sigma$ are introduced in $||.||_p^{(X)}$, i.e.
\begin{eqnarray}
|| x||_p^{(X)}
=
\Big(
\frac{|x_{xl}|^p}{\sigma_{xl}^p}
+
\frac{|x_{inl}|^p}{\sigma_{inl}^p}
+
\frac{|x_t|^p}{\sigma_t^p}
\Big)^{1/p}
,
\end{eqnarray}
eq. (\ref{eq:W0_app2}) imposes that the non-inverted $\sigma$ appear in $||.||_q^{(X)}$, i.e.
\begin{eqnarray}
\Big|\Big| \frac{\partial \varphi(x)}{\partial x}\Big|\Big|_q^{(X)}
=
\Big(
\sigma_{xl}^q\Big|\frac{\partial \varphi(x)}{\partial x_{xl}}\Big|^q
+
\sigma_{inl}^q\Big|\frac{\partial \varphi(x)}{\partial x_{inl}}\Big|^q
+
\sigma_t^q\Big|\frac{\partial \varphi(x)}{\partial x_t}\Big|^q
\Big)^{1/q}
\quad\mathrm{with}\quad
1/p+1/q=1
.
\nonumber\\
\end{eqnarray}

Note that the second equality in eq. (\ref{eq:W0_app2}) is valid when the Riesz representation theorem can be invoked in $X^*$ (\cite[]{Brezis1983} chapter IV), where $X^*=\mathcal{L}(X,\mathbb{R})$ denotes the (topological) dual of $X$.
This is the case for any norm in finite dimensional spaces like $X$ that is 3 dimensional here.
(For completeness, we remind that this is not the case in infinite dimensional space \cite[]{Rudin1991,Brezis1983}).

As the 1-Lipschitz constraint implies strong continuity, i.e. differentiability almost everywhere, the derivative $\partial\varphi(x)/\partial x$ is defined almost everywhere.
Constraining the derivative almost everywhere is sufficient for the Lagrange multipliers theoretical considerations of \ref{app:proofs}.

\section{KR adjoint-source (proofs)}
\label{app:proofs}

Throughout this Appendix, we use the notations of \S \ref{sec:ot-as}.

\subsection{Simplified case (thresholding constraint only)}

We first consider only the thresholding constraint in eqs. (\ref{eq:OT_7}) and (\ref{eq:OT_10}).
It implies $\forall x\in X: |\varphi(x)|\le \lambda$,
that can be split into two linear constraints $\varphi(x)\le \lambda$ and $\varphi(x)\ge -\lambda$.
Using the Lagrange multipliers method \cite[]{Bertsekas1996}, eq. (\ref{eq:OT_10}) can be rewritten
(dependencies to $\Delta f$ are implicit in the following to lighten the notations)
\begin{eqnarray}
&&
\tilde{W}_d
=
\max_{\varphi}
\min_{\alpha^+\ge 0,\alpha^-\ge 0}
L(\varphi,\alpha^+,\alpha^-)
\label{eq:proofs_1}\\
&&
L(\varphi,\alpha^+,\alpha^-)
=
\int_{X}
\varphi(x)\Delta f(x)d\mu(x)
-
\int_{X}
(\varphi(x)-\lambda)
\alpha^+(x)
d\mu(x)
-
\int_{X}
(-\varphi(x)-\lambda)
\alpha^-(x)
d\mu(x)
.
\nonumber
\end{eqnarray}
Minimization of eq. (\ref{eq:proofs_1}) with respect to $\alpha^+$ and $\alpha^-$ gives the Karush-Kuhn-Tucker (KKT) conditions \cite[]{Bertsekas1996}
\begin{eqnarray}
&
\varphi(x)=\lambda
&
\quad\mathrm{and}\quad
\alpha^+(x)> 0
\nonumber\\
\mathrm{or \hspace{1.5mm}}
&
\varphi(x)<\lambda
&
\quad\mathrm{and}\quad
\alpha^+(x)= 0
\nonumber\\
&
-\varphi(x)=\lambda
&
\quad\mathrm{and}\quad
\alpha^-(x)> 0
\nonumber\\
\mathrm{or \hspace{1.5mm}}
&
-\varphi(x)<\lambda
&
\quad\mathrm{and}\quad
\alpha^-(x)= 0
.
\label{eq:proofs_2}
\end{eqnarray}
We deduce
\begin{eqnarray}
&&
\alpha^+(x)> 0\Rightarrow \alpha^-(x)=0
\nonumber\\
&&
\alpha^-(x)> 0\Rightarrow \alpha^+(x)=0
.
\label{eq:proofs_3}
\end{eqnarray}
Maximization of eq. (\ref{eq:proofs_1}) with respect to $\varphi$ gives almost everywhere in $X$
\begin{eqnarray}
&&
\Delta f(x)=\alpha^+(x)-\alpha^-(x)
.
\label{eq:proofs_4}
\end{eqnarray}
Combining eqs. (\ref{eq:proofs_2})-(\ref{eq:proofs_4}),
we deduce
\begin{eqnarray}
&&
{x\in X \backslash X_{null}^{(\Delta f)} \mathrm{\hspace{1.5mm}with\hspace{1.5mm}} \Delta f(x)>0:}
\quad
\alpha^+(x)=\Delta f(x),\alpha^-(x)=0
,\varphi_{max}(x)=\lambda
\label{eq:proofs_5}\\
&&
{\hspace{3.1cm} \Delta f(x)<0:}
\quad
\alpha^-(x)=-\Delta f(x),\alpha^+(x)=0
,\varphi_{max}(x)=-\lambda
%\nonumber\\
%&&
%{$\forall x\in X_{null}^{(\Delta f)}$: $\varphi_{max}(x)$ values do not matter, remind \S \ref{sec:ot-as}}
.
\nonumber
\end{eqnarray}
This result defines $\varphi_{max}$ in $X \backslash X_{null}^{(\Delta f)}$,
where it obviously remains stable under infinitesimal perturbations of $\Delta f$.
Using the Lagrange multipliers to deduce this result in this simple case is somewhat ``oversized''.
However, the method becomes interesting when considering more involved constraints as below,
having the advantage to easilly generalize .
%We thus can deduce eq. (\ref{eq:OT_12_0}) in $X\backslash X_{null}^{(\Delta f)}$, that is sufficient as justified previously.

%Note that the constraint saturates everywhere in $X \backslash X_{null}^{(\Delta f)}$, i.e. in everywhere the subset that matters.

\subsection{Intermediary case (1-Lipschitz constraint only, 1D coordinate space)}

We here consider only the 1-Lipschitz constraint in eq. (\ref{eq:OT_10}), and a 1D coordinate space $X$ that represents the time coordinate in practice.
To lighten the notations, we consider a unit standard-deviation-like weight in eq. (\ref{eq:OT_7}).
From the considerations of \ref{app:proofs_0},
the 1-Lipschitz constraint can be rewritten almost-everywhere in $X$:
$
\Big|{\frac{\partial\varphi(x)}{\partial x}}\Big|\le 1
$.
This constraint can be split into two linear constraints ${\frac{\partial\varphi(x)}{\partial x}}\le 1$ and ${\frac{\partial\varphi(x)}{\partial x}}\ge -1$.
Using the Lagrange multipliers method \cite[]{Bertsekas1996}, eq. (\ref{eq:OT_10}) can be rewritten
\begin{eqnarray}
&&
\tilde{W}_d
=
\max_{\varphi}
\min_{\beta^+\ge 0,\beta^-\ge 0}
L(\varphi,\beta^+,\beta^-)
\label{eq:proofs_1b}\\
&&
L(\varphi,\beta^+,\beta^-)
=
\int_{X}
\varphi(x)\Delta f(x)d\mu(x)
\nonumber\\
&&\hspace{2.7cm}
-
\int_{X}
\Big(\frac{\partial\varphi(x)}{\partial x}-1\Big)
\beta^+(x)
d\mu(x)
-
\int_{X}
\Big(-\frac{\partial\varphi(x)}{\partial x}-1\Big)
\beta^-(x)
d\mu(x)
.
\nonumber
\end{eqnarray}
Minimization with respect to $\beta^+$ and $\beta^-$ gives the KKT conditions \cite[]{Bertsekas1996}
\begin{eqnarray}
&
\frac{\partial\varphi(x)}{\partial x}=1
&
\quad\mathrm{and}\quad
\beta^+(x)> 0
\label{eq:proofs_2b}\\
\mathrm{or \hspace{1.5mm}}
&
\frac{\partial\varphi(x)}{\partial x}<1
&
\quad\mathrm{and}\quad
\beta^+(x)= 0
\nonumber\\
&
-\frac{\partial\varphi(x)}{\partial x}=1
&
\quad\mathrm{and}\quad
\beta^-(x)> 0
\nonumber\\
\mathrm{or \hspace{1.5mm}}
&
-\frac{\partial\varphi(x)}{\partial x}<1
&
\quad\mathrm{and}\quad
\beta^-(x)= 0
.
\nonumber
\end{eqnarray}
We deduce
\begin{eqnarray}
&&
\beta^+(x)> 0\Rightarrow \beta^-(x)=0
\mathrm{\hspace{1.5mm}and\hspace{1.5mm}thus\hspace{1mm}}
\partial\beta^-(x)/\partial x=0
\label{eq:proofs_3b}\\
&&
\beta^-(x)> 0\Rightarrow \beta^+(x)=0
\mathrm{\hspace{1.5mm}and\hspace{1.5mm}thus\hspace{1mm}}
\partial\beta^+(x)/\partial x=0
.
\nonumber
\end{eqnarray}
The ``and thus'' is due to the strong continuity of $\varphi$.
For instance, if ${\partial\varphi(x)}/{\partial x}= 1$, it cannot ``jump'' to ${\partial\varphi(x')}/{\partial x'}= -1$ in the neighborhood of $x$.
Thus, if $\beta^+(x)> 0$, $\beta^-(x')$ must remain null in the neighborhood of $x$.

Considering $X=[x_{min},x_{max}]$, maximization with respect to $\varphi$ gives almost everywhere in $X$
\begin{eqnarray}
&&
\Delta f(x)
+\Big(\delta(x-x_{min})-\delta(x-x_{max})\Big)
\Big( \beta^+(x)-\beta^-(x) \Big)
\Big( \beta^+(x)-\beta^-(x) \Big)
\label{eq:proofs_3b_new}\\
&&
=
-\frac{\partial\beta^+(x)}{\partial x}+\frac{\partial\beta^-(x)}{\partial x}
.
\nonumber
\end{eqnarray}
Combining eqs. (\ref{eq:proofs_3b}) and (\ref{eq:proofs_3b_new}),
we can deduce that always one of the constraints must saturate almost everywhere in $X \backslash X_{null}^{(\Delta f)}$ (the edges $x_{min}$ and $x_{max}$ of $X$ have null measure).
More precisely, we have almost everywhere in $X \backslash X_{null}^{(\Delta f)}$
\begin{eqnarray}
\label{eq:proofs_5b}
{x\in X \backslash X_{null}^{(\Delta f)}:}
&
\hspace{0.42cm}
\frac{\partial\beta^+(x)}{\partial x}=-\Delta f(x)
\Rightarrow
\beta^+(x)>0,
\beta^-(x)=0,\frac{\partial\varphi_{max}(x)}{\partial x}=1
\nonumber\\
&
\mathrm{or\hspace{1.5mm}}
\frac{\partial\beta^-(x)}{\partial x}=\Delta f(x)
\Rightarrow
\beta^-(x)>0,\beta^+(x)=0,\frac{\partial\varphi_{max}(x)}{\partial x}=-1
.
\nonumber\\
\end{eqnarray}

Complementarilly, integrating eq. (\ref{eq:proofs_3b_new}) leads to
\begin{eqnarray}
\label{eq:proofs_4b}
&&
-\Delta F_{\beta}(x) =\beta^+(x)-\beta^-(x)
\\
&&
\Delta F_{\beta}(x)=\int_{x_{min}}^x\Delta f(x')d\mu(x')
%+ C_{\beta}
-\Big(\beta^+(x_{min})-\beta^-(x_{min})\Big)
.
\nonumber
\end{eqnarray}
Combining eqs. (\ref{eq:proofs_3b}) and (\ref{eq:proofs_4b}),
we deduce almost everywhere in $X \backslash X_{null}^{(\Delta F_{\beta})}$
\begin{eqnarray}
&&
{x\in X \backslash X_{null}^{(\Delta F_{\beta})} \mathrm{\hspace{1.5mm}with\hspace{1.5mm}} \Delta F_{\beta}(x)<0:}
\quad
\beta^+(x)=-\Delta F_{\beta}(x),\beta^-(x)=0,
\frac{\partial\varphi_{max}(x)}{\partial x}=1
\nonumber\\
%\label{eq:proofs_5b}\\
&&
{\hspace{3.25cm} \Delta F_{\beta}(x)>0:}
\quad
\beta^-(x)=\Delta F_{\beta}(x),\beta^+(x)=0,
\frac{\partial\varphi_{max}(x)}{\partial x}=-1
%\nonumber\\
%&&
%{$\forall x\in X_{null}$: $\varphi_{max}(x)$ values do not matter, remind \S \ref{sec:ot-as}}
.
\nonumber\\
\label{eq:proofs_5b_bis}
\end{eqnarray}
Eq. (\ref{eq:proofs_5b_bis}) is satisfied in $X\backslash X_{null}^{(\Delta F_{\beta})}$
and thus in $X \backslash X_{null}^{(\Delta f)}$
(as $X\backslash X_{null}^{(\Delta F_{\beta})}$ tends to be ``larger'' for seismic data).

Obviously, eq. (\ref{eq:proofs_3b_new}) (combined with eq. (\ref{eq:proofs_3b})), eq. (\ref{eq:proofs_5b}) or eq. (\ref{eq:proofs_5b_bis}) remain  stable under infinitesimal perturbations of $\Delta f$ in $X \backslash X_{null}^{(\Delta f)}$.
Also, eq. (\ref{eq:proofs_3b_new}) (combined with eq. (\ref{eq:proofs_3b})) is sufficient to prove that always one of the constraints must saturate almost everywhere in $X \backslash X_{null}^{(\Delta f)}$.
We introduced eqs. (\ref{eq:proofs_5b}) and eq. (\ref{eq:proofs_5b_bis}) to give complementary insight.

\subsection{Full case (1D coordinate space)}

We now combine the two constraints (thresholding and 1-Lipschitz), still in the 1D coordinate space case
and with unit standard-deviation-like weights.
Using the Lagrange multipliers method \cite[]{Bertsekas1996},
we write
\begin{eqnarray}
&&
\tilde{W}_d
=
\max_{\varphi}
\min_{\alpha^+ \ge 0,\alpha^-\ge 0,\beta^+\ge 0,\beta^-\ge 0}
L(\varphi,\alpha^+,\alpha^-,\beta^+,\beta^-)
\label{eq:proofs_1t}\\
&&
L(\varphi,\alpha^+,\alpha^-,\beta^+,\beta^-)
=
\int_{X}
\varphi(x)\Delta f(x)d\mu(x)
\nonumber\\
&&
\hspace{4cm}
-
\int_{X}
(\varphi(x)-\lambda)
\alpha^+(x)
d\mu(x)
-
\int_{X}
(-\varphi(x)-\lambda)
\alpha^-(x)
d\mu(x)
\nonumber\\
&&
\hspace{4cm}
-
\int_{X}
\Big(\frac{\partial\varphi(x)}{\partial x}-1\Big)
\beta^+(x)
d\mu(x)
-
\int_{X}
\Big(-\frac{\partial\varphi(x)}{\partial x}-1\Big)
\beta^-(x)
d\mu(x)
.
\nonumber
\end{eqnarray}
Minimization with respect to $\alpha^+$, $\alpha^-$, $\beta^+$ and $\beta^-$ gives KKT conditions \cite[]{Bertsekas1996}
\begin{eqnarray}
&
\varphi(x)=\lambda
&
\quad\mathrm{and}\quad
\alpha^+(x)> 0
\nonumber\\
\mathrm{or \hspace{1.5mm}}
&
\varphi(x)<\lambda
&
\quad\mathrm{and}\quad
\alpha^+(x)= 0
\label{eq:proofs_2t}\\
&
-\varphi(x)=\lambda
&
\quad\mathrm{and}\quad
\alpha^-(x)> 0
\nonumber\\
\mathrm{or \hspace{1.5mm}}
&
-\varphi(x)<\lambda
&
\quad\mathrm{and}\quad
\alpha^-(x)= 0
\nonumber\\
&
\frac{\partial\varphi(x)}{\partial x}=1
&
\quad\mathrm{and}\quad
\beta^+(x)> 0
\nonumber\\
\mathrm{or \hspace{1.5mm}}
&
\frac{\partial\varphi(x)}{\partial x}<1
&
\quad\mathrm{and}\quad
\beta^+(x)= 0
\nonumber\\
&
-\frac{\partial\varphi(x)}{\partial x}=1
&
\quad\mathrm{and}\quad
\beta^-(x)> 0
\nonumber\\
\mathrm{or \hspace{1.5mm}}
&
-\frac{\partial\varphi(x)}{\partial x}<1
&
\quad\mathrm{and}\quad
\beta^-(x)= 0
.
\nonumber
\end{eqnarray}
None of these constraints can saturate simultaneously on non-null measure sets,
as saturating the thresholding constraint implies a null derivative for $\varphi$
(and thus $\beta^+(x)=\beta^-(x)=0$)
and saturating the 1-Lipschitz constraints
implies a non-constant $\varphi$
(and thus $\alpha^+(x)=\alpha^-(x)=0$).
We deduce almost everywhere
\begin{eqnarray}
&&
\hspace{-0.5cm}
\alpha^+(x)> 0\Rightarrow \alpha^-(x)=0, 
\beta^+(x)=\partial\beta^+(x)/\partial x= 0, \beta^-(x)=\partial\beta^-(x)/\partial x=0
\label{eq:proofs_3t}\\
&&
\hspace{-0.5cm}
\alpha^-(x)> 0\Rightarrow \alpha^+(x)=0, 
\beta^+(x)=\partial\beta^+(x)/\partial x= 0, \beta^-(x)=\partial\beta^-(x)/\partial x=0
\nonumber\\
&&
\hspace{-0.5cm}
\beta^+(x)> 0\Rightarrow \beta^-(x)=\partial\beta^-(x)/\partial x=0, \alpha^+(x)= 0, \alpha^-(x)=0
\nonumber\\
&&
\hspace{-0.5cm}
\beta^-(x)> 0\Rightarrow \beta^+(x)=\partial\beta^+(x)/\partial x=0, \alpha^+(x)= 0, \alpha^-(x)=0
.
\nonumber
\end{eqnarray}
The justification of the configurations where the derivatives of $\beta^+$ and $\beta^-$ are null was already given after eq. (\ref{eq:proofs_3b}).

Considering $X=[x_{min},x_{max}]$, maximization with respect to $\varphi$ gives
\begin{eqnarray}
\label{eq:proofs_4t_2}
&&
\Delta f(x)
+\Big(\delta(x-x_{min}) - \delta(x-x_{max})\Big)\Big( \beta^+(x)-\beta^-(x) \Big)
\\
&&
=
\alpha^+(x)-\alpha^-(x)
-\frac{\partial\beta^+(x)}{\partial x}+\frac{\partial\beta^-(x)}{\partial x}
.
\nonumber
\end{eqnarray}
Combining eqs. (\ref{eq:proofs_3t}) and (\ref{eq:proofs_4t_2}),
we can deduce that always one of the constraints must saturate almost everywhere in $X \backslash X_{null}^{(\Delta f)}$, and that eq. (\ref{eq:proofs_4t_2}) remains stable under infinitesimal perturbations of $\Delta f$ in $X \backslash X_{null}^{(\Delta f)}$.

Just to gain further insight, we introduce
\begin{eqnarray}
\label{eq:proofs_5b_2}
&&
\Delta f_\beta(x)
=
\Delta f(x)
+\frac{\partial\beta^+(x)}{\partial x}-\frac{\partial\beta^-(x)}{\partial x}
\\
&&
\Delta F_{\beta,\alpha}(x)=
\int_{x_{min}}^x\Delta f(x')d\mu(x')
-\Big(\beta^+(x_{min})-\beta^-(x_{min})\Big)
-\int_{x_{min}}^x dx' \Big(\alpha^+(x')-\alpha^-(x')\Big)d\mu(x')
.
\nonumber
\end{eqnarray}
We then can deduce from eqs. (\ref{eq:proofs_3t})-(\ref{eq:proofs_5b_2})
\begin{eqnarray}
\label{eq:proofs_5_2)}
&&
{\forall x\in X \backslash X_{null}^{(\Delta f_\beta)} \mathrm{\hspace{1.5mm}with\hspace{1.5mm}} \Delta f_\beta(x)>0:}
\quad
\alpha^+(x)=\Delta f_\beta(x),\alpha^-(x)=0
,\varphi_{max}(x)=\lambda
\\
&&
{\hspace{3.4cm} \Delta f_\beta(x)<0:}
\quad
\alpha^-(x)=-\Delta f_\beta(x),\alpha^+(x)=0
,\varphi_{max}(x)=-\lambda
\nonumber\\
&&
{\forall x\in X \backslash X_{null}^{(\Delta F_{\beta,\alpha})} \mathrm{\hspace{1.5mm}with\hspace{1.5mm}} \Delta F_{\beta,\alpha}(x)<0:}
\quad
\beta^+(x)=-\Delta F_{\beta,\alpha}(x),\beta^-(x)=0
,\frac{\partial\varphi_{max}(x)}{\partial x}=1
\nonumber\\
%\label{eq:proofs_5b}\\
&&
{\hspace{3.7cm} \Delta F_{\beta,\alpha}(x)>0:}
\quad
\beta^-(x)=\Delta F_{\beta,\alpha}(x),\beta^+(x)=0
,\frac{\partial\varphi_{max}(x)}{\partial x}=-1
.
\nonumber
\end{eqnarray}
These equations define $\varphi_{max}$ quite similarly than in eqs. (\ref{eq:proofs_5}) and (\ref{eq:proofs_5b_bis}),
but they are here explicitly coupled.
An implicit coupling is also imposed through the necessary (strong) continuity of $\varphi_{max}$ and the fact that none of the constraints can saturate simultaneously,
thus that
\begin{eqnarray}
X_{null}=X_{null}^{(\Delta f_\beta)}\cup  X_{null}^{(\Delta F_{\beta,\alpha})}
\quad\mathrm{and}\quad
\varnothing=X_{null}^{(\Delta f_\beta)}\cap   X_{null}^{(\Delta F_{\beta,\alpha})}
.
\end{eqnarray}

Obviously, eq. (\ref{eq:proofs_4t_2}) (combined with eq. (\ref{eq:proofs_3t})) or eq. (\ref{eq:proofs_5_2)}) remain  stable under infinitesimal perturbations of $\Delta f$ in $X \backslash X_{null}^{(\Delta f)}$.

\subsection{Full case (2D coordinate space)}

We consider $X=[H^{min}_{inl},H^{max}_{inl}]\times[0,T]$ with $x=(x_{inl},x_t)^t$,
the results of \ref{app:proofs_0} with non-unit standard-deviation-like weights
and $d\rightarrow ||.||_1^{(X)}$. 
Eq. (\ref{eq:proofs_4t_2}) becomes, with obvious notations
\begin{eqnarray}
\label{eq:proofs_4t}
\Delta f(x)
&&
+\sigma_{inl}\Big( \delta(x_{inl}-H^{min}_{inl}) - \delta(x_{inl}-H^{max}_{inl})\Big)\Big( \beta^+_{inl}(x)-\beta^-_{inl}(x) \Big)
\\
&&
+\sigma_t \Big(\delta(x_t)-\delta(x_t-T) \Big)\Big( \beta^+_t(x)-\beta^-_t(x) \Big)
\nonumber\\
&&
=
\alpha^+(x)-\alpha^-(x)
-\sigma_{inl} \Big(\frac{\partial \beta^+_{inl}(x)}{\partial x_{inl}}-\frac{\partial \beta^-_{inl}(x)}{\partial x_{inl}}\Big)
-\sigma_t \Big(\frac{\partial \beta^+_t(x)}{\partial x_t}-\frac{\partial \beta^-_t(x)}{\partial x_t}\Big)
.
\nonumber
\end{eqnarray}
Combining eq. (\ref{eq:proofs_4t})  with the equivalent of eq. (\ref{eq:proofs_3t}) in the 2D coordinate space case,
we can deduce that always one of the constraints must saturate, almost everywhere in $X \backslash X_{null}^{(\Delta f)}$ and that eq. (\ref{eq:proofs_4t}) remains stable under infinitesimal perturbations of $\Delta f$ in $X \backslash X_{null}^{(\Delta f)}$.
The only subtlety is that, at a given position in $X \backslash X_{null}^{(\Delta f)}$ where the thresholding constraint does not saturate, only one of the Lipschitz constraints in the inline or time directions need to saturate, not necesarilly both.

 \newpage
 \clearpage
\bibliographystyle{apalike}
\bibliography{biblio_w}

\end{document}